\documentclass[12pt]{article}

\usepackage{jheppub} 
                     
\usepackage{slashed,bm,bbm}

\usepackage[T1]{fontenc} 

\usepackage{blkarray}

\usepackage[svgnames]{xcolor}

\newcommand{\LL}{\mathsf{L}}

\usepackage{tikz}
\usetikzlibrary{decorations.pathmorphing,arrows.meta,bending}

\def\nn{\nonumber\\ }
\def\rd{{\rm d}}
\def\abs#1{\left| #1 \right| }

\newcommand{\ba}[1]{\begin{align} #1 \end{align} }
\newcommand{\bs}[1]{\begin{split} #1 \end{split} }
\newcommand{\VEV}[1]{\left\langle #1 \right\rangle }

\def\tr{{\text{Tr}}}

\def\CD{{\cal D}}
\def\CJ{{\cal J}}
\def\CL{{\cal L}}
\def\CO{{\cal O}}

\def\cwp{Coleman-Weinberg potential}
\usepackage[normalem]{ulem}
\usepackage{caption}
\usepackage{subcaption}

\title{Renormalization Group Improvement of the Effective Potential: an EFT Approach
}

\author[a]{Aneesh V.~Manohar}
\author[b]{and Emily Nardoni}

\affiliation[a]{Department of Physics, University of California, San Diego, 9500 Gilman Drive,\\ La Jolla, CA 92093-0319, USA}
\affiliation[b]{Mani L.\ Bhaumik Institute for Theoretical Physics, Department of Physics and Astronomy, University of California, Los Angeles, CA 90095, USA}

\abstract{ 
We apply effective field theory (EFT) methods to compute the renormalization group improved effective potential for theories with a large mass hierarchy. Our method allows one to compute the effective potential in a systematic expansion in powers of the mass ratio, as well as to sum large logarithms of mass ratios using renormalization group evolution. The effective potential is the sum of one-particle irreducible diagrams (1PI) but information about which diagrams are 1PI is lost after matching to the EFT, since heavy lines get shrunk to a point. We therefore introduce a tadpole condition in place of the 1PI condition, and use the renormalization group improved value of the tadpole in computing the effective potential. We  explain why the effective potential computed using an EFT is not the same as the effective potential of the EFT.
We illustrate our method using the $O(N)$ model, a theory of two scalars in the unbroken and broken phases, and the Higgs-Yukawa model. Our leading-log result, obtained by integrating the one-loop $\beta$-functions, correctly reproduces the log-squared term in explicit two-loop calculations. Our method does not have a Goldstone boson infrared divergence problem.
}

\begin{document} 
\maketitle

\section{Introduction}

The effective potential $V(\phi)$ plays an important role in determining the vacuum of a quantum field theory. In a foundational paper~\cite{Coleman:1973jx}, Coleman and Weinberg computed the one-loop effective potential in a generic gauge theory and showed that in scalar QED,  for a range of parameters, the ground state of the theory is shifted from the classical vacuum due to radiative corrections. They also computed the renormalization group (RG) improved effective potential, obtained by RG evolution of the couplings.

In general, in a theory with multiple mass scales it is not possible to choose a single renormalization scale that removes all large logarithms in the effective potential. In this paper, we compute the renormalization group improved effective potential using techniques from effective field theory (EFT). 
In a theory with multiple scales, EFTs allow one to compute in an expansion in powers of the ratio of scales, $z \sim m_L/m_H \ll 1$, as well as to sum the  leading-log (LL)  renormalization group series $(\lambda \ln z)^n$, the next-to-leading-log series (NLL) $\lambda (\lambda \ln z)^n$,  etc. Here $m_L$ and $m_H$ are generic light and heavy scales, and $\lambda$ is a generic coupling constant. 
We will refer to the original theory whose effective potential is to be computed as the high-energy theory, and the theory given by integrating out $m_H$ as the low-energy theory, or EFT. We use the terms \cwp\ and effective potential interchangeably.

We apply our method to the $O(N)$ model in the broken phase, a theory with two scalar fields with widely separated vacuum expectation values in the unbroken and broken phases, and the Higgs-Yukawa model. There are several subtleties which are illustrated by these examples. The computation of the effective potential differs from usual EFT calculations in several important respects:
	\begin{itemize}
	\item The expectation value $\VEV{\phi}$ of a light field in the high energy theory is \emph{not} the same as  the expectation value  $\VEV{\phi}$ in the
	low-energy theory, because of matching corrections. Even though the same symbol $\phi$ is used, the two fields are not the same.
	\item As a result, the \cwp\ computed \emph{using} an EFT is not the same as computing the \cwp\ \emph{of} the EFT.  The distinction will become clear when we discuss the examples.
	\item The effective potential is obtained by computing one-particle irreducible (1PI) graphs in the high-energy theory. When heavy particles are integrated out, information about which graphs are one-particle irreducible is lost in the low-energy theory, since heavy particle lines are shrunk to a point.
We show that a vanishing tadpole condition in the high-energy theory is equivalent to only retaining 1PI graphs. We explain how  to match the tadpole condition to the EFT, and compute its RG improved value.
	\end{itemize}
Our method shows how to deal with these subtleties. We show that our results for power corrections and for the $(\lambda \ln z)^2$ terms in the LL series differ from those obtained in earlier attempts at RG improvement.
Our result at leading-log order is given by tree-level matching and one-loop running,  and can be compared with the two-loop (fixed order) computation of the \cwp~\cite{Ford:1991hw,Ford:1992pn,Martin:2001vx}. Our method 
correctly reproduces the $(\lambda \ln z)^2$ term in the two-loop effective potential by integrating the one-loop renormalization group equations (RGE). Importantly, there are tree-level matching corrections to the low-energy theory and tadpole contributions that have to be included to get the correct result. 

Our work departs from previous literature on the RG improvement of the effective potential in cases with multiple mass scales. 
One method which was proposed in Ref.~\cite{Einhorn:1983fc} introduces multiple independent renormalization scales $\mu$, such that the effective potential satisfies multiple renormalization group equations. This method was modified in Ref.~\cite{Ford:1996hd,Ford:1996yc}, and more recently in  Ref.~\cite{Steele:2014dsa}. 
Broadly, the difficulties of this approach are that it is hard to solve the multiple RGE at once, and also that it generically introduces logarithms of $m_L/m_H$ in the $\beta$-functions, which means that it does not allow one to sum all the higher-order logarithms using RG evolution.

An EFT inspired approach was first applied to this problem in Ref.~\cite{Bando:1992wy}, which studied the Higgs-Yukawa model that we consider in Sec.~\ref{sec:higgs-yukawa}.  A modified version of this approach was advocated in Ref.~\cite{Casas:1998cf}, which we discuss in more detail in Section~\ref{sec:comparison}. This method was generalized in  Ref.~\cite{Iso:2018aoa}, which applied it to the two-scalar case model discussed in Sec.~\ref{sec:unbroken}, and to the Higgs-Yukawa model. 
A recent attempt at applying the methods of Ref.~\cite{Bando:1992wy} to the two-scalar model is given in Ref.~\cite{Ookane:2019iwq}.
In Sec.~\ref{sec:comparison} we compare our method in various examples to these earlier results. The \cwp\ and RG improvement in curved spacetimes has been studied in Refs.~\cite{Buchbinder:2019bcc,Ribeiro:2019xgu}. The methods in this paper also apply to  curved spacetime calculations.

Other references using different approaches to those mentioned above are Ref.~\cite{Ford:1994dt} --- which conjectured  improved boundary conditions that summed the leading log series in some examples, but which gave no general prescription, and Ref.~\cite{Nakano:1993jq} --- which used a mass-dependent renormalization scheme. Ref.~\cite{Chataignier:2018aud} uses unmodified RGE with particular boundary conditions to resum logarithmic contributions at a given loop order (but not the leading logarithmic series). We will not comment further on these different approaches here. 

The outline of this paper is as follows. 
Section~\ref{sec:cwp} is a review of the \cwp, and Jackiw's functional integral method for computing it \cite{Jackiw:1974cv}. We show that only retaining 1PI diagrams is equivalent to a vanishing tadpole condition which can be implemented in the EFT. We discuss the relation between the one-loop \cwp\ and one-loop $\beta$-functions, and show how shift invariance leads to strong constraints on the form of the $\beta$-functions. We  show how the results in this section can be used to compute the renormalization of higher dimension operators, such as the $(H^\dagger H)^6$ term in the SMEFT Lagrangian.

We then illustrate our method by a sequence of examples, each of which illustrates new aspects of the method.
  In all cases, we compare with the explicit two-loop calculation, and show that our method correctly reproduces the two-loop term in the LL series. We start in Section~\ref{sec:on} by considering the $O(N)$ model in the broken phase. After introducing the power counting relevant to this first example,  we present the step-by-step guide to our procedure that is implemented in all the cases in Section~\ref{sec:overview}. A summary of these steps appears in Figure~\ref{fig:flowchart} of that section. We then  compute the RG improved \cwp\ for this case, and show how to implement the tadpole condition in the EFT.
  
Section~\ref{sec:unbroken} discusses the two-scalar model in the unbroken phase. This model in particular has off-diagonal terms in the mass matrix, and non-trivial tree-level matching contributions. 
Section~\ref{sec:broken} analyzes this model in the broken phase, with two widely separated vacuum expectation values (VEVs).
This example is closest to more realistic theories, such as a unified theory which has two VEVs which are widely separated --- one at the unification scale $\sim 10^{16}$\,GeV, and  the other at the electroweak scale $ \sim 246$\,GeV. We give plots of the RG improved \cwp\ and show the impact of RG summation and the tadpole contribution. 

Section~\ref{sec:higgs-yukawa} discusses the Higgs-Yukawa model, which features non-scalar fields and wave-function renormalization at one-loop. We analyze both the case where the boson is much heavier than the fermion(s), and the case where the boson is much lighter than the fermion(s).
 Sections~\ref{sec:on} and~\ref{sec:unbroken} contain most of the discussion necessary to understand the subtleties of our methods, and so our analysis in Sections~\ref{sec:broken} and~\ref{sec:higgs-yukawa} is more brief. 
Section~\ref{sec:comparison}  gives a comparison of our results with previous work. The $\beta$-functions for some common theories are given in the Appendices, along with solutions to the RGEs which are needed for our examples.

\section{The Coleman-Weinberg Potential}\label{sec:cwp}

This section reviews the classic computation of the effective potential and defines notation. 
After reminding the reader of the definition of the effective potential, we review in some detail Jackiw's method for computing the effective action in terms of one-particle-irreducible (1PI) graphs. We show that only computing 1PI graphs is equivalent to a vanishing tadpole condition. We then discuss the 1-loop renormalization group properties satisfied by the effective potential, and review the method of RG improving the effective potential in a problem with a single mass scale, which was given in the original paper~\cite{Coleman:1973jx}. Throughout we will highlight properties of the effective potential which are useful for the EFT computations in later sections.

\subsection{QFT 101}
\label{sec:qft101}

For the purposes of this section, let $\varphi^a$ denote a set of scalar fields, although we often suppress $a$ indices. The partition function is
	\begin{align}
	Z[J] &= e^{\frac{i}{\hbar} W[J]} = \int  \prod_a  D\varphi^a\ e^{\frac{i}{\hbar} \left( S(\varphi) + \int \rd^4x J_a \varphi^a\right)     }\  , \qquad   S(\varphi) = \int \rd^4x \ \CL(\varphi)\ . \label{eq:z}
	\end{align}
	This defines the connected generating functional $W[J]$, obtained by performing the path integral over the fields in the presence of  external sources $J_a$.

Define the classical background field $\hat{\varphi}$ as the expectation value of the field $\varphi$ in the presence of the current $J$. The field $\hat{\varphi}$ is the conjugate variable to $J(x)$,
\begin{align}
\hat{\varphi}(x) =  \VEV{\varphi(x)}_J = \frac{1}{Z[J]} \int D \varphi \ \varphi(x) \ e^{\frac{i}{\hbar} \left( S(\varphi) + \int \rd^4x J\varphi \right)}= \frac{\delta W[J]}{\delta J(x)}\ . \label{eq:background}
\end{align}
When $J=0$, $\hat{\varphi}(x)|_{J=0}\equiv \hat{\varphi}_0$ is the expectation value of $\varphi$ in the true vacuum of the theory.
The 1PI effective action is the Legendre transform of $W[J]$, 
	\begin{align}
	\Gamma[\hat{\varphi}] = W[J] - \int \rd^4x\ J(x) \hat{\varphi}(x)\ ,  \label{eq:effectiveaction}
	\end{align}
	and the effective potential is the first term in a derivative expansion of $\Gamma[\hat{\varphi}] $, i.e.\ the value of $\Gamma[\hat{\varphi}] $ for constant external fields,
	\begin{align}
	\Gamma[\hat{\varphi}]  &= -\int \rd^4x\ V_{\text{CW}} (\hat{\varphi}) + \int \rd^4 x \ \frac12 Z(\hat \varphi)  (\partial_\mu  \hat \varphi)^2 + \ldots \label{eq:effectiveaction2}
	\end{align}
Throughout we name the effective potential $V_{\text{CW}}$. 
The effective action $\Gamma[\hat{\varphi}]$ is  given by the sum of 1PI graphs. It is an extensive quantity, proportional to the volume of spacetime.
In particular, note that
	\begin{align}
	\frac{\delta \Gamma[\hat{\varphi}]}{\delta \hat{\varphi}(x)} = - J(x)\quad\Rightarrow\quad 	\frac{\delta \Gamma[\hat{\varphi}]}{\delta \hat{\varphi}(x)}\bigg|_{J=0} = 0\ . \label{eq:variation}
	\end{align}
From now on, we assume that $\hat{\varphi}$ is independent of $x$.  Then, we can remove the $\dots$ from \eqref{eq:effectiveaction2} and pull  $V_{\text{CW}} (\hat{\varphi})$ out of the integral, leaving an overall factor of spacetime volume. Equation~\eqref{eq:variation} implies
	\begin{align}
	 \frac{\partial V_{\text{CW}}(\hat{\varphi})}{\partial \hat{\varphi}} \bigg|_{\hat{\varphi}_0 }=0\ . \label{eq:og}
	\end{align}
The value of $\hat{\varphi}$ for which the minimum occurs is the expectation value of $\varphi$ in the true vacuum, and the value of $V_{\text{CW}}(\hat{\varphi}_0)$ gives the vacuum energy density in the ground state.

\subsection{Jackiw's Method}
\label{sec:jackiw}

We summarize here Jackiw's method for  computing the effective potential~\cite{Jackiw:1974cv}. We present the results in a manner that makes clear why the effective potential is the sum of 1PI diagrams, and also show how to adapt Jackiw's method to an EFT.

The first step is to shift the scalar field by its background value $\varphi (x)= \hat{\varphi}(x)+\varphi_q(x)$. The functional integral is now over the ``quantum field'' $\varphi_q$ since the functional integral measure is invariant under shifts, ${D} \varphi = {D} \varphi_q$. Requiring $\VEV{\varphi} = \hat{\varphi}$ is equivalent to requiring $\VEV{\varphi_q}=0$.

The effective action $\Gamma[\hat{\varphi}]$ is computed from the functional integral Eq.~\eqref{eq:z} by subtracting $J \hat\varphi$ from $W[J]$, using Eq.~\eqref{eq:effectiveaction}.  Since the functional integral is over $\varphi_q$, and $J$ and $\hat \varphi$ are independent of $\varphi_q$, the term $J  \hat \varphi$ can be taken out of the functional integral,
	\begin{align}
	Z[J] &= e^{\frac{i}{\hbar} W[J]} = e^{\frac{i}{\hbar} \int \rd^4 x J \hat \varphi }\int  \prod_a  {D}\varphi_q^a\ e^{\frac{i}{\hbar} \int \rd^4 x \left[ {\CL}(\hat{\varphi} + \varphi_q ) + J  \varphi_q  \right]    } \,, \label{eq:z1}
	\end{align}
so that
	\begin{align}
	e^{\frac{i}{\hbar} \Gamma[ \hat \varphi ] } &= \int  \prod_a  D\varphi_q^a\ e^{\frac{i}{\hbar} \int \rd^4 x \left[ {\CL}( \hat{\varphi} +\varphi_q ) + J  \varphi_q  \right]    }  
	= e^{\frac{i}{\hbar}\left( W[J]-  \int \rd^4 x  J  \hat \varphi  \right)   }  
	\,. \label{eq:z2}
	\end{align}

The shifted Lagrangian can be expanded in powers of $\varphi_q$,
	\begin{align}
 {\CL}( \hat{\varphi} +\varphi_q )   &= - \hat{\Lambda} (\hat \varphi)	 - \hat{\sigma}(\hat \varphi)	 \varphi_q + \frac12 i \CD^{-1}(\hat{\varphi}) \hat \varphi_q^2 
 - \frac16 \hat{\rho} (\hat \varphi) \varphi_q^3 - \frac{1}{24}\hat{\lambda}(\hat \varphi) \varphi_q^4 + \ldots \label{eq:lshift}	
 \end{align}
The couplings without hats are couplings before the shift, and couplings with hats are those after the shift. Eq.~\eqref{eq:lshift} defines notation we will use in the rest of this paper for the couplings in the shifted theory.
 If the theory is renormalizable, the series terminates at order $\varphi_q^4$.  We have defined $\hat{\Lambda}(\hat{\varphi})$ as (minus) the constant term in this expansion, and it is equal to the potential evaluated on the background fields plus the cosmological constant,
	\begin{align}
	\hat{\Lambda}(\hat{\varphi})=  V(\hat{\varphi}) + \Lambda\ . \label{eq:cc1}
	\end{align}
Following Ref.~\cite{Coleman:1973jx}, the matrix of second derivatives of the scalar potential is
	\begin{align}
	W_{ab} (\hat{\varphi}) = \frac{\partial^2 V(\hat{\varphi})}{\partial \varphi^a \partial \varphi^b}\ , \quad \text{with eigenvalues}\ w_i(\hat{\varphi})\ . \label{eq:w}
	\end{align}
	If the kinetic terms are diagonal, the inverse propagator is $i\CD^{-1}_{ab} = - \partial^2 \delta_{ab}  - W_{ab}$. 
	
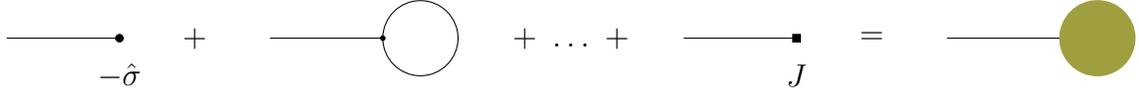
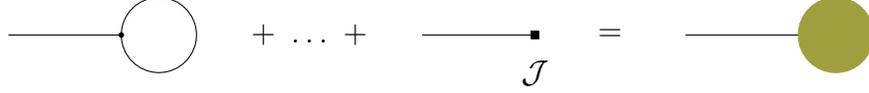
\begin{figure}
	\centering
	\begin{subfigure}[b]{1\textwidth}
	\centering
	\begin{tikzpicture}
	\draw (-0.5,0) -- (1,0) ;
	\filldraw (1,0) circle (0.05);
	\node at (1,-0.5) [align=center] {$-\hat{\sigma}$};
	\node at (2,0) [align=center] {$+$};
	\begin{scope}[shift={(3.5,0)}]
	 \draw (-0.5,0) -- (1,0);
	 \draw (1.5,0) circle (0.5);
	\filldraw (1,0) circle (0.03);
	\node at (3.5,0) [align=center] {$+ \ \ldots \ + $};
	 \end{scope}
	\begin{scope}[shift={(9,0)}]
	 \draw (-0.5,0) -- (1,0);
	\filldraw (0.95,-0.05) rectangle (1.05,0.05);
	\node at (1,-0.5) [align=center] {$J$};
	\node at (2,0) [align=center] {$=$};
	\end{scope}
	\begin{scope}[shift={(12.5,0)}]
	\draw (-0.5,0) -- (1,0);
	\filldraw[Olive!75] (1.5,0) circle (0.5);
	 \end{scope}
	\end{tikzpicture}
	\caption{The tadpole graphs, including those from the linear term in $\mathcal{L}(\varphi_q+\hat \varphi)$ and from $J$.}
	\vspace{0.3cm}
	\label{fig:tadpole} 
	\end{subfigure}
	\begin{subfigure}[b]{1\textwidth}
	\centering
	\begin{tikzpicture}
	\begin{scope}[shift={(3.5,0)}]
	 \draw (-0.5,0) -- (1,0);
	 \draw (1.5,0) circle (0.5);
	\filldraw (1,0) circle (0.03);
	\node at (3.5,0) [align=center] {$+ \ \ldots \ + $};
	 \end{scope}
	\begin{scope}[shift={(9,0)}]
	 \draw (-0.5,0) -- (1,0);
	\filldraw (0.95,-0.05) rectangle (1.05,0.05);
	\node at (1,-0.5) [align=center] {$\CJ$};
	\node at (2,0) [align=center] {$=$};
	\end{scope}
	\begin{scope}[shift={(12.5,0)}]
	\draw (-0.5,0) -- (1,0);
	\filldraw[Olive!75] (1.5,0) circle (0.5);
	 \end{scope}
	\end{tikzpicture}
	\caption{The tadpole graphs after dropping $\hat{\sigma}$ and replacing $J$ by $\CJ$.}
	\label{fig:2} 
	\end{subfigure}
	\caption{$\langle \varphi \rangle_q$ is given by the sum of all tadpole graphs. Regardless of how the source $J$ is decomposed, the total tadpole contribution vanishes.}
	\label{fig:tadpole1}
	\end{figure}

 The functional integral Eq.~\eqref{eq:z2} is given by the sum of vacuum graphs of the theory, with the subsidiary condition that $\VEV{\varphi_q}=0$. Diagrammatically, $\VEV{\varphi_q}$ is the sum of tadpole graphs, Fig.~\ref{fig:tadpole}.
The graphs include the linear term $ \hat{\sigma} \varphi_q$ in the shifted Lagrangian \eqref{eq:lshift}, as well the source $J$. The source is adjusted to make the total tadpole vanish.
 At tree-level, this is accomplished by simply setting $J=\hat{\sigma}$. Let us implement this tree-level condition by splitting
 	\begin{align}
	J = \hat{\sigma} + \CJ \ , \label{eq:j}
	\end{align}
	where $\CJ$ is to be formally treated as starting at one-loop order. Eq.~\eqref{eq:z2} can then be written as
	\begin{align}
	e^{\frac{i}{\hbar} \Gamma[ \hat \varphi ] } &= \int  \prod_a  D\varphi^a\ e^{\frac{i}{\hbar} \int \rd^4 x  \hat{\CL}  } \, .\label{eq:z3}
	\end{align}
We have defined $\hat{\CL}$ in the exponent in Eq.~\eqref{eq:z3} as the shifted Lagrangian with the linear term dropped, plus $\CJ \varphi_q$:
	\begin{align} 
	\begin{split}
	\hat{\CL} &\equiv {\CL}(\varphi_q +\hat{\varphi} ) + \hat{\sigma}(\hat \varphi)\varphi_q + \CJ  \varphi_q \\
	&=- \hat{\Lambda} (\hat \varphi)	+ \CJ \varphi_q+ \frac12 i \CD^{-1}(\hat{\varphi}) \hat \varphi_q^2 
 - \frac16 \hat{\rho} (\hat \varphi) \varphi_q^3 - \frac{1}{24} \hat{\lambda}(\hat \varphi) \varphi_q^4   + \ldots 
	\end{split}
	\label{eq:cl}
	\end{align}
The tadpole graphs are thus {modified} as shown in Fig.~\ref{fig:2}, with $J$ replaced by $\CJ$, and the $\hat{\sigma}$ graph dropped.
	Equation \eqref{eq:z3} is the expression derived by Jackiw~\cite{Jackiw:1974cv}.

	Let us set aside the subtleties with the source $\CJ$ for the moment, and give 
the effective potential to one-loop order. Performing the path integral over $\varphi_q$, we obtain 
	\begin{align}
	V_{\text{CW}} (\hat{\varphi})= \hat{\Lambda}(\hat{\varphi}) + \frac{i \hbar}{2} \tr \ln i \CD^{-1} (\hat{\varphi})+ \CO(\hbar)^2 \ .
	\end{align}
	The tree level piece is just the potential in the Lagrangian evaluated on the background fields plus the cosmological constant, Eq.~\eqref{eq:cc1}. 
The one-loop contribution in the $\overline{\text{MS}}$ scheme including scalar, fermion and gauge loops is
	\begin{align}
	\begin{split}
	V_{\text{CW}}(\hat{\varphi}) &= V_{\text{tree}}(\hat{\varphi}) + \hbar V_{1\text{-loop}}(\hat{\varphi}) + \CO(\hbar^2)\ \\
	V_{\text{tree}} &= \hat{\Lambda}(\hat{\varphi}) = V(\hat{\varphi}) + \Lambda(\hat{\varphi})\ \\
	V_{1\text{-loop}} &= \frac{1}{64 \pi^2} \biggl\{  \tr\, W^2 \left[ \ln \frac{W}{\mu^2} - \frac32 \right] - 2 \, \tr\, \left( M_F^\dagger M_F \right)^2  \left[ \ln \frac{M_F^\dagger M_F }{\mu^2} - \frac32 \right] \\
&  \hspace{1cm} +  3 \tr\, M_V^4 \left[ \ln \frac{M_V^2}{\mu^2} - \frac56 \right] \biggr\}\ ,
	\end{split} \label{2.1} 
	\end{align}
where $W(\hat \varphi) \equiv W_{ab}(\hat \varphi)$  defined in Eq.~\eqref{eq:w}, $M_F$ and $M_V^2$ are the mass matrices for real scalar fields, Weyl fermions, and gauge bosons, respectively, and the traces are over flavor indices. To reiterate, the mass matrices are computed in a background external field, and so depend on $\hat{\varphi}$. 
	
\begin{figure}
\begin{center}
\begin{tikzpicture}
\draw (0,0) circle (0.75);
\end{tikzpicture}
\end{center}
\caption{\label{fig:3} One-loop contribution to the effective action. }
\end{figure}
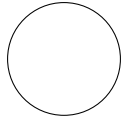
	
\begin{figure}
\centering
	\begin{subfigure}[b]{1\textwidth}
	\centering
	\begin{tikzpicture}
	\begin{scope}[shift={(3,0)}]
	\draw (0,0) circle (0.75);
	\draw (-0.75,0) -- (0.75,0);
	\filldraw (-0.75,0) circle (0.03);
	\filldraw (0.75,0) circle (0.03);
	 \end{scope}
	\begin{scope}[shift={(5.5,0)}]
	\draw (0,0.5) circle (0.5);
	\draw (0,-0.5) circle (0.5);
	\filldraw (0,0) circle (0.03);
	 \end{scope}
	\begin{scope}[shift={(8.5,0)}]
	 \draw (-0.5,0) -- (0.5,0);
	 \draw (1,0) circle (0.5);
	 \draw (-1,0) circle (0.5);
	 \filldraw (-0.5,0) circle (0.03);
	 \filldraw (0.5,0) circle (0.03);
	 \end{scope}
	\begin{scope}[shift={(12.5,0)}]
	 \draw (-0.5,0) -- (0.5,0);
	 \draw (-1,0) circle (0.5);
	  \filldraw (-0.5,0) circle (0.03);
	\filldraw (0.45,-0.05) rectangle (0.55,0.05);
	 \end{scope}
	\begin{scope}[shift={(14.5,0)}]
	 \draw (-0.5,0) -- (0.5,0);
	 \draw (1,0) circle (0.5);
	  \filldraw (0.5,0) circle (0.03);
	\filldraw (-0.45,-0.05) rectangle (-0.55,0.05);
	\end{scope}
	\begin{scope}[shift={(17,0)}]
	 \draw (-0.5,0) -- (0.5,0);
	\filldraw (0.45,-0.05) rectangle (0.55,0.05);
	\filldraw (-0.45,-0.05) rectangle (-0.55,0.05);
	\end{scope}
	\end{tikzpicture}
	\caption{Two loop contribution to the effective action including the source term $\CJ$, which is formally of one-loop order.  }\vspace{0.3cm}
	\label{fig:4} 
\end{subfigure}	
\begin{subfigure}[b]{1\textwidth}
	\centering
	\begin{tikzpicture}
	\begin{scope}[shift={(3,0)}]
	\draw (0,0) circle (0.75);
	\draw (-0.75,0) -- (0.75,0);
	\filldraw (-0.75,0) circle (0.03);
	\filldraw (0.75,0) circle (0.03);
	 \end{scope}
	\begin{scope}[shift={(5.5,0)}]
	\draw (0,0.5) circle (0.5);
	\draw (0,-0.5) circle (0.5);
	\filldraw (0,0) circle (0.03);
	 \end{scope}
	\begin{scope}[shift={(8.5,0)}]
	 \draw (-0.5,0) -- (0.5,0);
	 \filldraw[Olive!75] (1,0) circle (0.5);
	 \filldraw[Olive!75] (-1,0) circle (0.5);
	 \end{scope}
	\end{tikzpicture}
	\caption{The same graphs as in (a), but using instead the total tadpole contribution (see Fig.~\ref{fig:tadpole1}).}
	\label{fig:5}
\end{subfigure}
\caption{The two-loop contribution to the effective action. The black rectangles are the $\CJ$ vertex, and are formally of one-loop order. The shaded blob is  the total tadpole contribution.}
\label{fig:f45}
\end{figure}
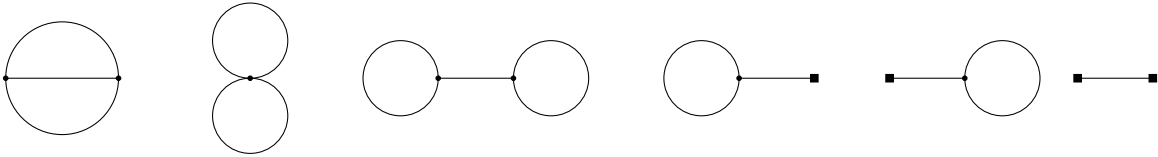
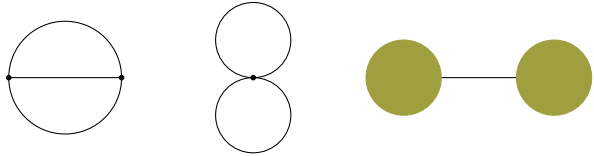

\begin{figure}
\begin{center}

\begin{tikzpicture}[
blob/.style={circle,draw=black!50,fill=black!50,minimum size = 0.75cm},
tad/.style={circle,draw=Olive!75,fill=Olive!75,minimum size = 0.75cm},
scale=0.5
]
\begin{scope}[rotate=90]
\node[rotate=90] at (90:4.5) (four) [tad] {};
\node[rotate=90] at (-30:2) (one) [blob] {};
\node[rotate=90] at (90:2) (two) [blob] {};
\node[rotate=90] at (210:2) (three) [blob] {};
\draw [line width=0.2mm] (one) to [out=90,in=-30] (two);
\draw [line width=0.2mm] (two) to [out=210,in=90] (three);
\draw [line width=0.2mm] (two) to [out=270,in=30] (three);
\draw [line width=0.2mm] (three) to [out=-30,in=210] (one);
\draw [line width=0.2mm] (two) to [out=90,in=-90] (four);
\end{scope}

\end{tikzpicture}

\end{center}
\caption{A one-particle reducible contribution to the effective action, which vanishes due to the tadpole constraint. \label{fig:6}}
\end{figure}
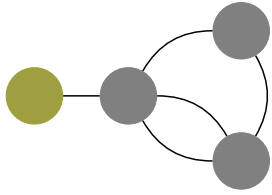

We now return to understanding the 1PI-nature of $V_{\text{CW}}$.   Graphically, the functional integral Eq.~\eqref{eq:z3} is the sum of vacuum graphs of $\varphi_q$ with vertices which depend on {$\hat \varphi$}. The action starts at quadratic order in $\varphi_q$. The one-loop and two-loop graphs are shown in Fig.~\ref{fig:3} and Fig.~\ref{fig:f45}, respectively. $\CJ$ is counted as being one-loop, and so the last three graphs in Fig.~\ref{fig:4} with  insertions of $\CJ$ are treated as two-loop graphs.
Comparing with the tadpole graphs Fig.~\ref{fig:2}, we see that the last four graphs in Fig.~\ref{fig:4} can be combined, so that the two-loop contribution is given diagrammatically as in Fig.~\ref{fig:5}.
By construction, the total tadpole contribution vanishes, so that $\VEV{\varphi_q}=0$. This means we can drop the last diagram in Fig.~\ref{fig:5}, leaving only the two \emph{one-particle irreducible graphs}, {one} with two insertions of the $\varphi^3_q$ interaction, and one with one insertion of the $\varphi_q^4$ interaction. The argument clearly holds at higher order. In general, one-particle reducible lines such as those in Fig.~\ref{fig:6} end in tadpoles, which vanish.
We thus obtain the result of Jackiw~\cite{Jackiw:1974cv}: 

\begin{quote}
To compute $\Gamma[\hat \varphi]$, shift $\varphi \to \hat \varphi + \varphi_q$, drop the terms linear in $\varphi_q$ in the action, and compute 1PI vacuum graphs.
\end{quote}

The use of an explicit source term $\CJ$ and a vanishing tadpole condition rather than summing 1PI graphs allows us to avoid the problem mentioned in the Introduction that 1PI graphs in the low-energy theory are not the same as 1PI graphs in the high-energy theory, since heavy particle lines have been shrunk to a point. The tadpole condition can be matched onto the EFT by constructing the EFT in the presence of a source $\CJ$, as shown in Sec.~\ref{sec:on}.
It is worth noting that $\CJ$ cancels the full tadpole, and so has both infinite $1/\epsilon^r$ and finite pieces.

\subsection{Effective Potential for a Subset of the Fields}
\label{sec:some}

In a theory with multiple scalar fields, we are sometimes interested in the effective action as a function of only some of the scalar fields. Divide the scalar fields into two groups, $\varphi$ and $\chi$. As reviewed in Section \ref{sec:qft101}, we can compute the effective action $\Gamma[\hat \varphi, \hat \chi]$ by introducing sources for $\varphi$ and $\chi$, and computing the Legendre transform. We can instead compute $\Gamma[\hat\varphi]$ by introducing sources and taking the Legendre transform only for $\varphi$, but not for $\chi$. How are  $\Gamma[\hat \varphi, \hat \chi]$ and $\Gamma[\hat\varphi]$ related? From the definition of the effective action in Eq.~\eqref{eq:z} -- \eqref{eq:effectiveaction}, one can easily see that $\Gamma[\hat\varphi]$ can be obtained by introducing sources for both $\varphi$ and $\chi$, with the source $J_\chi$ for $\chi$ set to zero. From Eq.~\eqref{eq:variation}, we see that this implies
\begin{align}
\Gamma[\hat \varphi] =   \Gamma[\hat \varphi,\hat \chi] \biggr|_{\frac{\partial \Gamma[\hat \varphi,\hat \chi] }{\partial \hat \chi} = 0}\,.
\label{2.17}
\end{align}
I.e.\ one finds the value of $\hat \chi$ that extremizes $ \Gamma[\hat \varphi,\hat \chi]$ for a fixed value of $\hat \varphi$, and substitutes this value (which is a function of $\hat \varphi$) back into $ \Gamma[\hat \varphi,\hat \chi]$. The \cwp\ in the theory with $\chi$ integrated out is $ \Gamma[\hat \varphi] $, and is related to the \cwp\ in the theory with both fields by Eq.~\eqref{2.17}.

\subsection{$\beta$-functions and One-loop Running}
\label{sec:running}

Let us now address the renormalization group running of the effective potential.\footnote{For a discussion of the RG improved potential in massive theories, see References~\cite{Kastening:1991gv,Ford:1992mv}.}
Implicit in the computations of Section \ref{sec:jackiw} is the fact that all computations are done in a renormalized theory. We use dimensional regularization in this paper, so all couplings are evaluated at a scale $\mu$. Since the scalar field can have an anomalous dimension, $\varphi$ implicitly also has a $\mu$ label $\varphi(\mu)$.

The all-orders potential obeys the renormalization group equation
\begin{align}
\left[  \frac{\partial}{\partial t} + \beta_i \frac{\partial}{\partial \lambda_i} - \gamma_\varphi \varphi \frac{\partial}{\partial \varphi} \right] V_{\text{CW}} &= 0\,.
\label{2.3}
\end{align}
The couplings $\lambda$, $m^2$, etc.\ are denoted generically as $\{\lambda_i\}$, and satisfy the $\beta$-function equations
\begin{align}
\frac{\rd \lambda_i}{\rd t} &= \beta_i( \{ \lambda_i \})\,,
\label{2.5}
\end{align}
where we have defined
\begin{align}
t &\equiv \frac{1}{16 \pi^2} \ln \frac{\mu}{\mu_0}\,,
\label{1.2}
\end{align}
and $\mu_0$ is any fixed reference scale. We will interchangeably use $\mu$ and $t$ as the coupling constant argument.
Using $t$ instead of $\ln \mu$ multiplies all the $\beta$-functions and $\gamma_\varphi$ by $16 \pi^2$ relative to the $\ln \mu$ derivatives. 
$\gamma_\varphi$ is the anomalous dimension of $\varphi$,
\begin{align}
\frac{\rd \varphi}{\rd t}  = - \gamma_\varphi\, \varphi\,.
\label{2.5a}
\end{align}

At one-loop order the total potential is $V_{\text{CW}}=V_{\text{tree}}+V_{1\text{-loop}}$, the $\beta$-functions start at one-loop order, and $V_{\text{tree}}$ has no explicit $\mu$-dependence, so Eq.~\eqref{2.3} reduces to
\begin{align}
& \left[ \beta_i \frac{\partial}{\partial \lambda_i} - \gamma_\varphi \varphi \frac{\partial}{\partial \varphi} \right] V_{\text{tree}}  +  \frac{\partial}{\partial t} V_{1\text{-loop}}
 = 0\, \nn
\implies & \left[ \beta_i \frac{\partial}{\partial \lambda_i} - \gamma_\varphi \varphi \frac{\partial}{\partial \varphi} \right] V_{\text{tree}}  
 = \frac12  \tr\, W^2 -  \, \tr\, \left( M_F^\dagger M_F \right)^2  +  \frac32 \tr\, M_V^4 \,.
\label{2.6}
\end{align}
From Eq.~\eqref{2.6}, we can immediately read off the one-loop $\beta$-functions for the terms in the scalar potential by matching powers of $\varphi$ on both sides of the equation. Eq.~\eqref{2.6} can also be applied to theories with interactions beyond dimension four, and we use it to derive the anomalous dimension for the SMEFT operator $(H^\dagger H)^3$ in Appendix~\ref{sec:smeft}.
 
As an example, consider the Higgs-Yukawa theory discussed in Sec.~\ref{sec:higgs-yukawa}, with Lagrangian
\begin{align}
\CL &= \frac12 (\partial_\mu \varphi)^2 + \sum_k i \, \overline \psi_k\,  \slashed{\partial}\, \psi_k - (m_F + g \varphi) \overline \psi_k \psi_k - V(\varphi) 
\label{2.2}
\end{align}
and scalar potential
\begin{align}
V(\varphi) &= \Lambda + \sigma \varphi + \frac{ m_B^2}{2} \varphi^2 + \frac{ \rho}{6} \varphi^3 + \frac{\lambda}{24}  \varphi^4 \,.
\label{2.4}
\end{align}
The sum on $k$ is over $N_F$ fermions. The theory has an exact $U(N_F)$ symmetry, and all couplings are taken to be real.
The couplings are $\{\lambda_i\}=\{ \Lambda,\sigma,M^2,\rho,\lambda,m,g\}$ and
\begin{align}
W(\varphi) &= {\frac{\partial^2 V}{\partial \varphi^2} } =  m_B^2 +  \rho \varphi + \frac{1}{2} \lambda \varphi^2\,, \qquad
M_F(\varphi) = m_F + g \varphi \,,
\label{2.7}
\end{align}
are the scalar and Dirac fermion mass matrices.
Eq.~\eqref{2.6} and Eq.~\eqref{2.1} give
\begin{align}
& \beta_\Lambda + \left(\beta_\sigma - \gamma_\varphi \right)  \varphi + \frac12 \left(\beta_{m_B^2} -2 \gamma_\varphi \right) \varphi^2 + \frac16 \left( \beta_{\rho}  -3 \gamma_\varphi \right) \varphi^3 + \frac{1}{24} \left( \beta_{\lambda} -4 \gamma_\varphi \right) \varphi^4 \nn
&=  \frac{1}{2} \left(m_B^2 +  \rho \varphi + \frac{1}{2} \lambda \varphi^2 \right)^2 - 2 N_F \left( m_F + g \varphi \right)^4 \,,
\label{2.15}
\end{align}
so that
\begin{align}
\begin{split}
\beta_\Lambda &= \frac12 m_B^4 - 2 N_F m_F^4 \,, \\
\beta_\sigma &=  m_B^2 \rho  -8 N_F g m_F^3 + \gamma_\varphi \sigma\,, \\
\beta_{m_B^2} & =\lambda   m_B^2 +  \rho^2 - 24 N_F g^2 m_F^2 + 2 \gamma_\varphi  m_B^2\,, \\
\beta_{\rho}  &= 3 \lambda   \rho -48 N_F g^3 m_F +  3 \gamma_\varphi  \rho\,, \\
\beta_{\lambda} &= 3 \lambda^2 -48 N_F g^4 + 4 \gamma_\varphi \lambda \,.
\end{split}
\label{2.8}
\end{align}
The anomalous dimension $\gamma_\varphi$ requires computing graphs with external momentum. In this theory, the only graph which contributes is shown in Fig.~\ref{fig:7}, which gives
\begin{figure}
\begin{center}
\begin{tikzpicture}[x=0.75cm,y=0.75cm]
\draw (0,0) circle (1);
\draw[dashed]  (-2.25,0) -- (-1,0);
\draw[dashed]  (1,0) -- (2.25,0);
\filldraw (-1,0) circle (0.05);
\filldraw (1,0) circle (0.05);
\end{tikzpicture}
\end{center}
\caption{\label{fig:7} One-loop wavefunction renormalization for $\varphi$. The solid line is the fermion $\psi_k$ and the dashed line is the scalar $\varphi$.}
\end{figure}
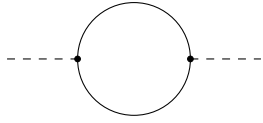
\begin{align}
\gamma_\varphi &= 2 g^2 N_F \,.
\label{2.9}
\end{align}
The $\beta$-functions for couplings containing fermions require computing one-loop graphs with external fermion fields, and are given as
\begin{align}
\gamma_\psi &= \frac12 g^2\,, \qquad
\beta_{m_F}  = 3 g^2 m_F\,, \qquad
\beta_{g}  = \left(3+2 N_F \right) g^3 \,.
\label{2.11}
\end{align}
We will need these expressions in Sec.~\ref{sec:higgs-yukawa}. The solution of the RG equations Eq.~\eqref{2.8} in the theory without fermions will be needed later, and is given in Appendix~\ref{app:soln}.

For the computation of the \cwp\ using the method in Sec.~\ref{sec:jackiw}, one also needs the Lagrangian after the shift $\varphi = \hat \varphi + \varphi_q$. The Higgs-Yukawa Lagrangian in Eq.~\eqref{2.4} retains the same form after this shift, but with new couplings $\hat m_F$, etc.\ which are functions of $\hat \varphi$, given by
\begin{align}
\begin{split}
\hat m_F &= m_F + g \hat \varphi \,,  \\
\hat g &= g \,, \\
\hat \Lambda &= \Lambda + \sigma \hat \varphi + \frac12 m_B^2 \hat \varphi^2 + \frac16 \rho \hat \varphi^3 + \frac1{24} \lambda \hat \varphi^4\,, \\
\hat \sigma &= \sigma + m_B^2 \hat \varphi + \frac12 \rho \hat \varphi^2 + \frac16 \lambda \hat \varphi^3 \,, \\
\hat m_B^2 &= m_B^2 + \rho \hat \varphi + \frac12 \lambda \hat \varphi^2 \,, \\
\hat \rho &= \rho + \lambda \hat \varphi \,, \\
\hat \lambda &= \lambda\,.
\end{split}
\label{2.12}
\end{align}
Since the shifted Lagrangian has the same form as the original one, the $\beta$-functions for the hatted couplings are given by the same functions Eq.~\eqref{2.8}--\eqref{2.11} as before, with all couplings replaced by their hatted values.

There is another way to think of the RG evolution of the hatted parameters in Eq.~\eqref{2.12}. Rather than treat (e.g.)  $\hat \Lambda$ as a single object, one can instead compute its RG evolution by evolving $\Lambda$, $\sigma$, etc.\ that appear on the r.h.s. of Eq.~\eqref{2.12}, and then computing $\hat \Lambda$ using the evolved couplings. One can think of this procedure as ``running and then shifting'' instead of  ``shifting and then running.'' If we also evolve $\hat \varphi$ according to its anomalous dimension then the two methods give the same answer, since one is simply making a linear shift in an integration variable in the functional integral. The ``shift and run'' vs. ``run and shift'' evolution will prove useful in comparing our results to those in the literature.

Shift invariance is a non-trivial constraint on RG evolution. As an amusing aside, consider the case of  Higgs-Yukawa theory. Suppose one assumes that the RG equations take the form of Eq.~\eqref{2.8}--\eqref{2.11} but with arbitrary coefficients for each term (except for the $\gamma_\phi$ and $\gamma_\psi$ dependence, which is fixed by the number of fields, and which we will also keep explicit). For example, suppose we guessed 
\begin{align}
\beta_{m_B^2} & = \widetilde c_1 \lambda   m_B^2 + \widetilde c_2 \rho^2  + \widetilde c_3  g^2 m_F^2 + 2 \gamma_\varphi  m_B^2 \,,
\label{2.30}
\end{align}
and similarly for the other $\beta$-functions.
In total this introduces 13 arbitrary coefficients $\widetilde{c}_i$, as well as two anomalous dimensions $\gamma_\phi$ and $\gamma_\psi$. Demanding shift invariance as defined in the previous paragraph severely constrains the RGE, and fixes 10 of the 13 coefficients. The resulting equations take the form
\renewcommand{\arraystretch}{1}
\begin{align}
& \frac{\rd}{\rd t} \begin{bmatrix} 
\phi \\ \psi \\ \Lambda \\ \sigma \\ m_B^2 \\ \rho \\ \lambda \\ m_F \\ g
\end{bmatrix}
&=c_1  \begin{bmatrix}
0 \\ 0 \\
\frac12 m_B^4   \\ 
m_B^2 \rho  \\ 
\lambda   m_B^2 + \rho^2 \\
3 \lambda   \rho  \\
3 \lambda^2 \\
0 \\
0
\end{bmatrix} +
c_2
\begin{bmatrix}
0 \\ 0 \\
 m_F^4  \\ 
4 g m_F^3 \\ 
12 g^2 m_F^2  \\
24 g^3 m_F  \\
24g^4  \\
0 \\
 0
\end{bmatrix} + c_3
\begin{bmatrix}
0 \\ 0  \\ 0 \\ 0 \\ 0 \\ 0 \\ 0 \\
g^2 m_F  \\
g^3
\end{bmatrix}
+
\begin{bmatrix}
\gamma_\phi \\ \gamma_\psi \\ 0 \\ \gamma_\phi \sigma \\ 2 \gamma_\phi m_B^2 \\ 3 \gamma_\phi \rho  \\ 4 \gamma_\phi  \lambda \\
2 \gamma_\psi m_F \\ \left( \gamma_\phi + 2 \gamma_\psi \right) g
\end{bmatrix}
\label{2.14}
\end{align}
There are only three independent coefficients $c_{1,2,3}$ in the RG equations, as well as the anomalous dimensions $\gamma_\phi$ and $\gamma_\psi$. The $c_1$ term is an overall normalization for the scalar sector;  the relative coefficients are completely fixed by shift invariance. The $c_2$ term is an overall factor for all  terms involving a fermion loop, and the $c_3$ term is an overall factor for  terms which arise from scalar corrections to the fermion sector. Note that we did not group the terms into these three categories; shift invariance automatically fixed all the relative coefficients within these three sectors. Then by explicit computation, one determines $c_1=1$, $c_2=-2N_F$, $c_3=2$.

\subsection{Summing Logarithms}
\label{sec:log}

We now review the method for computing the RG improved \cwp\  introduced in the original paper~\cite{Coleman:1973jx}. The discussion also explains which terms in the \cwp\ are computed in this paper. 

Let us begin with a single  
 real scalar field with $\phi \to -\phi$ symmetry,
\begin{align}
\CL &= \frac12 (\partial_\mu \phi)^2 -  \frac{m^2}{2} \phi^2 - \frac{\lambda}{24}  \phi^4 -\Lambda\,.
\label{4.1}
\end{align}
The \cwp\ to one-loop order is
\begin{align}
V_{\text{CW}}(\hat{\phi},\lambda(\mu_0),m(\mu_0),\mu_0) &=\Lambda +   \frac{m^2(\mu_0) }{2}\hat{\phi}^2 + \frac{\lambda(\mu_0) }{24} \hat{\phi}^4 + \frac{ W^2 }{64 \pi^2}
\left[ \ln \frac{W}{\mu_0^2} - \frac32 \right] \,,
\label{4.2}
\end{align}
where, as in Eq.~\eqref{eq:w}, we have identified
\begin{align}
W &= {\left. \frac{\partial^2 V}{\partial \phi^2} \right|_{\hat \phi}} = m^2(\mu_0) + \frac{\lambda(\mu_0)}{2} \hat{\phi}^2 
\label{4.3}
\end{align}
as the $1 \times 1$ matrix of second derivatives of the potential. This problem has a single scale, which is given by $W$.

Fixed order perturbation theory breaks down if $ \abs{\ln W/\mu_0^2} \gg 1$. The validity of perturbation theory can be restored by choosing a scale $\mu^2$ of order $W$, so that the logarithm is not large. Renormalization group invariance of the potential implies
\begin{align}
V_{\text{CW}}(\hat{\phi}(\mu_0),\lambda(\mu_0),m(\mu_0),\mu_0) &= V_{\text{CW}}(\hat{\phi}(\mu),\lambda(\mu),m(\mu),\mu) \,,
\label{4.4}
\end{align}
so that
\begin{align}
V_{\text{CW}}(\hat{\phi} ,\lambda(\mu_0),m(\mu_0),\mu_0) &= V_{\text{CW}}(e^{- \int_{\mu_0}^{\mu} \frac{\rd \mu^\prime}{\mu^\prime} \gamma_\phi(\mu^\prime) } \ \hat{\phi} ,\lambda(\mu),m(\mu),\mu) \,.
\label{4.5}
\end{align}
In this particular problem $\gamma_\phi=0$, but we will need the anomalous dimension scaling in other examples. By a suitable choice of $\mu$, one can compute the desired expression on the l.h.s.\ of Eq.~\eqref{4.5} by instead computing the r.h.s. 

Writing the couplings $\lambda(\mu)$ in terms of their original values $\lambda(\mu_0)$ as given explicitly in Appendix~\ref{app:soln}, one sees that the expression on the r.h.s.\ sums the leading-log series $(\lambda(\mu_0) \LL)^n$, where $\LL =(\ln W/\mu_0^2)/(16 \pi^2) $. Generically the structure of the perturbation series has the form
\begin{align}
V &=
\begin{blockarray}{cccc}
\text{LL} & \text{NLL} & \text{NNLL} \\[5pt]
\begin{block}{(ccc)c}
\framebox{1}  & & & \text{tree} \\[5pt]
\framebox{$\lambda \LL$} & \framebox{$\lambda$}  & &  \text{1-loop}  \\[5pt]
\framebox{$\lambda^2 \LL^2$} & \lambda^2 \LL & \lambda^2 & \text{2-loop}  \\[5pt]
\framebox{$\lambda^3 \LL^3$} & \lambda^3 \LL^2 & \ldots & \text{3-loop} \\[5pt]
\vdots & & \\[5pt]
\end{block}
\end{blockarray}
\label{4.6}
\end{align}
relative to the tree-level value, which is denoted by $1$. The first column gives the leading-log (LL) series, the second column gives the next-to-leading-log (NLL) series, etc. The first row is the tree-level contribution, the second row is the one-loop contribution, and so on.  As is well-known, the LL series is given by integrating the one-loop RGE, the NLL series by integrating the two-loop RGE, etc. 

Now suppose that there are multiple mass scales in the matrix $W$. For instance, we could have two eigenvalues of the $W$-matrix with $w_H \gg w_L$.  The one-loop contribution to the \cwp\ now contains two types of logarithms, $\ln w_H/\mu^2$ and $\ln w_L/\mu^2$. Unlike the single-scale case in Eq.~\eqref{4.1}, one cannot simultaneously make both logarithms small; if we choose $\mu^2 \sim w_H$, we are left with large logarithms of the form $\ln w_H/w_L$. 

In this paper, these large logarithms are summed by constructing an EFT and using RG evolution between the scales $w_H$ and $w_L$. As in any EFT, one has to start with a power counting expansion. We introduce a power counting parameter $w_L/w_H\sim z{^2} \ll 1$ (to be defined in each of the examples we present), and provide a method to systematically compute the Coleman-Weinberg potential as a power series in $z$, known as the power counting expansion. As is well-known, the LL series is given by tree-level matching and one-loop running, the NLL series is given by one-loop matching and two-loop running, etc. To be able to RG improve the potential requires being able to compute the entire first column in Eq.~\eqref{4.6}, not just the $\lambda \LL$ term.
In particular, we will compute the boxed terms of Eq.~\eqref{4.6}, i.e.\ the LL series as well as the non-log terms at one-loop. 
Our results differ at order $(\lambda \LL)^2$ from those presented in the literature for RG resummation of the \cwp, but agree with explicit two-loop computations.

\section{EFT Method Applied to the $O(N)$ Model}
\label{sec:on}

We introduce the EFT method for computing the \cwp, using the $O(N)$ model as a test case. A summary of the method appears in Section~\ref{sec:overview}.

\subsection{Preliminaries and Power Counting}
\label{sec:preliminaries}

In the unbroken phase, the Lagrangian is given by
	\begin{align}
	\CL &= \frac 12 \left( \partial_\mu \bm{\phi \cdot} \partial^\mu \bm{ \phi}\right) - \frac{m^2}{2} \left( \bm{\phi \cdot \phi} \right) - \frac{\lambda}{24}   \left( \bm{\phi \cdot \phi} \right)^2\ ,
	\label{eq:Lonun}
	\end{align}
where $\bm{\phi}$ is an $N$-component vector.  This phase has effectively a single mass scale. 
More interesting for our purposes is the broken phase, in which the mass-squared acquires a minus sign, and classically $\bm{\phi}$ can get a vacuum expectation value (VEV) $v^2 = {-}6m^2/\lambda$. 
Let us parameterize the Lagrangian in the broken phase as
	\begin{align}
	\CL &= \frac 12 \left( \partial_\mu \bm{\phi \cdot} \partial^\mu \bm{ \phi}\right)- \frac{\lambda}{24}   \left( \bm{\phi \cdot \phi}- v^2 \right)^2 - \Lambda\ ,
	\label{eq:Lon}
	\end{align}
where we have included a cosmological constant $\Lambda$. 

The \cwp\ for this theory is conventionally written as a function of $\bm{\phi \cdot \phi}$ by $O(N)$ symmetry. Nevertheless, we emphasize that the potential being computed is the potential of the field $\bm{\phi}$, not the composite field $\bm{\phi \cdot  \phi}$. To compute the latter would require introducing source terms for $\bm{\phi \cdot \phi}$ in the functional integral Eq.~\eqref{eq:z}.

Expanding in fluctuations about a VEV for one component of $\bm{\phi}$, we take  
	\begin{align}
	\bm{\phi} = (\hat \phi + \chi_q) \bm{\hat n} + \bm{\phi}_q\ .
	\label{eq:shifton}
	\end{align}
Here $\bm{\hat{n}}$ is a unit vector pointing in a fixed direction,  $\chi_q$ is a single field representing the quantum fluctuation about the VEV $\hat{\phi}$ in the $\bm{\hat{n}}$ direction (the radial Higgs mode), and $\bm{\phi}_q$ is an $N_{\text{GB}}\equiv (N-1)$-component vector orthogonal to $\bm{\hat{n}}$ (the Goldstone modes). 
We thus obtain the scalar mass-squared matrix   
	\begin{align}
	W &= \begin{bmatrix} W_{\chi \chi} & W_{\chi\phi} \mathbbm{1}_{1\times N_{\text{GB}}}  \\ W_{\phi\chi} \mathbbm{1}_{ N_{\text{GB}}\times 1} & W_{\phi \phi}  \mathbbm{1}_{N_{\text{GB}}\times N_{\text{GB}}} \end{bmatrix} \,, \qquad N_{\text{GB}} \equiv N-1\ ,
	\label{eq:won}
	\end{align}
with
	\begin{align}
	W_{\chi \chi} &= \frac{\lambda}{6} \left(3\hat \phi^2-v^2\right) \,, \qquad W_{\phi \phi} = \frac{\lambda}{6} \left(\hat \phi^2-v^2\right) \,,\qquad W_{\phi\chi} = W_{\chi\phi} = 0 \,.
	\label{eq:wcompon}
	\end{align}
In the $O(N)$ theory, $W$ given in \eqref{eq:won} is diagonal.  In the examples considered in the  following sections $W$ will not be diagonal.
At the classical minimum of the potential $\hat{\phi}^2=v^2$, and the Goldstone modes are massless. 

The shifted Lagrangian defined in Eq.~\eqref{eq:cl} is
\begin{align}
\hat{\CL} &= \frac12 \partial_\mu \chi_q \partial^\mu \chi_q + \frac 12 \left( \partial_\mu \bm{\phi}_q \bm{\cdot} \partial^\mu \bm{ \phi}_q \right)- \frac{\lambda}{24} \,  \chi_q^4  
 - \frac{\lambda}{12}  \, \chi_q^2 \left( \bm{\phi}_q \bm{ \cdot \phi}_q \right) - \frac{\lambda}{24}  \,  \left( \bm{\phi}_q \bm{ \cdot \phi}_q \right) ^2 \nn
& - \frac{\lambda}{6}  \, \hat\phi \, \chi_q^3  - \frac{\lambda}{6}  \, \hat\phi \, \chi_q\,  \left( \bm{\phi}_q \bm{ \cdot \phi}_q \right)  -\frac12 W_{\chi \chi}\, \chi_q^2 -\frac12 W_{\phi \phi}  \, \left( \bm{\phi}_q \bm{ \cdot \phi}_q \right)   +  \CJ_\chi \chi_q  -\hat \Lambda\ .
\label{7.1a}
\end{align}
The $\beta$-functions for these couplings can be read off from Eq.~\eqref{2.6} and are given in Eq.\eqref{A5.13}--\eqref{A5.14}.

The \cwp\ at one loop (fixed order) follows from Eq.~\eqref{2.1},
	\begin{align} 
	\begin{split}
	V_{\text{CW}} &= V_{\text{tree}} + \hbar V_{1-\text{loop}}\ ,\\
	V_{\text{tree}} &= \hat{\Lambda}(\hat{\phi}) =  \frac{\lambda}{24} \left( \hat{\phi}^2 - v^2 \right)^2 + \Lambda  \ , \\
	V_{1\text{-loop}} &= \frac{W_{\chi\chi}^2 }{64\pi^2} \left( \ln \frac{W_{\chi\chi}}{\mu^2} - \frac{3}{2} \right) +  \frac{N_{\text{GB}} W_{\phi\phi}^2}{64\pi^2} \left( \ln \frac{W_{\phi\phi}}{\mu^2} - \frac{3}{2} \right)   \ .
	\end{split} 
	\label{eq:onfo}
	\end{align}
Notice that the first term in $V_{1\text{-loop}}$ comes from integrating out the heavy field $\chi_q$ to quadratic order in the functional integral over $\hat{\CL}$. 
 For future reference, we name this contribution
	\begin{align}
	V_{\text{match}} =  \frac{W_{\chi\chi}^2 }{64\pi^2} \left( \ln \frac{W_{\chi\chi}}{\mu^2} - \frac{3}{2} \right) \ . \label{eq:vmatch}
	\end{align}

The zeroth order step in our procedure is to introduce a power counting parameter. 
 If $\hat \phi$ is very different from $v$, then all particles have comparable masses and RG improvement is not necessary, since the \cwp\ can be computed reliably using the known fixed-order expressions in Eq.~\eqref{eq:onfo}. Thus for our power counting, we take $v\sim 1$ and $\hat{\phi}\sim 1$ with $\hat{\phi}^2 - v^2 \sim z^2$, such that $W_{\chi\chi}\sim 1$ and $W_{\phi\phi}\sim z^2$. A nice feature of the $O(N)$ model in the broken phase is that this hierarchy of scales,
with the Goldstone modes being much lighter than the radial mode,  is natural. To summarize, our power counting is
	\begin{align}
	\hat{\phi} &\sim 1\,, & 
	( \hat{\phi}^2  - v^2 ) & \sim z^2\,, &
	W_{\chi\chi} & \sim 1 + \CO(z^2)\,, & 
	W_{\phi\phi} & \sim z^2\ .
	\end{align}
In particular, the power counting expansion depends on the values of background fields. 
We also split Eq.~\eqref{eq:wcompon} as 
\begin{align}
W_{\chi \chi} &= m_\chi^2 +\frac{ \lambda}{2} \left(\hat \phi^2-v^2\right) \,,\qquad  m_\chi^2 \equiv \frac{\lambda v^2}{3}\,,
\label{7.4z}
\end{align}
and treat $m_\chi^2 \sim 1$, and the second term in $W_{\chi \chi} $ as order $z^2$.
This allows us to RG improve the effective potential in a region near the classical VEV. 

To power count terms in the shifted Lagrangian Eq.~\eqref{7.1a}, we use $\chi_q \sim 1$ and $\bm{\phi}_q\sim z$, which are determined by the power counting of their propagators.
 Then, the $\chi_q$ contribution to the \cwp\ is order unity, and the $\bm{\phi}_q$ contribution is order $z^4$, as can be seen from the fixed order expressions in Eq.~\eqref{eq:onfo}. In particular, $V_{\text{tree}}$ is $\CO(z^4)$, and all non-trivial resummation effects start at order $z^4$. Thus, we compute the \cwp\  to this order. In some examples we will consider later in Sec.~\ref{sec:unbroken}, the new effects start at order $z^6$.

\subsection{A Step-by-Step Overview}
\label{sec:overview}

\begin{figure}[h!]
\begin{center}
\begin{tikzpicture}[
param/.style={circle,draw=red!50,fill=red!50},
obs/.style={rectangle,draw=blue!24,fill=blue!24,rounded corners},
ans/.style={rectangle,draw=red!24,fill=red!24,rounded corners},
low/.style={rectangle,draw=cyan!24,fill=cyan!24,rounded corners},
high/.style={rectangle,draw=blue!24,fill=blue!24,rounded corners},
almost/.style={rectangle,draw=orange!24,fill=orange!24,rounded corners},
almosts/.style={rectangle,draw=orange!24,fill=orange!24,rounded corners},
tad/.style={rectangle,draw=Olive!45,fill=Olive!45,rounded corners}
]
\node at (6,5) (start) [high] { \vbox{\hbox to 3cm{ \hfil $ \mathcal{L}(\phi,\chi,\mu_0)$ \hfil }\hbox to 3 cm {\hfil $ \lambda_i(\mu_0),\  \Lambda(\mu_0)$ \hfil}}};
\node at (2,2) (run) [high] { \vbox{\hbox to 3cm{ \hfil $ \mathcal{L}(\phi,\chi,\mu_H)$ \hfil }\hbox to 3 cm {\hfil $ \lambda_i(\mu_H),\  \Lambda(\mu_H)$ \hfil}}};
\node at (10,2) (shift) [high] { \vbox{\hbox to 4.1cm{ \hfil $ \hat{\mathcal{L}}(\phi_q,\chi_q,\mu_0)$ \hfil }\hbox to 4 cm {\hfil $ \hat\lambda_i(\hat\phi,\hat \chi,\mu_0), \ \hat\Lambda(\hat\varphi,\hat \chi,\mu_0)$ \hfil}}};
\node at (6,-1) (runshift) [high] { \vbox{\hbox to 4.5cm{ \hfil $ \hat{\mathcal{L}}(\phi_q,\chi_q,\mu_H)$ \hfil }\hbox to 4.5 cm {\hfil $ \hat\lambda_i(\hat\phi,\hat \chi,\mu_H), \ \hat\Lambda(\hat\phi,\hat \chi,\mu_H)$ \hfil}}};
\node at (6,-4) (match) [low] { \vbox{\hbox to 4.5 cm{ \hfil $ \mathcal{L}_{\text{EFT}}(\phi_q,\mu_H)$ \hfil }\hbox to 4.5 cm {\hfil $ \widetilde \lambda_i(\hat\phi,\hat \chi,\mu_H),\  \widetilde \Lambda(\hat\phi,\hat \chi,\mu_H)$ \hfil}}};
\node at (6,-7) (low) [almost] { \vbox{\hbox to 4.5 cm{ \hfil $ \mathcal{L}_{\text{EFT}}(\phi_q,\mu_L)$ \hfil }\hbox to 4.5 cm {\hfil $ \widetilde \lambda_i(\hat\phi,\hat \chi,\mu_L),\  \widetilde \Lambda(\hat\phi,\hat \chi,\mu_L)$ \hfil}}};
\node at (6,-9.7) (ans) [ans] { \hbox to 3 cm{ \hfil $V_{\text{CW}}(\hat{\phi},\hat{\chi},\mu_L)$ \hfil} };
\draw [-{>[scale=1.5, length=5, width=3,flex]},line width=0.4mm] (start) to[out=180,in=90] (run);
\draw [-{>[scale=1.5, length=5, width=3,flex]},line width=0.4mm] (start) to[out=0,in=90] (shift);
\draw [-{>[scale=1.5, length=5, width=3,flex]},line width=0.4mm] (run) to[out=-90,in=180] (runshift);
\draw [-{>[scale=1.5, length=5, width=3,flex]},line width=0.4mm] (shift) to[out=-90,in=0] (runshift);
\draw [-{>[scale=1.5, length=5, width=3,flex]},line width=0.4mm] (runshift) to[out=-90,in=90] (match);
\draw [-{>[scale=1.5, length=5, width=3,flex]},line width=0.4mm] (match) to[out=-90,in=90] (low);
\draw [-{>[scale=1.5, length=5, width=3,flex]},line width=0.4mm] (low) to[out=-90,in=90] (ans);

\draw (11.25,4) node [align=center] { \vbox{\hbox to 3cm{ \hfil $\phi=\hat \phi + \phi_q$ \hfil }\hbox to 3 cm {\hfil $\chi = \hat \chi + \chi_q$ \hfil}\hbox to 3 cm {\hfil drop linear \hfil}}};
\draw (1,0) node [align=center] { \vbox{\hbox to 3cm{ \hfil $\phi=\hat \phi + \phi_q$ \hfil }\hbox to 3 cm {\hfil $\chi = \hat \chi + \chi_q$ \hfil}\hbox to 3 cm {\hfil drop linear \hfil}}};
\draw (11,0) node [align=center] {\vbox{\hbox to 3cm{ \hfil $\mu_0$ \hfil }\hbox to 3 cm {\hfil $\downarrow$ \hfil} \hbox to 3 cm {\hfil $\mu_H$ \hfil}}};
\draw (1,4) node [align=center] {\vbox{\hbox to 3cm{ \hfil $\mu_0$ \hfil }\hbox to 3 cm {\hfil $\downarrow$ \hfil} \hbox to 3 cm {\hfil $\mu_H$ \hfil}}};
\draw (7.75,-2.5) node [align=center] {integrate out $\chi_q$};
\draw (7,-5.5) node [align=center] {\vbox{\hbox to 3cm{ \hfil $\mu_H$ \hfil }\hbox to 3 cm {\hfil $\downarrow$ \hfil} \hbox to 3 cm {\hfil $\mu_L$ \hfil}}};
\draw (7.75,-8.5) node [align=center] {integrate out $\phi_q$};

\node at (1.4,-9.7) (new) [tad] { \hbox to 2 cm{ \hfil $\langle \chi_q \rangle \overset{!}{=}0$ \hfil} };
\draw [-{>[scale=1.5, length=5, width=3,flex]},line width=0.4mm] (new) to[out=0,in=180] (ans);

\end{tikzpicture}
\end{center}
\caption{Flowchart for the EFT method of RG improving the \cwp. The general procedure is similar to other EFT computations. The new features are the field shifts in the high-energy theory, and the tadpole condition for the heavy field which is evaluated in the low-energy theory. \label{fig:flowchart}}
\end{figure}
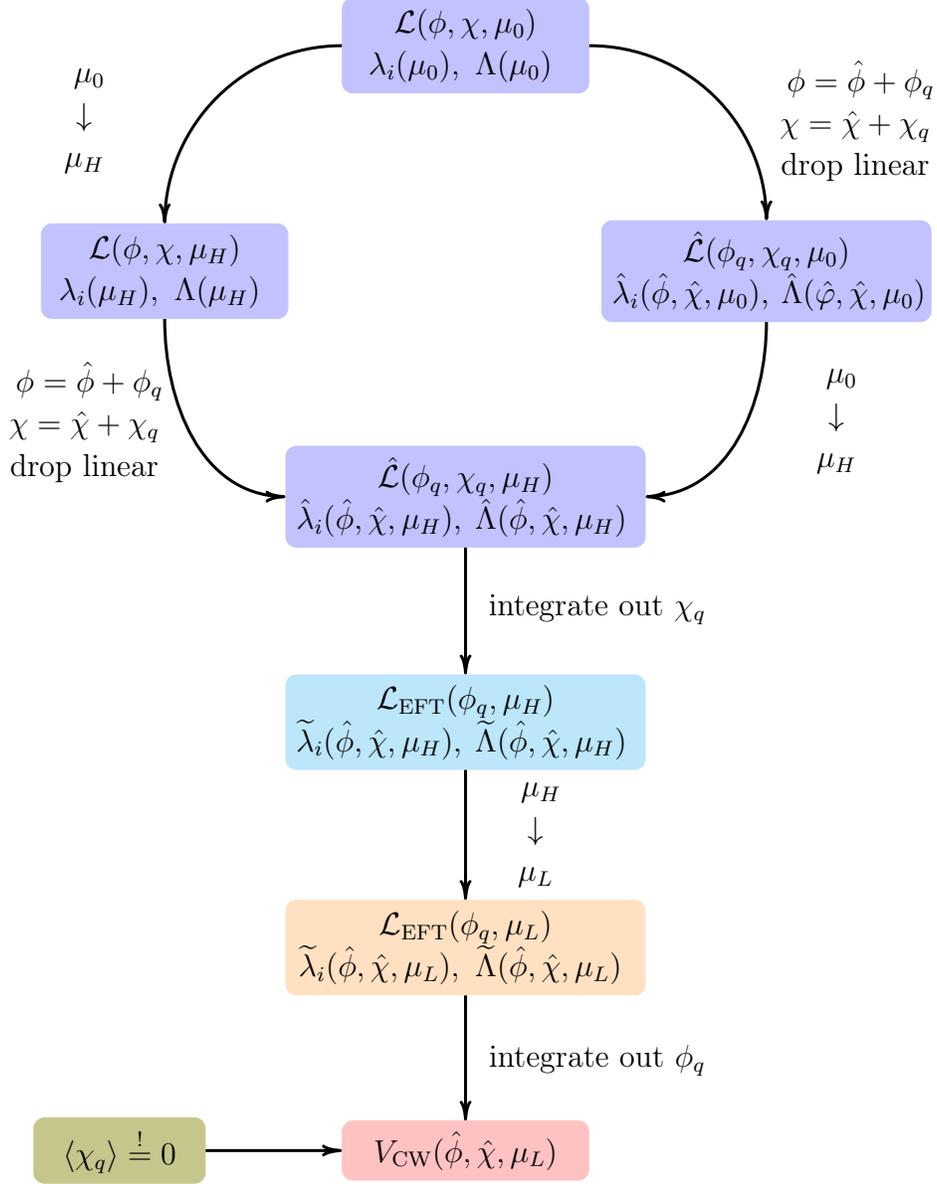

Let us now give a step-by-step overview of our procedure, postponing detailed computation to subsequent sections. These steps are summarized in the (color-coded) flowchart in Figure~\ref{fig:flowchart}.

\newcommand{\specialitem}[3][black]{%
  \item[%
    \colorbox{#2}{\textcolor{#1}{\makebox[1em]{#3}}}%
  ]
}

\begin{itemize}

 \specialitem{blue!24}{1.}  Run the Lagrangian from the starting scale $\mu_0$ to the heavy scale $\mu_H^2\sim W_{\chi\chi}$ using the  $\beta$-functions in the high-energy theory. From our discussion around Eq.~\eqref{2.6}, at one loop the high-energy $\beta$ functions satisfy
	\begin{align}
	 \beta_{\lambda_i}\frac{dV}{d\lambda_i} - \gamma_{\phi }\phi \frac{dV}{d\phi}  = \frac{1}{32} \tr W^2\ ,
	\end{align}
	where in this example $\{ \lambda_i \} = \{\lambda, v, \Lambda\}$. 

Because shifting and running the fields commute, as emphasized in Section \ref{sec:running}, we can first run the Lagrangian to $\mu_H$ and then shift the fields. Upon shifting, the functional integral is over Eq.~\eqref{eq:cl} (in this case, Eq.~\eqref{7.1a}) with couplings and fields renormalized at $\mu_H$.  
There is nothing new in this part of the calculation, so we start our examples at the scale $\mu_H$.

In the $O(N)$ model $\gamma_\phi$ is zero at one-loop, but it is nonzero at two-loop order. It would be nonzero at one-loop if we introduce gauge or Yukawa interactions into the theory. Shifting the field at $\mu_H$ gives the potential  in terms of the expectation value of $\bm{\phi}$ renormalized at $\mu_H$.

 \specialitem{cyan!24}{2.}  Integrate out $\chi_q$ at the scale $\mu_H$. Below the matching scale, the theory is described by an EFT, whose Lagrangian we denote $\CL_{\text{EFT}}$. There are no large logs in the matching.
The \cwp\ is given by the vacuum graph Fig.~\ref{fig:3} computed using the quadratic action in quantum fields. To include all the one-loop terms and perform the LL resummation, we need two contributions: 

\begin{itemize}

\item Tree level matching of $\hat{\CL}$ in Eq.~\eqref{eq:cl} (Eq.~\eqref{7.1a} in the $O(N)$ example) onto the EFT Lagrangian. 
This matching involves expanding the  tree-level $n$-point functions of the high-energy theory in the power counting parameter $z$, and matching onto the EFT $n$-point functions at the scale $\mu_H$. Denoting the EFT couplings with tildes as $\{ \widetilde{\lambda}_i(\mu) \}$, this determines the couplings $\{ \widetilde{\lambda}_i(\mu_H)\}$ as a function of the high-energy couplings $\{ \lambda_i(\mu_H)\}$. For example, 
$V_{\text{tree}}(\mu_H)$ gives a contribution to the EFT cosmological constant $\widetilde{\Lambda}(\mu_H)$.

\item One-loop matching to the cosmological constant in the EFT from integrating out $\chi_q$. 
This matching involves computing the one-loop vacuum graphs, expanding in the power counting parameter $z$, and dropping all divergent terms~(see \cite{Manohar:2018aog}). These graphs will contribute to the cosmological constant $\widetilde{\Lambda}$ of the EFT, since they are independent of $\bm{\phi}_q$. 
The result of this computation is precisely $V_{\text{match}}(\mu_H)\subset \widetilde{\Lambda}(\mu_H)$, with $V_{\text{match}}$ given in Eq.~\eqref{eq:vmatch} for the $O(N)$ example. All the logarithms in this expression are of the form $\ln |W_{\chi\chi}/\mu_H^2|$, so there are no large logs. This term would usually be neglected in the LL calculation, but is included here because we compute the full one-loop correction to the potential. 

Note that in examples with non-diagonal $W$, this one-loop matching computation is more involved. In such a case, the \cwp\ is a sum over contributions of the eigenvalues of $W$, and this matching computation corresponds to the contribution of the larger eigenvalue, expanded in $z$. We will explore this point in more detail in Section~\ref{sec:interlude}.

\item Computation of higher order terms in the perturbation expansion Eq.~\eqref{4.6} requires matching to the EFT Lagrangian $\hat{\CL}$ beyond tree-level, e.g.\ the NLL series requires one-loop matching. At this order, one also obtains finite matching corrections from graphs such as Fig.~\ref{fig:7} to the kinetic energy terms in the EFT Lagrangian. These can be absorbed into rescaling the EFT quantum field $\phi_q$. Such rescalings are one reason why shifting $\phi$ in the high-energy theory is not equivalent to shifting $\phi$ in the EFT. In the two-scalar example in Sec.~\ref{sec:unbroken}, there are contributions to the kinetic energy term and terms with higher derivatives in the EFT Lagrangian from expanding the $\chi_q$ propagator in a power series in $p^2/m_\chi^2$. However, these terms are of higher order than what is required for the results in this paper (see Eq.~\eqref{9.9}).

\end{itemize}  

As a result of this matching, we obtain in particular the boundary value of the EFT cosmological constant, 
	\begin{align}
	\widetilde{\Lambda}(\mu_H) = \hat{\Lambda}(\mu_H) + V_{\text{match}}(\mu_H) + (\text{tadpole}) \ , \label{eq:ccmatch}
	\end{align}
	as well as the boundary values of the other $\widetilde{\lambda}_{i}(\mu_H)$.  The $(\text{tadpole})$ term in Eq.~\eqref{eq:ccmatch} is a function of the source $\CJ$ as defined in Eq.~\eqref{eq:j}, and will be given explicitly in examples.

 \specialitem{orange!24}{3.}  Run the EFT couplings from the heavy scale $\mu_H$ to the light scale $\mu_L^2\sim \widetilde{m}^2$ using the RGE defined by the EFT Lagrangian,
	\begin{align}
 \beta_{\widetilde{\lambda}_i} \frac{dV_{\text{EFT}}}{d\widetilde{\lambda}_i} - \widetilde{\gamma}_\phi \phi  \frac{dV_{\text{EFT}}}{d\phi} = \frac{1}{2 } ( W_{\text{EFT}})^2 \ .
	\end{align}
Here $\widetilde{m}^2$ is the quadratic coupling in the EFT.
The entire LL series is obtained by running the cosmological constant $\widetilde{\Lambda}$ from $\mu_H$ to $\mu_L$.

 \specialitem{red!24}{4.} Perform the functional integral over $\bm{\phi}_q$.  The one-loop contribution is the graph in Fig.~\ref{fig:3}, and only involves  terms in $\CL_{\text{EFT}}$ to quadratic order---i.e. the cosmological constant $\widetilde{\Lambda}$ and the quadratic mass term $\widetilde{m}^2$. The computation of the one-loop determinant follows the usual steps reviewed in Section~\ref{sec:jackiw}, resulting in 
	\begin{align}
	V_{\text{CW}}(\mu_L) = \widetilde{\Lambda}(\mu_L) +  \frac{\widetilde{m}^4(\mu_L)}{64\pi^2} \left( \ln \frac{\widetilde{m}^2(\mu_L)}{\mu_L^2} - \frac{3}{2} \right)\,. \label{eq:vcwlow}
	\end{align}
With the power counting relevant to the $O(N)$ model, this is the \cwp\ computed through $\CO(z^4)$.
The second term in Eq.~\eqref{eq:vcwlow} is not part of the LL series, but is included since we are computing the entire one-loop contribution.

Note that even though only the quadratic EFT couplings appear in Eq.~\eqref{eq:vcwlow}, the RG running between $\mu_H$ and $\mu_L$ in Step 3 involves all the interactions of the quantum $\bm{\phi}_q$ theory, including the non-quadratic terms. This is why we needed to compute the entire $\CL_{\text{EFT}}$ to $\CO(z^4)$.

 \specialitem{Olive!45}{5.} 
We emphasize that in Step 2 one must match using  Eq.~\eqref{eq:cl} (Eq.~\eqref{7.1a} in the $O(N)$ example), which is the shifted Lagrangian that includes the one-loop sources $\CJ$ as defined in Eq.~\eqref{eq:j}. The computation outlined in the previous steps has not yet fixed the tadpole $\CJ_\chi$ associated to the heavy field. The final step is thus to compute the RG improved tadpole $\CJ_\chi$ by requiring that $\langle \chi_q\rangle$ vanishes, with $\chi_q$ the quantum field in the original high-energy theory. RG improvement of the tadpole is necessary to sum the full LL series of the effective potential. The required tadpole is that of the \emph{heavy} field, but computed in the low-energy theory. This {computation} is possible because we have matched the source $\CJ_\chi$ onto the EFT.

\end{itemize}

After completing these steps and obtaining Eq.~\eqref{eq:vcwlow}, we compare our RG improved result to the known two-loop results by expanding the RG series in a power series in  $\ln {\mu_L}/{\mu_H}$, in order to check that we correctly capture the leading-log terms. To reiterate, the entire LL contribution comes from the cosmological constant $\widetilde{\Lambda}$ in the low-energy theory.

An important comment is that at no point have we computed the \cwp\ of $\CL_{\text{EFT}}$.  In particular, there is no sense in which we shift $\bm{\phi}_q$ in the low energy theory after integrating out the heavy field to compute a one-loop potential. As we see explicitly in  examples, after integrating out $\chi_q$, running the couplings and shifting the fields no longer commute. It is crucial that we shift $\phi$ in the original theory \emph{before} integrating out any fields.

\subsection{Matching to the EFT at $\mu_H$}\label{sec:matchon}

Let us now fill in the step-by-step details for the $O(N)$ model. We begin with the matching computation of Step 2, in which we match Eq.~\eqref{7.1a} to $\CL_{\text{EFT}}$ at the heavy scale. Parameterize the EFT Lagrangian by
\begin{align}
\CL_{\text{EFT}} &=\frac 12 \left( \partial_\mu \bm{\phi}_q \bm{\cdot} \partial^\mu \bm{ \phi}_q \right)  - \frac{\widetilde \lambda}{24}   \left( \bm{\phi}_q \bm{ \cdot \phi}_q \right) ^2 -\frac{\widetilde m^2}{2} \left( \bm{\phi}_q \bm{ \cdot \phi}_q \right)  -\widetilde  \Lambda + \mathcal{O}(z^5)\,.
\label{7.12}
\end{align}
The EFT couplings will be denoted by a tilde to distinguish them from couplings in the original high-energy theory Eq.~\eqref{eq:Lon} or in the shifted theory Eq.~\eqref{7.1a}.
In this case, the EFT will describe the light Goldstone bosons. 

There can be finite wavefunction corrections to the EFT Lagrangian from matching, which can be absorbed into a rescaling of the EFT field $\bm{\phi}_q$. There is no such correction at one-loop order in the $O(N)$ model.

The one-loop matching computation is given by computing the one-loop $\chi$ bubble in Fig.~\ref{fig:3}, which gives the heavy-field contribution to the \cwp. In the $O(N)$ example, this calculation is particularly simple because the scalar mass matrix Eq.~\eqref{eq:won} does not couple the heavy field $\chi_q$ to the light fields $\bm{\phi}_q$. Such couplings are present in the examples considered in Sections~\ref{sec:unbroken}-\ref{sec:broken}, and the one-loop matching in such cases is more involved. For the case at hand, this
 matching contribution is simply given by Eq.~\eqref{eq:vmatch}, which we repeat here:
\begin{align}
V_{\text{match}}(\mu_H) &=  \frac{1}{64 \pi^2}  W_{\chi \chi}^2
  \left[ 
 \ln \frac{W_{\chi \chi}}{\mu_H^2}     -\frac32   \right] \,.  \label{7.7}
\end{align}
This is  order $1$ in the $z$ power counting. The matching is performed at a scale $\mu_H^2 \sim W_{\chi \chi}$, so there are no large logarithms in this expression. $V_{\text{match}}$ adds to the cosmological constant in the low-energy theory below $\mu_H$, since Eq.~\eqref{7.7} has no quantum fields.

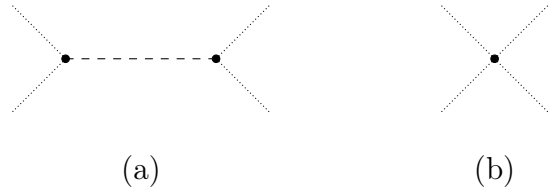
\begin{figure}
\begin{center}
\begin{tikzpicture}
\draw[dashed] (-1,0) -- (1,0);
\filldraw (-1,0) circle (0.05);
\filldraw (1,0) circle (0.05);
\draw[densely dotted] (1,0) -- +(45:1);
\draw[densely dotted] (1,0) -- +(-45:1);
\draw[densely dotted] (-1,0) -- +(135:1);
\draw[densely dotted] (-1,0) -- +(225:1);
\draw (0,-1.5) node [align=center] {(a)};
\end{tikzpicture}
\hspace{2cm}
\begin{tikzpicture}
\filldraw (0,0) circle (0.05);
\draw[densely dotted] (0,0) -- +(45:1);
\draw[densely dotted] (0,0) -- +(-45:1);
\draw[densely dotted] (0,0) -- +(135:1);
\draw[densely dotted] (0,0) -- +(225:1);
\draw (0,-1.5) node [align=center] {(b)};
\end{tikzpicture}
\end{center}
\caption{\label{fig:8} (a) Tree-level matching onto the EFT. The dashed line is $\chi_q$, and the dotted lines are $\phi_q$. (b) The same interaction in the low-energy EFT. Unlike graph (a), graph (b) is 1PI.}
\end{figure}

The tree-level matching to order $z^4$ is given by the graph shown in Fig.~\ref{fig:8}(a), leading to the EFT interaction shown in Fig.~\ref{fig:8}(b). To compute this graph, write the terms linear in $\chi_q$ in Eq.~\eqref{7.1a},
\begin{align}
\hat{\CL} &\supset - X_1(\phi_q) \chi_q \ ,\qquad X_1(\phi_q) = \frac{1}{6} \lambda \, \hat\phi \,   \left( \bm{\phi}_q \bm{ \cdot \phi}_q \right)  -  \CJ_\chi\ ,
\label{7.8}
\end{align}
where $X_1 \sim z^2$.
Fig.~\ref{fig:8}(a) gives the tree-level matching
\begin{align}
\CL_{\text{EFT}}\supset \frac12 X_1 \frac{1}{W_{\chi \chi}-p^2} X_1 = \frac12 X_1 \frac{1}{W_{\chi \chi}} X_1 + \frac12 X_1 \frac{p^2}{W_{\chi \chi}^2} X_1 +\ldots
\label{7.9}
\end{align}
where $p$ is the momentum flowing through the $\chi_q$ line, and $1/2$ is the symmetry factor due to the two identical vertices $X_1$. Since $p$ is a momentum in the EFT, and is of order $z$, to order $z^4$ we only need the first term. The EFT Lagrangian is then
\begin{align}
\CL_{\text{EFT}} &=\frac 12 \left( \partial_\mu \bm{\phi}_q \bm{\cdot} \partial^\mu \bm{ \phi}_q \right)  - \frac{\lambda}{24}   \left( \bm{\phi}_q \bm{ \cdot \phi}_q \right) ^2 -\frac12 W_{\phi \phi}  \, \left( \bm{\phi}_q \bm{ \cdot \phi}_q \right)  - \left[ \hat \Lambda + V_{\text{match}}(\mu_H) \right] \nn
& + \frac1{2 W_{\chi\chi}^2}  \left(\frac{\lambda  \hat\phi}{6}     \left( \bm{\phi}_q \bm{ \cdot \phi}_q \right)  -  \CJ_\chi
 \right)^2 + \mathcal{O}(z^5)\,.
\label{7.11}
\end{align}
Comparing with Eq.~\eqref{7.12}, we identify the boundary values of the EFT couplings to the desired order in $z$ as
\begin{align}
\begin{split}
\widetilde \lambda \overset{\mu=\mu_H }=&\lambda \left( 1 - \frac{\lambda \hat{\phi}^2 }{3 W_{\chi\chi}} \right)= \frac{ \lambda^2}{2 W_{\chi \chi}^2}
\left( \hat \phi^2-v^2\right)
 \sim \CO(z^2)\,,\\
\widetilde m^2 \overset{\mu=\mu_H}=&   W_{\phi \phi} +  \frac{ \lambda \hat \phi \CJ_\chi}{3 W_{\chi\chi}} =  \frac16 \lambda \left(\hat \phi^2 -v^2\right) +  \frac{ \lambda \hat \phi \CJ_\chi }{3 W_{\chi\chi}}  + \CO(z^3) \,,\\
&\equiv \widetilde m^2_- +   \frac{ \lambda \hat \phi \CJ_\chi }{3 W_{\chi\chi}}  + \CO(z^3) \,,\\
\widetilde \Lambda \overset{\mu=\mu_H}=&\hat  \Lambda  + V_{\text{match}} - \frac{ \left( \CJ_\chi \right)^2 }{2 W_{\chi\chi}} \,,  \\
=&  \Lambda+  \frac{\lambda}{24}   \left( \hat \phi^2- v^2 \right)^2  + V_{\text{match}}- \frac{ \left( \CJ_\chi \right)^2 }{2 W_{\chi\chi}} + \CO(z^5)\,.
\label{7.4}
\end{split}
\end{align}
The relations Eq.~\eqref{7.4} hold with both sides evaluated at the matching scale $\mu=\mu_H$.  As promised, the EFT cosmological constant $\widetilde{\Lambda}(\mu_H)$ takes the form Eq.~\eqref{eq:ccmatch}. 

In this example the quartic coupling $\widetilde{\lambda}(\mu_H)$ is $\CO(z^2)$, and thus only contributes to the EFT Lagrangian at order $z^6$, since  it multiplies
$\left( \bm{\phi}_q \bm{ \cdot \phi}_q \right) ^2$ which is order $z^4$. That $\widetilde \lambda$ is order $z^2$ is not an accident; the low-energy fields are Goldstone bosons when $\hat \phi \to v$, and only have derivative couplings at this point.

At this stage, we can illustrate the point that the field shift must be made in the high-energy theory, not in the EFT.  To be able to shift the field in the low-energy theory would require starting with the parameters in Eq.~\eqref{7.4} with $\hat \phi=0$, and then obtaining the $\hat \phi$ dependence in the parameters by a shift of the EFT fields $\bm{\phi}_q$ in the EFT Lagrangian Eq.~\eqref{7.12}. 
A shift of $\bm{\phi}_q$ in the EFT would introduce a $O(N-1)$ vector, whereas $\hat \phi$ in Eq.~\eqref{7.4} is an $O(N-1)$ singlet, and these two objects don't have the same quantum numbers.
The former would be used for computing the \cwp\ \emph{of} the EFT. This is not the same as the \cwp\ of the original theory, computed using an EFT, which is the quantity we {wish to} compute.

\subsection{RG Improvement}\label{sec:muLon}

We now perform Steps 3 and 4 for the $O(N)$ model. 
We sum the large logarithms by RG evolution of the low-energy theory from $\mu=\mu_H$ to a low scale $\mu=\mu_L$ of order the light scalar masses.
Here the couplings in the low-energy are evolved to a scale $\mu_L$ of order $\widetilde m^2$, the mass of $\bm{\phi}_q$.  The $\beta$-functions are given in Eq.~\eqref{A5.19}, with the initial values at $\mu_H$ given in Eq.~\eqref{7.4}. This case is particularly simple since $\widetilde{m}^2$ does not run at this order, i.e.  $\widetilde{m}^2(\mu_L) = \widetilde{m}^2(\mu_H)$, since $\widetilde \lambda$ is zero to this order and the EFT is a free theory. The running of the cosmological constant is
\begin{align}
\widetilde{\Lambda}(\mu_L) &= \widetilde{\Lambda}(\mu_H) + \frac12 N_{\text{GB}} \widetilde{m}^4(\mu_H) t  \,,\qquad t = \frac{1}{16\pi^2} \ln \frac{\mu_L}{\mu_H}\,,
\label{eq:onimprove}
\end{align}
where $N_{\text{GB}} = N-1$ is the number of Goldstone bosons.

At the low-scale $\mu_L$, the light field $\bm{\phi}_q$ is integrated out. The one-loop contribution is the graph in Fig.~\ref{fig:3}, and only involves the quadratic action.  The \cwp\ computed from the low-energy theory is
\begin{align}
V_{\text{CW}}(\mu_L) &= \widetilde \Lambda(\mu_L) + \frac{N_{\text{GB}}}{64 \pi^2} \widetilde m^4(\mu_L) \left[ \ln \frac{ \widetilde m^2(\mu_L) }{\mu_L^2} - \frac32 \right] \,.
\label{7.16}
\end{align}
The first term is the tree-level contribution, and the second term is the one-loop contribution.

To make contact with the standard one-loop fixed order result, ignore any RG evolution. Then the \cwp\  is given by Eq.~\eqref{7.16}, with $\widetilde \Lambda(\mu_L) \to \widetilde \Lambda(\mu_H)$ given in Eq.~\eqref{7.4},  $V_{\text{match}}(\mu_H)$ given in Eq.~\eqref{7.7}, and $\mu_H=\mu_L=\mu$. The $( \CJ_\chi )^2$ term in $\widetilde \Lambda$ and $\widetilde m^2$ should be dropped since it is formally of two-loop order. Then, $W_{\chi\chi}$ and $\widetilde m^2 \to W_{\phi\phi}$ are the mass-squareds of the heavy and light fields, and Eq.~\eqref{7.16} reduces to
\begin{align}
V_{\text{CW}} &= \Lambda +  \frac{\lambda}{24}   \left( \hat \phi^2- v^2 \right)^2 + \frac{W_{\chi \chi}^2}{64 \pi^2}  
  \left[ 
 \ln \frac{W_{\chi \chi}}{\mu^2}     -\frac32   \right]+ \frac{N_{\text{GB}}}{64 \pi^2} W_{\phi \phi}^2 \left[ \ln \frac{ W_{\phi \phi} }{\mu^2} - \frac32 \right]\,.
\label{7.16a}
\end{align}
 This is the expected result Eq.~\eqref{eq:onfo}.

\subsection{Tadpole Improvement}\label{sec:tadpoleon}

Our final result  Eq.~\eqref{7.16} still contains $\CJ_\chi$ {(through the dependence of $\widetilde \Lambda$ on  $\CJ_\chi$)} which has not been determined. It is fixed by requiring that $\VEV{\chi_q}$ vanishes, where $\chi_q$ is the quantum field in the original high-energy theory. From Eq.~\eqref{eq:z3}, one sees that this expectation value is determined by differentiation w.r.t.\  $\CJ_\chi$. This derivative can be done in the EFT, to compute the tadpole of the heavy field $\chi_q$ using the low-energy theory.

\begin{figure}
\begin{center}
\begin{tikzpicture}[scale=1.5]
 \draw[densely dotted] (1,0) circle (0.5);
\filldraw (0.5,0) circle (0.05);
 \end{tikzpicture}
\end{center}
\caption{\label{fig:11} One-loop tadpole graph for the heavy field $\chi$ in the low-energy theory.}
\end{figure}
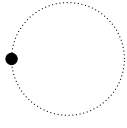

Differentiating the EFT Lagrangian at $\mu=\mu_H$ w.r.t.\ $\CJ_\chi(\mu_H)$ using Eq.~\eqref{7.4}, we see that $\chi_q(\mu_H)$ in the high-energy theory matches onto
\begin{align}
\chi_q(\mu_H) & \to -\frac{ \lambda (\mu_H) \hat \phi }{6 W_{\chi\chi}(\mu_H)} [\bm{\phi}_q \bm{\cdot \phi}_q]_{\mu_H}  +  \frac{ \CJ_\chi(\mu_H)}{ W_{\chi\chi}(\mu_H)} \,.
\label{7.17}
\end{align}
Here we are using a bracket notation $[\CO]$ to emphasize that $\CO$ is a composite operator that needs additional renormalization relative to its {constituent}  fields. 
Then, the tadpole condition becomes
\begin{align}
\CJ_\chi(\mu_H) &=\frac16 \lambda(\mu_H) \hat \phi   \ [ \bm{\phi}_q \bm{\cdot \phi}_q]_{\mu_H}\,,
\label{7.18}
\end{align}
where the r.h.s.\ is computed in the low-energy theory. At one-loop, the graph for $[ \bm{\phi}_q \bm{\cdot \phi}_q]$  is shown in Fig.~\ref{fig:11}, and gives
\begin{align}
\VEV{\left[ \bm{\phi}_q \bm{\cdot \phi}_q\right]}_{\mu} &= N_{\text{GB}}  \frac{ \widetilde m^2(\mu)}{16\pi^2}  \left[ \ln \frac{\widetilde m^2(\mu)}{\mu^2} - 1 \right]\,,
\label{7.30a}
\end{align}
so that
\begin{align}
\CJ_\chi(\mu_H) &= \frac{\lambda(\mu_H)\hat{\phi}}{6}  N_{\text{GB}}  \frac{ \widetilde m^2(\mu_H)}{16\pi^2} \left[ \ln \frac{\widetilde m^2(\mu_H)}{\mu_H^2} - 1 \right]\,.
\label{7.30}
\end{align}
 This equation determines $\CJ_\chi(\mu_H)$ implicitly, since $\widetilde{m}^2(\mu_H)$ depends on $\CJ_\chi(\mu_H)$. The result Eq.~\eqref{7.30} contains a large logarithm, since $\widetilde m$ is the light mass, and is formally order $\lambda \LL$.

Eq.~\eqref{7.30} is sufficient for comparing with existing two-loop calculations. However, to sum the LL series we need the RG improved tadpole.
One can sum the LL series for the tadpole by computing the anomalous dimension of the composite operator $\left[ \bm{\phi}_q \bm{\cdot \phi}_q \right]$ in the low-energy theory.
A simple computation shows that
\begin{align}
\frac{\rd}{\rd t} \left[ \bm{\phi}_q \bm{\cdot \phi}_q \right]  &= - \frac13 (N_{\text{GB}} +2) \widetilde \lambda \left[ \bm{\phi}_q \bm{\cdot \phi}_q \right] -2 N_{\text{GB}} \widetilde m^2 \,,
\label{7.19bb}
\end{align}
which since $\widetilde{\lambda}$ is zero to the order to which we're working, reduces to
\begin{align}
\frac{\rd}{\rd t} \left[ \bm{\phi}_q \bm{\cdot \phi}_q \right]  &= -2 N_{\text{GB}} \widetilde m^2 \,.
\label{7.19b}
\end{align}
Integrating this equation gives
\begin{align}
\left[ \bm{\phi}_q \bm{\cdot \phi}_q \right]_{\mu_H}  &= \left[ \bm{\phi}_q \bm{\cdot \phi}_q \right]_{\mu_L} + 2 N_{\text{GB}} \widetilde m^2(\mu_H) t \,,\qquad t = \frac{1}{16\pi^2} \ln \frac{\mu_L}{\mu_H}\,. \label{eq:rgimp}
\end{align}
Another way to obtain the same result is to differentiate the Lagrangian at $\mu_L$ w.r.t.\ $\CJ_\chi(\mu_H)$ using Eq.~\eqref{eq:onimprove} and Eq.~\eqref{7.4}.

The RG improved tadpole is determined by substituting Eq.~\eqref{eq:rgimp} into Eq.~\eqref{7.18},
\begin{align}
\CJ_\chi(\mu_H)  &=\frac{\lambda(\mu_H) \hat{\phi}}{6}  \left( \left[ \bm{\phi}_q\cdot \bm{\phi}_q\right]_{\mu_L}  + 2 N_{\text{GB}} \widetilde{m}^2(\mu_H) t  \right) \,.
\label{7.43}
\end{align}
Here $\widetilde m^2$ depends on $\CJ_\chi(\mu_H)$. 
Replacing $\widetilde m^2$ by $\widetilde m^2_-$ using Eq.~\eqref{7.4} and solving for $\CJ_\chi(\mu_H)$, we obtain
\begin{align}
\CJ_\chi(\mu_H)  &=\frac{\lambda(\mu_H) \hat{\phi}}{6}  \left( \left[ \bm{\phi}_q\cdot \bm{\phi}_q\right]_{\mu_L}  + 2 N_{\text{GB}} \widetilde{m}^2_-(\mu_H) t  \right)
\left(1 - \frac{\lambda^2(\mu_H) \hat \phi^2}{9 W_{ \chi \chi }(\mu_H)} N_{\text{GB}}  t \right)^{-1}  .
\label{7.43A}
\end{align}
The quantity $ \left[ \bm{\phi}_q\cdot \bm{\phi}_q\right]_{\mu_L}$ is given by Eq.~\eqref{7.30a} with $\mu = \mu_L$, which still depends on $\CJ_\chi(\mu_H)$ through $\widetilde{m}^2(\mu_L)$. The LL expression is given by dropping
$ \left[ \bm{\phi}_q\cdot \bm{\phi}_q\right]_{\mu_L}$, and the LL plus one-loop expression is given by using Eq.~\eqref{7.43A} and substituting the LL value of $ \CJ_\chi(\mu_H)$ into $\widetilde m^2(\mu_L)$ in Eq.~\eqref{7.30a}.
The first term in the LL series for the tadpole from Eq.~\eqref{7.43A} is
\begin{align}
\CJ_\chi(\mu_H)  &\approx \frac13N_{\text{GB}} \lambda(\mu_H) \hat{\phi}   \widetilde{m}^2_-(\mu_H) t  \,.
\label{7.43B}
\end{align}
We will use this expression in the following subsection to compare with the two-loop result.

\subsection{Comparison with the Two-loop Result}\label{sec:twoloopon}

We can now compare our RG improved result to the known two-loop calculation by expanding the RG series in a power series in $t = (\ln {\mu_L}/{\mu_H})/(16 \pi^2) $.
The cosmological constant contribution is
\begin{align}
\widetilde \Lambda(\mu_L) &=\widetilde \Lambda(\mu_H) + \frac12 N_{\text{GB}} \widetilde m^4(\mu_H) t  \,,
\label{7.46a}
\end{align}
where we used Eq.~\eqref{eq:onimprove}. The entire LL contribution is from the cosmological constant. Using Eq.~\eqref{7.4} for $\widetilde m^2(\mu_H) $ and $\widetilde \Lambda(\mu_H) $,
and dropping $V_{\text{match}}$ (which has no large logarithms), this becomes
\begin{align}
\widetilde \Lambda(\mu_L) &= \left[ \hat \Lambda(\mu_H)  - \frac{ \left(\CJ_\chi(\mu_H) \right)^2}{2 W_{\chi\chi}(\mu_H)} \right] + \frac12 N_{\text{GB}} 
\left[W_{\phi \phi} (\mu_H)+  \frac{\lambda(\mu_H) \hat \phi  \, \CJ_\chi (\mu_H)}{3 W_{\chi\chi}(\mu_H)}  \right]^2 t   \,.
\label{7.45}
\end{align}
Keeping the $(\lambda \LL)$ and $(\lambda \LL)^2 $ terms in the leading log series, and remembering from Eq.~\eqref{7.43B} that $\CJ_\chi \sim (\lambda \LL)$,  gives
\begin{align}
\widetilde \Lambda(\mu_L) &\approx
\frac{ N_{\text{GB}} }{2} \left[W_{\phi \phi} (\mu_H) \right]^2 t +  N_{\text{GB}} 
 W_{\phi \phi} (\mu_H)   \frac{\lambda(\mu_H) \hat \phi  \, \CJ_\chi(\mu_H) }{3 W_{\chi\chi}(\mu_H)}   t   - \frac{ \left(\CJ_\chi (\mu_H)\right)^2}{2 W_{\chi\chi}(\mu_H)}  \,,
\label{7.46}
\end{align}
where all terms except the first are order $t^2$.
To the order to which we are working we can substitute Eq.~\eqref{7.43B} for $\CJ_\chi(\mu_H)$ with $\widetilde m^2_-(\mu_H)\to W_{\phi\phi}(\mu_H)$. 
The terms combine in an interesting way --- the second term has twice the value and the opposite sign of the third term. The same sign flip happens in the other examples we consider.
The result is
\begin{align}
\widetilde \Lambda(\mu_L) &\approx \left[W_{\phi \phi} (\mu_H) \right]^2  \Bigg(
\frac{N_{\text{GB}}}{2}   t +  N_{\text{GB}}^2
    \frac{\lambda^2(\mu_H) \hat \phi^2  }{18 m_\chi^2(\mu_H)}    t^2\bigg) \,.
\label{7.44}
\end{align}
The $t^2$ terms in Eq.~\eqref{7.44} arise from tadpole contributions depending on $\CJ$.

Our result can be compared with explicit two-loop computations of the \cwp.
The contribution to the two-loop potential from the first graph in Fig.~\ref{fig:5}, in the notation of Ref.~\cite{Martin:2001vx}, is given in terms of a function $f_{SSS}(m_1^2,m_2^2,m_3^2)$, and that from the second graph in terms of $f_{SS}(m_1^2,m_2^2)$, where  $m_i^2$ are the masses of particles in the  internal lines. The integral $f_{SS}(m_1^2,m_2^2)$ factors into two one-loop integrals, 
\begin{align}
f_{SS}(m_1^2 ,m_2^2) &= \left[ m_1^2 \left( \ln \frac{m_1^2}{\mu^2}-1\right) \right]\left[ m_2^2 \left( \ln \frac{m_2^2}{\mu^2}-1\right) \right] \,.
\end{align}
Our problem has only two particle masses, so we need $f_{SSS}(m^2,m^2,m^2)$ with three equal masses, or $f_{SSS}(m_1^2,m_1^2,m_2^2)$, where two masses are equal. The expressions for $f_{SSS}$ in these special cases are much simpler than the general case, and were evaluated in  Ref.~\cite{Ford:1991hw},
\begin{align}
\begin{split}
f_{SSS}(m_1^2,m_1^2,m_2^2) = -\frac{\Delta+2}{2} m_1^2 \Big\{& -5 + \frac{2 \Delta \ln \Delta}{\Delta +2} \left[2 - \ln \frac{m_1^2}{\mu^2}\right] 
+ 4 \ln \frac{m_1^2}{\mu^2} \\
&- \ln^2 \frac{m_1^2}{\mu^2} - 8 \Omega(\Delta) \Big\} \,,
\end{split}
\end{align}
which is given in terms of $\Delta \equiv {m_2^2}/{m_1^2}$.
The formula for $f_{SSS}(m_1^2,m_1^2,m_2^2)$ depends on a function $\Omega(\Delta)$. Since we are interested in summing logarithms for a large mass ratio, we need the expansions of $\Delta$ around $\Delta=\infty$ and $\Delta=0$.  We have computed these expansions to second order in Appendix~\ref{app:omega}.

Using these pieces, we can compute the $\LL^2$ term in the two-loop calculation of Ref.~\cite{Ford:1991hw}, and check that
our LL result Eq.~\eqref{7.44} for the $\LL^2$ term agrees with this computation.
This agreement provides a highly non-trivial check of our method. It requires, in particular,  shifting $\phi$ in the high-energy theory before matching onto the low-energy theory, and also requires including the tadpole
contributions.

\section{Two Scalar Fields in the Unbroken Phase}\label{sec:unbroken}

We now illustrate the EFT method for computing the \cwp\ in a theory with two real scalar fields. In this section we consider the power counting relevant for the unbroken phase of this theory.

\subsection{Preliminaries and Power Counting}
\label{sec:preliminariesunbroken}

We take as our starting point the Lagrangian for two real scalar fields $\chi$ and $\phi$,
\ba{
	\CL  &= \frac{1}{2} (\partial_\mu \chi)^2 + \frac{1}{2} (\partial_\mu \phi)^2 - \frac{\lambda_\chi}{24} \chi^4  - \frac{\lambda_3}{4}  \chi^2  \phi^2   -  \frac{\lambda_\phi}{24} \phi^4 -\frac{m_\chi^2}{2} \chi^2 - \frac{m_\phi^2}{2} \phi^2 - \Lambda\ ,
\label{9.1}
}
and assume widely separated scales $m_\chi \gg m_\phi$. Since the power counting is different depending on the sign of the mass-squared terms, we  first do the computation in the unbroken phase, where $m_\chi^2$ and $m_\phi^2$ are both positive, and in Section \ref{sec:broken} we analyze the broken phase where the scalars receive VEVs.

Shifting the fields $\chi = \hat \chi + \chi_q$, $\phi = \hat \phi + \phi_q$, the shifted Lagrangian  Eq.~\eqref{eq:cl} over which the functional integral is performed is 
	\begin{align}
	\begin{split}
\hat{\CL} &=  \frac{1}{2} (\partial_\mu \chi_q)^2 +  \frac{1}{2} (\partial_\mu \phi_q)^2   - \frac{\lambda_\chi}{24}\chi_q^4  - \frac{\lambda_3}{4}  \chi^2_q  \phi^2_q  -  \frac{\lambda_\phi}{24} \phi^4_q  - \frac{\lambda_\chi \hat \chi}{6} \chi_q^3 - \frac{\lambda_3 \hat \phi}{2} \chi_q^2 \phi_q  \\  
& - \frac{\lambda_3 \hat \chi}{2} \chi_q  \phi_q^2 - \frac{\lambda_\phi \hat \phi}{6} \phi_q^3  -\frac12  W_{\chi\chi} \chi_q^2    -\frac12 W_{\phi\phi}  \phi^2_q  -\frac{1}{2} (W_{\chi\phi} + W_{\phi\chi}) \phi_q \chi_q\\
	& + \CJ_\chi \chi_q + \CJ_\phi \phi_q - \hat{\Lambda} \ .
	\end{split}
	\label{9.2}
	\end{align}
The entries of the scalar mass-squared matrix $W$ are given by
	\ba{\bs{
	W_{\chi\chi}&= m_\chi^2 +\frac12 \lambda_\chi \hat \chi^2 +\frac12 \lambda_3 \hat \phi^2    \ , \\
	W_{\phi\phi}&=m_\phi^2 + \frac12 \lambda_\phi \hat \phi^2 +\frac12 \lambda_3 \hat \chi^2 \ , \\
	W_{\phi\chi}&=W_{\chi\phi} = \lambda_3 \hat{\phi} \hat{\chi}\ ,
	}
	\label{9.3}}
	with 
\begin{align}
W &= \begin{bmatrix}W_{\chi\chi}  &  W_{\chi\phi}\\
W_{\phi\chi} &W_{\phi\phi} 
\end{bmatrix}\ .
\label{9.50}
\end{align}
The tree-level \cwp\ is the cosmological constant Eq.~\eqref{eq:cc1},
\ba{
V_{\text{tree}} = \hat \Lambda&=  \frac{\lambda_\chi}{24} \hat \chi^4  + \frac{\lambda_3}{4} \hat  \chi^2 \hat  \phi^2   + \frac{\lambda_\phi}{24} \hat  \phi^4 + \frac{m_\chi^2}{2} \hat \chi^2 +  \frac{m_\phi^2}{2} \hat \phi^2 + \Lambda\ .
\label{9.4}
}
The one-loop \cwp\ is given in terms of the eigenvalues $w_\pm$ of $W$,
	\begin{align}
	w_\pm = \frac{W_{\chi\chi}}{2}\left( 1+ \frac{W_{\phi \phi}}{W_{\chi\chi}} \pm  \sqrt{\frac{4 W_{\phi \chi}^2 }{W_{\chi\chi}^2}+ \left(1-\frac{W_{\phi\phi}}{W_{\chi\chi}} \right)^2 }\right)\ ,  \label{eq:eigs}
	\end{align}	
as
	\begin{align}
	V_{1\text{-loop}}  =  \frac{w_-^2}{64\pi^2} \left[ \ln \frac{w_-}{\mu^2} - \frac{3}{2}  \right] + \frac{w_+^2}{64\pi^2} \left[ \ln \frac{w_+}{\mu^2}  - \frac{3}{2} \right] \ . \label{eq:v1loopscalar}
	\end{align}

Let us introduce a power counting scheme in order to systematically construct the EFT by integrating out the heavy field $\chi_q$. 
We take $m_\chi \sim 1$ and $m_\phi \sim z$, with $z\ll 1$ so that the two masses are widely separated, and assume the  couplings $\lambda_\chi$, $\lambda_\phi$, $\lambda_3$ are order 1. To retain the mass hierarchy, we assume that the VEVs scale like $\hat{\chi}\sim \hat{\phi}\sim z$, i.e.\ our power counting is
\begin{align}
m_\chi^2\sim 1\ ,\qquad m_\phi^2 \sim z^2\ ,\qquad \hat{\chi}\sim \hat{\phi}\sim z\ .
\end{align}
This power counting is valid near the origin, so we can compute the RG improved potential near the classical vacuum.
The $W$-matrix entries in Eq.~\eqref{9.3} scale as
\begin{align}
W_{\chi \chi}  \sim 1\,, \qquad
W_{\phi \phi} \sim z^2 \ ,\qquad 
W_{\phi \chi} = W_{\chi \phi} \sim z^2\,.
\label{9.6}
\end{align}
The first term for $W_{\chi \chi}$ in Eq.~\eqref{9.3} is order 1, and the remaining terms are order $z^2$. This scaling leads to a hierarchy of the eigenvalues $w_\pm$, with $w_+\sim 1$ and $w_-\sim z^2$. In this example, we will keep terms to order $z^6$ in the \cwp, since  non-trivial effects first occur at this order. They will occur at order $z^4$ in the broken phase studied in Section~\ref{sec:broken}.

In particular, let us check the scaling of Eq.~\eqref{eq:v1loopscalar}. $V_{\text{match}}$  is equal to the contribution of the larger eigenvalue expanded in $z$, 
	\ba{\bs{
V_{\text{match}}(\mu) &=	\frac{w_+^2}{64\pi^2} \left( \ln \frac{w_+}{\mu} - \frac{3}{2} \right) \\
&= \frac{1}{64\pi^2} \Bigg\{W_{\chi\chi}^2 \left( \ln \frac{W_{\chi\chi}}{\mu^2} - \frac{3}{2} \right)  + 2  W_{\phi \chi}^2 \left( \ln \frac{W_{\chi\chi}}{\mu^2} - 1   \right)\\
	&\qquad + \frac{2 W_{\phi\chi}^2 W_{\phi\phi}}{W_{\chi\chi}} \left(  \ln \frac{W_{\chi\chi}}{\mu^2}  -1\right)  \Bigg\}+ \CO(z^8)
	}\label{eq:lplus1}
	}
where the first term is order $1$, the second term is order $z^4$, and the last term is  order $z^6$. 
The other eigenvalue gives a contribution that starts at order $z^4$, 
	\ba{ 
\frac{w_-^2}{64\pi^2} \left( \ln \frac{w_-}{\mu} - \frac{3}{2} \right) =	\frac{\left(W_{\phi\phi} - \frac{W_{\phi \chi}^2}{W_{\chi\chi}}\right)^2 }{64\pi^2}\left[  \ln \frac{ \left(W_{\phi\phi} - \frac{W_{\phi \chi}^2}{W_{\chi\chi}}\right)}{\mu^2}  - \frac{3}{2} \right ] + \CO(z^8) \,.\label{eq:lminus1}
	}
We return to these expanded expressions in Section~\ref{sec:interlude}, where we explain how each of the terms in Eq.~\eqref{eq:lplus1} and Eq.~\eqref{eq:lminus1} arise from our approach.

\subsection{Matching to the EFT at $\mu_H$}\label{sec:matchunbroken}

With the preliminaries in place, we can carry out the step-by-step procedure outlined in Section \ref{sec:overview}. 
In Step 1, we evolve the high-energy theory Eq.~\eqref{9.1} to a scale $\mu_H$ comparable to the heavy mass $\mu_H^2 \sim W_{\chi \chi}$, and then shift the fields. 
This gives $\CL(\hat \chi , \hat \phi)$ in terms of the expectation value of fields renormalized at $\mu_H$.  The high energy \cwp\ $V_{\text{CW}}$ in terms of fields renormalized at the reference scale $\mu_0$ is given by using Eq.~\eqref{4.5} where $\gamma_{\chi,\phi}$ are computed in the high-energy theory Eq.~\eqref{9.1}. In this example, the field anomalous dimensions are zero at one-loop.

\begin{table}
\begin{align*}
\renewcommand{\arraystretch}{1.5}
\begin{array}{c|c|c|c|c|c}
\ \text{\,B\,} \ &\  \text{\,U\,}\  & \text{Graph} & \ S \ & \text{Integrand} & V \\
\hline
&  & &  & \\
1\  & 1\  & \raisebox{-0.75cm}{\begin{tikzpicture}
\pgfmathsetmacro{\rad}{0.75}
\draw[dashed] (0,0) circle (\rad);
\draw(-1,0) node [align=center] {$\chi$};
\draw(1,0) node [align=center] {\phantom{$\chi$}};
\end{tikzpicture}} & \frac12 & - \ln (p^2- W_{\chi \chi}) & \frac{1}{32\pi^2} W_{\chi \chi}^2 \left[- \frac{1}{2\epsilon} + \frac12 \ln \frac{W_{\chi \chi}}{\mu^2} - \frac34 \right] 
\\[25pt]
z^2\  & z^4\  & \raisebox{-0.75cm}{\begin{tikzpicture}
\pgfmathsetmacro{\rad}{0.75}
\draw[dashed] (0,0) ++ (90:\rad) arc (90:270:\rad);
\draw[densely dotted] (0,0) ++ (-90:\rad) arc (-90:90:\rad);
\filldraw (0,-\rad) circle (0.05);
\filldraw (0,\rad) circle (0.05);
\draw(-1,0) node [align=center] {$\chi$};
\draw(1,0) node [align=center] {$\phi$};
\end{tikzpicture}} & \frac12 & \frac{W_{\chi \phi} W_{\phi \chi}}{(p^2-W_{\chi \chi}) p^2} &
\frac{1}{32\pi^2} W_{\chi \phi} W_{\phi \chi} \left[ -\frac{1}{\epsilon} + \ln \frac{W_{\chi \chi}}{\mu^2} - 1 \right]
\\[25pt]
z^4 \ & z^6\   & \raisebox{-0.75cm}{\begin{tikzpicture}
\pgfmathsetmacro{\rad}{0.75}
\draw[dashed] (0,0) ++ (90:\rad) arc (90:270:\rad);
\draw[densely dotted] (0,0) ++ (-90:\rad) arc (-90:90:\rad);
\filldraw (0,-\rad) circle (0.05);
\filldraw (0,\rad) circle (0.05);
\draw (\rad,0) circle (0.05);
\draw(-1,0) node [align=center] {$\chi$};
\draw(1,0) node [align=center] {$\phi$};
\end{tikzpicture}} & \frac12 & \frac{W_{\chi \phi} W_{\phi \chi}}{(p^2-W_{\chi \chi}) p^2} \frac{W_{\phi\phi}}{p^2} &
\frac{1}{32\pi^2} \frac{W_{\chi \phi} W_{\phi \chi} W_{\phi \phi}}{W_{\chi \chi}} \left[ -\frac{1}{\epsilon} + \ln \frac{W_{\chi \chi}}{\mu^2} - 1 \right]
\\[25pt]
z^4\ & z^8\  &\raisebox{-0.75cm}{\begin{tikzpicture}
\pgfmathsetmacro{\rad}{0.75}
\draw[dashed] (0,0) ++ (90:\rad) arc (90:180:\rad);
\draw[dashed] (0,0) ++ (-90:\rad) arc (-90:0:\rad);
\draw[densely dotted] (0,0) ++ (0:\rad) arc (0:90:\rad);
\draw[densely dotted] (0,0) ++ (180:\rad) arc (180:270:\rad);
\filldraw (0,-\rad) circle (0.05);
\filldraw (0,\rad) circle (0.05);
\filldraw (\rad,0) circle (0.05);
\filldraw (-\rad,0) circle (0.05);
\draw (135:1) node [align=center] {$\chi$};
\draw (-45:1) node [align=center] {$\chi$};
\draw (45:1) node [align=center] {$\phi$};
\draw (225:1) node [align=center] {$\phi$};
\end{tikzpicture}} & \frac14 &  \frac{ W_{\chi \phi}^2 W_{\phi \chi}^2}{(p^2-W_{\chi \chi})^2 p^4} &
\frac{1}{32\pi^2} \frac{W_{\chi \phi}^2 W_{\phi \chi}^2}{W_{\chi \chi}^2} \left[ \frac{1}{2\epsilon} - \frac12 \ln \frac{W_{\chi \chi}}{\mu^2} + 1 \right]
\\[25pt]
\end{array}
\end{align*}
\caption{\label{tab:1} One loop matching for the cosmological constant to order $z^6$. The open circle is $W_{\phi \phi}$ and the black dot is $W_{\phi \chi}$. $S$ is the symmetry factor for the diagram. The integrand is given in the third column. The matching contribution to $V$ is given by the last column, dropping $1/\epsilon$ terms. We give the $z$-scaling in both the unbroken (U) and broken (B) phases. In the unbroken phase we keep contributions through $\CO(z^6)$, and in the broken phase we keep them through $\CO(z^4)$.
}
\end{table}

In Step 2, we integrate out $\chi_q$ and match onto the EFT. The EFT Lagrangian is parameterized as
\begin{align}
\CL_{\text{EFT}}  &=   \frac{1}{2} (\partial_\mu \phi_q)^2    -\widetilde{\sigma} \phi_q -\frac{\widetilde{m}^2}{2} \phi^2_q  -  \frac{\widetilde{\rho}}{6} \phi_q^3 - \frac{ \widetilde{\lambda}}{24} \phi_q^4     
    - \widetilde{\Lambda} + \mathcal{O}(z^8 )\ .
\label{9.12a}
\end{align}
The one-loop matching onto the EFT Lagrangian is given by computing the one-loop $\chi$ bubble,  expanding in the power counting parameter $z$, and dropping all divergent terms. The graphs are shown in Table~\ref{tab:1}. Since we include power corrections up to order $z^4$, we have to include insertions of power suppressed operators in the $\chi$ loop, with all infrared scales such as $W_{\phi \phi}$ expanded out.
In the unbroken case, only the first three graphs contribute, the fourth is  necessary for the broken phase analysis in Section~\ref{sec:broken}. These contributions add up precisely to $V_{\text{match}}(\mu_H)$ given in Eq.~\eqref{eq:lplus1}, which contributes to the cosmological constant in the EFT since it has no quantum fields. Because the matching is performed at $\mu_H^2 \sim W_{\chi \chi}$, there are no large logarithms in this expression. It is worth noting that the matching computation only involves logarithms of $W_{\chi \chi}/\mu^2$. Even though $\phi$ particles enter in  loop graphs, there are no logarithms of $W_{\phi \phi}$ because it is an infrared scale which has been expanded out in a power series.\footnote{See Ref.~\cite{Manohar:2018aog} for a more extensive discussion of logarithms in matching conditions.}

The tree-level matching to order $z^6$ is given by the graph shown in Fig.~\ref{fig:8}. To compute this graph, we follow the same steps outlined in the $O(N)$ example. Write the terms linear in $\chi_q$ in Eq.~\eqref{9.2} as
\begin{align}
\hat{\CL} &\supset - X_1(\phi_q) \chi_q \ ,\qquad
X_1(\phi_q) = W_{\phi \chi} \phi_q + \frac12 \lambda_3 \hat \chi \phi_q^2 -  \CJ_\chi\ .
\label{9.10}
\end{align}
 Here $X_1 \sim z^3$ since $W_{\phi\chi} \sim z^2$, $\hat \chi \sim z$,  and $\phi_q \sim z$.
Fig.~\ref{fig:8} then gives the tree-level matching
\begin{align}
\CL_{\text{EFT}} &\supset \frac12 X_1 \frac{1}{W_{\chi \chi}-p^2} X_1 = \frac12 X_1 \frac{1}{W_{\chi \chi}} X_1 + \frac12 X_1 \frac{p^2}{W_{\chi \chi}^2} X_1 +\ldots
\label{9.9}
\end{align}
where $p$ is the momentum flowing through the $\chi_q$ line, which is order $z$ since it is a momentum in the EFT, and $1/2$ is the symmetry factor. To order $z^6$, we only need the first term. The EFT Lagrangian is then 
\begin{align}
\begin{split}
\CL_{\text{EFT}} &=   \frac{1}{2} (\partial_\mu \phi_q)^2   -\frac12 W_{\phi\phi}  \phi^2_q   - \frac{\lambda_\phi \hat{\phi}}{3!} \phi_q^3  - \frac{\lambda_\phi}{4!} \phi_q^4   - \hat \Lambda - V_{\text{match}}(\mu_H) \\
	&+ \frac12 \frac{1}{W_{\chi \chi}} \left( W_{\phi \chi} \phi_q + \frac12 \lambda_3 \hat \chi \phi_q^2 -  \CJ_\chi 
 \right)^2 + \CJ_\phi \phi_q + \mathcal{O}(z^7) \ .
 \end{split}
\label{9.11}
\end{align}
Using Eq.~\eqref{9.3}, and keeping the leading term $W_{\chi \chi} \to m_\chi^2$ to order $z^6$, we identify the matching conditions for the EFT couplings at the scale $\mu_H$ as
\begin{align}
\begin{split}
\widetilde \lambda \overset{\mu=\mu_H}{=}&  \lambda_\phi-  \frac{3 \lambda_3^2 \hat \chi ^2 }{ m_\chi^2}  + \CO(z^4)  \,, \\
\widetilde \rho  \overset{\mu=\mu_H}{=}& \hat \phi \left[ \lambda_\phi- \frac{3 \lambda_3^2  \hat \chi^2}{ m_\chi^2 }\right]  + \CO(z^5)\,,\\
\widetilde m^2  \overset{\mu=\mu_H}{=}&  m_\phi^2 + \frac12 \lambda_\phi  \hat \phi^2 + \frac12 \lambda_3 \hat \chi^2  - \frac{ \lambda_3^2 \hat \chi^2 \hat \phi^2}{m_\chi^2} + \frac{ \lambda_3 \hat \chi  \CJ_\chi }{m_\chi^2}   + \CO(z^6)  \,, \\
{\equiv}\ &  \widetilde m_{-}^2 +  \frac{ \lambda_3 \hat \chi  \CJ_\chi }{m_\chi^2}  + \CO(z^6)  \,, \\
\widetilde \sigma \overset{\mu=\mu_H}{=}& -\CJ_\phi + \frac{ \lambda_3 \hat \chi \hat \phi \CJ_\chi}{m_\chi^2} \,   + \CO(z^7)\,,  \\
\widetilde \Lambda  \overset{\mu=\mu_H}{=}&   \hat \Lambda + V_{\text{match}} -  \frac{ \left(\CJ_\chi  \right)^2}{2 m_\chi^2} + \CO(z^8) \,.
\end{split}
\label{9.13}
\end{align}

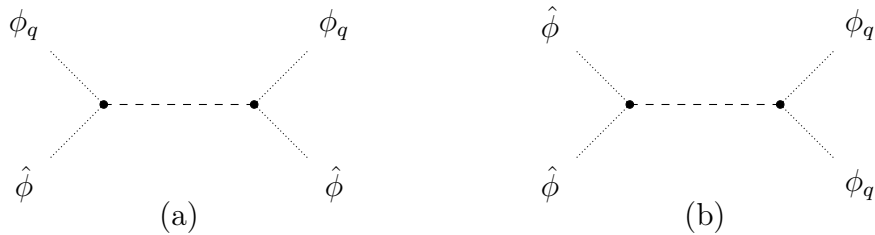
\begin{figure}
\begin{center}
\begin{tikzpicture}
\draw[dashed] (-1,0) -- (1,0);
\filldraw (-1,0) circle (0.05);
\filldraw (1,0) circle (0.05);
\draw[densely dotted] (1,0) -- +(45:1);
\draw[densely dotted] (1,0) -- +(-45:1);
\draw[densely dotted] (-1,0) -- +(135:1);
\draw[densely dotted] (-1,0) -- +(225:1);
\draw (0,-1.5) node [align=center] {(a)};
\draw (1,0)+(45:1.5)  node [align=center] {$\phi_q$};
\draw (1,0)+(-45:1.5)  node [align=center] {$\hat \phi$};
\draw (-1,0)+(135:1.5)  node [align=center] {$\phi_q$};
\draw (-1,0)+(225:1.5)  node [align=center] {$\hat \phi$};
\end{tikzpicture}
\hspace{2cm}
\begin{tikzpicture}
\draw[dashed] (-1,0) -- (1,0);
\filldraw (-1,0) circle (0.05);
\filldraw (1,0) circle (0.05);
\draw[densely dotted] (1,0) -- +(45:1);
\draw[densely dotted] (1,0) -- +(-45:1);
\draw[densely dotted] (-1,0) -- +(135:1);
\draw[densely dotted] (-1,0) -- +(225:1);
\draw (0,-1.5) node [align=center] {(b)};
\draw (1,0)+(45:1.5)  node [align=center] {$\phi_q$};
\draw (1,0)+(-45:1.5)  node [align=center] {$\phi_q$};
\draw (-1,0)+(135:1.5)  node [align=center] {$\hat \phi $};
\draw (-1,0)+(225:1.5)  node [align=center] {$\hat \phi$};
\end{tikzpicture}
\end{center}
\caption{\label{fig:9} (a)  Graphs allowed in the high-energy theory. (b) Graphs not allowed in the high-energy theory. }
\end{figure}

We emphasize that the low-energy Lagrangian Eq.~\eqref{9.12a} is manifestly {\it not} given by taking the EFT Lagrangian and shifting $\phi=\hat \phi+\phi_q$. The effect of a scalar field shift has already been written in Eq.~\eqref{2.12}.  To illustrate this, start with the couplings in Eq.~\eqref{9.13} with $\hat \phi=0$, and then shift the field $\phi_q \to \hat \phi + \phi_q$. The shifted value of $\widetilde \rho$ is $\widetilde \lambda \hat \phi$, which matches the value in Eq.~\eqref{9.13}.
However, the mass term after the shift is
\begin{align}
\widetilde m^2 &\overset{\text{shift}}{=} m_\phi^2 + \frac{ \lambda_\phi}{2} \hat \phi^2 + \frac{\lambda_3}{2} \hat \chi^2   - \frac{3 \lambda_3^2 \hat \chi^2 \hat \phi^2}{2m_\chi^2} + \frac{ \lambda_3 \hat \chi  \CJ_\chi }{ m_\chi^2}   \,,
\label{9.14}
\end{align}
where the third term has coefficient $-3/2$ instead of $-1$ as in Eq.~\eqref{9.13}.
The reason for this difference can be seen by looking at the contribution from Fig.~\ref{fig:8} to $\widetilde \lambda$ in the EFT.  If we first match onto the EFT and then shift, the $\hat \phi^2 \phi_q^2$ term in the Lagrangian is given by replacing two external lines in Fig.~\ref{fig:8}(b) by $\hat \phi$ and two by $\phi_q$. There are 6 possible ways of doing this. Now, let us make the same replacement in Fig.~\ref{fig:8}(a). The six possible ways are divided into two types: four possibilities with topology Fig.~\ref{fig:9}(a) with $\hat \phi$ on opposite ends of the $\chi$ line, and two with topology Fig.~\ref{fig:9}(b), with $\hat \phi$ on the same side of the $\chi$ line. The left vertex in Fig.~\ref{fig:9}(b) is  $\chi_q \hat \phi^2$ which is not present in the high-energy theory because it is cancelled by the classical piece of the source, i.e.\ we have no terms linear in the quantum fields. Thus only 4 out of 6 choices contribute, and the correct coefficient is $2/3$ of that given by shifting in the EFT, i.e. $(2/3) \cdot (-3/2) = -1$. This explains the difference between Eq.~\eqref{9.14} and Eq.~\eqref{9.13}, and is a reflection of the fact that information about which diagrams are 1PI is lost in the EFT matching.

\begin{figure}
\begin{center}
\begin{tikzpicture}
\begin{scope}[shift={(3,0)}]
\draw[densely dotted] (0,0) circle (0.75);
\draw[dashed] (-0.75,0) -- (0.75,0);
\filldraw (-0.75,0) circle (0.05);
\filldraw (0.75,0) circle (0.05);
\draw (0,-1.5) node [align=center] {(a)};
 \end{scope}
\begin{scope}[shift={(6,0)}]
 \draw[dashed] (-0.5,0) -- (0.5,0);
 \draw[densely dotted] (1,0) circle (0.5);
 \draw[densely dotted] (-1,0) circle (0.5);
 \filldraw (-0.5,0) circle (0.05);
 \filldraw (0.5,0) circle (0.05);
 \draw (0,-1.5) node [align=center] {(b)};
\end{scope}
 \begin{scope}[shift={(10,0)}]
\draw[densely dotted] (0,0.5) circle (0.5);
\draw[densely dotted] (0,-0.5) circle (0.5);
\filldraw (0,0) circle (0.05);
\draw (0,-1.5) node [align=center] {(c)};
 \end{scope}
 \end{tikzpicture}
\end{center}
\caption{\label{fig:10} Two loop contribution from a $\chi \phi^2$ interaction. Graphs (a) and (b) in the high-energy theory reduce to (c) in the EFT.}
\end{figure}
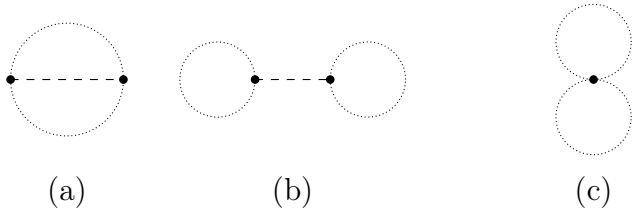

There is a related problem for the two-loop contribution to the \cwp. A $\chi \phi^2$ interaction leads to the two-loop contributions shown in Fig.~\ref{fig:10}(a,b), which reduce to Fig.~\ref{fig:10}(c) in the EFT. However, the 1PI nature of the \cwp\ implies that only Fig.~\ref{fig:10}(a) must be included in computing $V$, not 
 Fig.~\ref{fig:10}(b). The two possible contractions of the $\phi$ lines cannot be distinguished in the EFT, since the $\chi$ line has been shrunk to a point. This reason is why we use the tadpole condition discussed in Sec.~\ref{sec:jackiw} rather than the 1PI structure of graphs to compute $V_{\text{CW}}$.

\subsection{RG Improvement}
\label{sec:muLunbroken}

We move on to Steps 3 and 4. 
The couplings in the low-energy theory are evolved using the RGE to a low scale $\mu_L$ of order $\widetilde m^2$, the mass of $\phi_q$. The RGE are given in Eq.~\eqref{B.1} with solutions Eq.~\eqref{B.2}. In particular, we need
\begin{align}
\begin{split}
\widetilde{m}^2(\mu_L)
&=\widetilde{m}^2(\mu_H) \left\{ \eta^{-1/3} \left[1- \frac12 \xi \right] +\frac12 \xi  \eta^{-1} \right\} \,,\\
\widetilde{\Lambda}(\mu_L)   &= \widetilde{\Lambda}(\mu_H)+ \frac{\widetilde{m}^4(\mu_H)}{2\widetilde{\lambda}(\mu_H)} \biggl\{ \frac13 \left( 3 - 6 \xi +2  \xi^2 \right) -\frac12
   \xi \left(\xi -2 \right) \eta^{-1/3} \\
   &\qquad\qquad\qquad\qquad\qquad\qquad -\frac{1}{4} \left(\xi -2\right)^2\eta^{1/3} + \frac1{12} \xi^2 \eta^{-1} \biggr\}\ ,
   \end{split}
\label{B.2a}
\end{align}
where
\begin{align}
t = \frac{1}{16\pi^2} \ln \frac{\mu_L}{\mu_H}\,, \qquad
\eta = 1- 3 \widetilde{\lambda}(\mu_H) t \,, \qquad
\xi = \frac{ \widetilde{\rho}^2(\mu_H) }{ \widetilde{\lambda}(\mu_H) \widetilde{m}^2(\mu_H)} =  \frac{ \hat{\phi}^2 \widetilde{\lambda}(\mu_H) }{  \widetilde{m}^2(\mu_H)} \,.
\label{B.3a}
\end{align}

At the scale $\mu_L$, the quantum field $\phi_q$ is integrated out, and we obtain the \cwp\
\begin{align}
V_{\text{CW}} &= \widetilde \Lambda(\mu_L) + \frac{1}{64 \pi^2} \widetilde m^4(\mu_L) \left[ \ln \frac{ \widetilde m^2(\mu_L) }{\mu_L^2} - \frac32 \right]\ .
\label{9.16}
\end{align}
The one-loop term does not have any large logarithms if $\mu_L^ 2\sim \widetilde m^2$.

\subsection{Interlude: A Fixed Order Perspective}
\label{sec:interlude}

Before completing Step 5, let us take a moment to gain some intuition as to how our result Eq.~\eqref{9.16} matches onto the one-loop  fixed  order expression in Eq.~\eqref{eq:v1loopscalar}. 
In doing so, we will understand how the large-eigenvalue and small-eigenvalue terms in $V_{1{\text{-loop}}}$ in Eq.~\eqref{eq:v1loopscalar} arise from expanding in $z$ the functional determinant that results from doing the path integral over the quantum fields. This subsection can be read as a more detailed perspective on computing $V_{\text{CW}}$ from our Steps 2 and 4, where we work only at fixed order in perturbation theory and thus ignore the steps that involve RG evolution.  We have postponed this discussion until now since it is most interesting in the case where $W$ is not diagonal.

A preliminary comment is that our RG improved result Eq.~\eqref{9.16} manifestly evaluates to the fixed one-loop order expression $V_{\text{CW}} = V_{\text{tree}} + V_{1\text{-loop}}$ upon taking $\mu=\mu_H=\mu_L$. Dropping the $\CJ$ terms since they are formally of higher loop order, and taking $\eta = 1$ in Eq.~\eqref{B.2a}, we see that
	\ba{
	\widetilde{m}^2 &\to  \widetilde{m}_-^2 = W_{\phi\phi} - \frac{W_{\phi\chi}^2}{W_{\chi\chi}}  \,,\qquad \widetilde{\Lambda} \to V_{\text{match}} + \hat{\Lambda}\,.
	}
The quantity $\widetilde{m}_-^2$ (which was first defined in Eq.~\eqref{9.13}) is precisely the combination that appeared in the expansion of the small eigenvalue $w_-$ in the one-loop \cwp\ in Eq.~\eqref{eq:lminus1}. We have already noted that $V_{\text{match}}$ gives the expansion of the large eigenvalue in Eq.~\eqref{eq:lplus1}, and that $\hat{\Lambda}$ is the tree-level \cwp\ in Eq.~\eqref{9.4}. 
Therefore, in this limit we recover the fixed order answer.

We can understand why this  is the case by starting from the functional integral Eq.~\eqref{eq:z3}. The quadratic action in the quantum fields in momentum space is
\begin{align}
\hat{\mathcal{L}} &= \frac12 \begin{bmatrix} \chi_q & \phi_q \end{bmatrix}
\Delta
\begin{bmatrix} \chi_q \\ \phi_q \end{bmatrix} + \CJ^T
\begin{bmatrix} \chi_q \\ \phi_q \end{bmatrix}  - \hat \Lambda \,,
\label{32.1}
\end{align}
with
\begin{align}
\Delta &=
\begin{bmatrix}p^2 - W_{\chi \chi} & - W_{\chi \phi} \\ -W_{\phi \chi} & p^2 - W_{\phi \phi}  \end{bmatrix}\,,\qquad
  \CJ = \begin{bmatrix} \CJ_\chi \\ \CJ_{\phi} \end{bmatrix} \,.
\label{32.3}
\end{align}
The effective action is
\begin{align}
e^{\frac{i}{\hbar} \Gamma} &= \int D\chi_q D \phi_q \ e^{\frac{i}{\hbar} \int d^4x  \hat{\mathcal{L}}} \nn
&= \left[ \det \Delta \right]^{-1/2} \exp \left[ -i \int \rd^4 x  \left( \hat \Lambda +  \CJ^{T} \Delta^{-1} \CJ \right) \right]\,.
\label{32.2}
\end{align}
$\CJ$ is adjusted to that $\VEV{\chi_q}=0$ and $\VEV{\phi_q}=0$, and is formally of one-loop order, so the terms quadratic in $\CJ$ can be dropped to one-loop order, and we drop them for the rest of this subsection.

The one-loop contribution is given by $\left[ \det \Delta \right]^{-1/2} $. Diagonalizing $W$,
\begin{align}
\Delta &=
\begin{bmatrix}p^2 - w_+ &  0 \\  0 &p^2-w_-  \end{bmatrix}\,,\qquad
 \det \Delta = \det \left(p^2 - w_+ \right) \det \left(p^2-w_-\right)\,.
\label{32.4}
\end{align}
The functional determinant is  well-known, and evaluates to
\begin{align}
\frac12 \ln \det \left(p^2 - w \right) = i \int \rd^4 x  \ \frac{w^2}{64 \pi^2} \left[ \ln \frac{w}{\mu^2} - \frac32\right]
\label{32.5}
\end{align}
giving Eq.~\eqref{eq:v1loopscalar} for the \cwp\ in the full theory at one loop.

To compare with the EFT method, the integrals over $\chi_q$ and $\phi_q$ are performed sequentially. The EFT result is reminiscent of the determinant identity
for a block matrix
\begin{align}
\det \Delta = \det \begin{bmatrix} A & B \\ B^T & C \end{bmatrix} &= \det A \det \left( C - B^T A^{-1} B \right)\,.
\label{32.6}
\end{align}
Performing the integral over $\chi_q$ gives
\begin{align}
\frac12 \ln \det A &= \frac12 \ln \det \left(p^2 - W_{\chi \chi} \right) =  i \int \rd^4 x  \frac{W_{\chi \chi}^2}{64 \pi^2} \left[ \ln \frac{W_{\chi \chi}}{\mu^2} - \frac32\right]\,,
\label{32.7}
\end{align}
which  is the contribution of the first graph in Table~\ref{tab:1} for $V_{\text{match}}$.

The remaining functional integral is over $\phi_q$. The quadratic part is
\begin{align}
\Delta_\phi \equiv C - B^T A^{-1} B &= p^2- W_{\phi \phi} - W_{\phi \chi} \frac{1}{p^2 - W_{\chi \chi}} W_{\chi \phi}\,,
\label{32.8}
\end{align}
which is precisely the form of the matching Eq.~\eqref{9.9} onto the EFT.  We also need to compute $\ln \det \Delta_\phi = \tr \ln \Delta_\phi$. The trace in momentum space is the integral $\rd ^4 p$. The momentum integral gets contributions from $p^2 \sim W_{\chi \chi} \sim 1$ and $p^2 \sim W_{\phi \phi} \sim z^2$.
For $p^2 \sim W_{\chi \chi}$,
\begin{align}
\ln \Delta_\phi &= \int \frac{\rd^4 p}{(2\pi) ^4}\biggl\{ \ln p^2 + \ln \left[ 1 - \frac{W_{\phi \phi}}{p^2} - W_{\phi \chi} \frac{1}{p^2 \left(p^2 - W_{\chi \chi}\right)} W_{\chi \phi} \right]  \biggr \} \nn
 &= \int \frac{\rd^4 p}{(2\pi) ^4}\biggl\{ \ln p^2 - \sum_n  \left[ \frac{W_{\phi \phi}}{p^2} + W_{\phi \chi} \frac{1}{p^2 \left(p^2 - W_{\chi \chi}\right)} W_{\chi \phi} \right]^n
 \biggr\}\ .
\label{32.9}
\end{align}
The first term vanishes in dimensional regularization, and the remaining terms are the matching contributions in Table~\ref{tab:1}.

For $p^2 \sim W_{\phi \phi}$, 
\begin{align}
\ln \Delta_\phi  &= \int \frac{\rd^4 p}{(2\pi) ^4} {\ln} \biggl\{ p^2- \left[ W_{\phi \phi} - W_{\phi \chi} \left( \frac{1}{W_{\chi \chi}} + \frac{p^2}{W_{\chi \chi}^2} + \ldots \right) W_{\chi \phi} \right]
  \biggr \} \,, \nn
 & {= \int \frac{\rd^4 p}{(2\pi) ^4} {\ln} \left[ p^2- \left( W_{\phi \phi} - \frac{W_{\phi \chi} W_{\chi \phi}}{W_{\chi \chi}}\right)  \right] + \ldots \ . }
\label{32.10}
\end{align}
The expression $W_{\phi \phi} - {W_{\phi \chi} W_{\chi \phi}}/{W_{\chi \chi}}$ is the mass $\widetilde m^2$ in the EFT after matching, and so {the low-energy contribution of $\ln \Delta_\phi$} gives the $w_-$ contribution to the \cwp.

Let us summarize this discussion: We start with a two-field problem with a hierarchy of mass scales $m_\chi^2 \sim W_{\chi\chi} \gg m_\phi^2 \sim W_{\phi\phi}$, which leads to a hierarchy of  mass eigenvalues $w_+  \gg w_-$.  $V_{\text{match}}$ is the contribution to the one-loop \cwp\ coming from the larger $w_+$ eigenvalue. Graphs that contribute to  $V_{\text{match}}$ in a $z$ expansion arise from (1) performing the functional integral over $\chi$, and (2) expanding the remaining functional determinant in the power counting expansion. Meanwhile, the contribution of $w_-$ to the one-loop \cwp\ comes from expanding the functional determinant from integrating out $\phi_q$ for small EFT momentum $p^2\sim z^2$. This gives the fixed one-loop order \cwp, which matches the result of applying our Steps 2 and 4. As we have explained, RG improvement takes more work, and is the content of the remaining steps.

\subsection{Tadpole Improvement}
\label{sec:tadpoleunbroken}

It remains to compute  $\CJ_\phi$ and $\CJ_\chi$. Moving on to Step 5, we fix these by requiring that $\VEV{\chi_q}$ and $\VEV{\phi_q}$ vanish, where $\chi_q$ and $\phi_q$ are fields in the original high-energy theory. From Eq.~\eqref{eq:z3}, these expectation values are determined by differentiation w.r.t. $\CJ_\phi$ and $\CJ_\chi$.
Differentiating the EFT Lagrangian w.r.t.\ $\CJ_\phi$ brings down a factor of $\phi_q$ in the EFT, as can be seen from the parameters Eq.~\eqref{9.13}, so setting the $\phi_q$ tadpole to zero in the high-energy theory is equivalent to setting it to zero in the low-energy theory. We do not need this calculation for our result, since the running of the cosmological constant and the terms in Eq.~\eqref{9.16} do not depend on $\widetilde \sigma$, which is the only parameter that depends on $\CJ_\phi$.

Differentiating the EFT Lagrangian w.r.t.\ $\CJ_\chi(\mu_H)$ using Eq.~\eqref{9.13}, we see that $\chi_q(\mu_H)$ in the high-energy theory matches on to
\begin{align}
\chi_q(\mu_H)\to -\frac{ \lambda_3(\mu_H) \hat \chi }{2 m_\chi^2(\mu_H)} {\left[ \phi_q^2 \right]_{\mu_H} } -\frac{ \lambda_3(\mu_H) \hat \chi \hat \phi}{m_\chi^2(\mu_H)} \phi_q (\mu_H)  +  \frac{ \CJ_\chi(\mu_H)}{ m_\chi^2(\mu_H)} \ ,
\label{9.17}
\end{align}
and the tadpole condition becomes
\begin{align}
\CJ_\chi(\mu_H) &= \frac12 \lambda_3(\mu_H) \hat \chi  \VEV{ {\left[ \phi_q^2 \right] } }_{\mu_H} +   \lambda_3(\mu_H) \hat \chi  \hat \phi \VEV{ \phi_q }_{\mu_H }
\label{9.18}
\end{align}
where the r.h.s.\ is computed in the low-energy theory. Since $\VEV{\phi_q}=0$ in the low-energy theory by the $\phi$ tadpole condition, we only need the first term. The one-loop graph to be computed is shown in Fig.~\ref{fig:11}, and contains a large logarithm.

One can sum the LL series for the tadpole, as in Section~\ref{sec:tadpoleon}, by computing the anomalous dimension of the composite operator $\left[ \phi_q^2 \right]$ in the low-energy theory, 
\begin{align}
\frac{\rd}{\rd t} \left[\phi_q^2 \right] &= - \widetilde  \lambda  \left[ \phi_q^2 \right] -2 \widetilde \rho\, \phi_q - 2 \widetilde m^2\ ,
\label{9.19bb}
\end{align}
and integrating to relate { $\left[ \phi_q^2 \right]_{\mu_H}$ and $\left[ \phi_q^2 \right]_{\mu_L}$. } An alternate derivation is to differentiate the Lagrangian renormalized at $\mu_L$ with respect to $\widetilde m^2(\mu_H)$ using the RGE solutions Eq.~\eqref{B.2}, which gives the same result:
\begin{align}
\begin{split}
\left[ \phi_q^2 \right]_{\mu_H} &= \eta^{-1/3}\left[ \phi_q^2 \right]_{\mu_L} + 2 \frac{\widetilde \rho(\mu_H)}{\widetilde \lambda(\mu_H)} \left( \eta^{-1/3}-1\right) \left[ \phi_q \right]_{\mu_L} \\
& + \frac{\widetilde \rho^2(\mu_H)} {\widetilde \lambda^2(\mu_H)} 
\left( \eta^{1/3}-1\right)^2 \eta^{-1/3} 
-  \frac{2\widetilde m^2(\mu_H)} {\widetilde \lambda (\mu_H)} \left( \eta^{1/3}-1\right) \,,
\end{split}
\\
\left[ \phi_q \right] _ {\mu_H} &= \left[ \phi_q \right]_{\mu_L} \,,
\label{9.19b}
\end{align}
where $\eta=1-3 \lambda (\mu_0)/(16 \pi^2) \ln \mu/\mu_0$.
Substituting in Eq.~\eqref{9.18},
\begin{align}
\begin{split}
\CJ_\chi(\mu_H) &=\frac12  \lambda_3(\mu_H) \hat \chi  \eta^{-1/3}  \VEV{ \left[ \phi_q^2 \right] }_{\mu_L} + \lambda_3(\mu_H) \hat \chi   \frac{\widetilde \rho(\mu_H)}{\widetilde \lambda(\mu_H)} \left( \eta^{-1/3}-1\right) \VEV{ \phi_q }_{\mu_L} \\
& + \frac12  \lambda_3(\mu_H) \hat \chi \frac{\widetilde \rho^2(\mu_H)} {\widetilde \lambda^2(\mu_H)} 
\left( \eta^{1/3}-1\right)^2 \eta^{-1/3} -  \lambda_3(\mu_H) \hat \chi\frac{\widetilde m^2(\mu_H)} {\widetilde \lambda (\mu_H)} \left( \eta^{1/3}-1\right) \\
& +   \lambda_3(\mu_H) \hat \chi  \hat \phi \VEV{ \phi_q }_{\mu_L}
\end{split}
\label{9.19a}
\end{align}
which can also be obtained by differentiating the EFT Lagrangian at $\mu_L$ w.r.t.\ $\CJ(\mu_H)$.
Using $\widetilde \rho(\mu_H) = \widetilde \lambda(\mu_H) \hat \phi$ and $\VEV{\phi_q}_{\mu_L}=0$, the RG improved tadpole is
\begin{align}
\begin{split}
\CJ_\chi(\mu_H)  &= \frac12 \lambda_3(\mu_H) \hat \chi \eta^{-1/3} \VEV{ \left[ \phi_q^2 \right] }_{\mu_L} +  \frac{\lambda_3(\mu_H)}{2\widetilde \lambda(\mu_H)  } \frac{ \left( \eta^{1/3}-1\right)^2}{\eta^{1/3}} \hat \chi  \hat \phi^2 \\
&  -  \frac{\lambda_3(\mu_H) }{\widetilde \lambda(\mu_H)} \left( \eta^{1/3}-1\right)\hat \chi\, \widetilde m^2(\mu_H)  \ .
\end{split}
\label{9.32}
\end{align}

The low-energy matrix element of $\left[ \phi_q^2 \right]_\mu$ is
\begin{align}
 \VEV{\left[ \phi_q^2 \right]}_\mu &= \frac{1}{16\pi^2}\  \widetilde m^2 \left[ \ln \frac{\widetilde m^2}{\mu^2} - 1 \right]\,,
\label{9.30}
\end{align}
from Fig.~\ref{fig:11}.
Eq.~\eqref{9.32} with the substitution of Eq.~\eqref{9.30} can be solved to obtain $\CJ_\chi(\mu_H) $. $\widetilde m^2$ depends linearly on $\CJ_\chi(\mu_H) $, as given in Eq.~\eqref{9.13}. Taking the linear term in $\CJ_\chi(\mu_H) $ from the third term to the l.h.s.\ and solving gives
\begin{align}
\CJ_\chi(\mu_H)  & = \biggl\{ 1 + \frac{3 \lambda^2_3(\mu_H)}{ \widetilde \lambda(\mu_H)\lambda_\chi (\mu_H)v_\chi^2(\mu_H) } \hat \chi^2
\biggr\}^{-1} \biggl\{\frac{\lambda_3(\mu_H) \hat \chi }{32\pi^2} \eta^{-1/3}   \widetilde m^2(\mu_L) \left[ \ln \frac{\widetilde m^2(\mu_L)}{\mu_L^2} - 1 \right]   \nn
&+\frac{\lambda_3(\mu_H)}{2\widetilde \lambda(\mu_H)  } \frac{ \left( \eta^{1/3}- 1\right)^2}{\eta^{1/3}} \hat \chi  \hat \phi^2  -  \frac{\lambda_3(\mu_H) }{\widetilde \lambda(\mu_H)} \left( \eta^{1/3}-1\right)\hat \chi\, \widetilde m^2_-(\mu_H)  \biggr\}
\label{9.33b}
\end{align}
where the  $\VEV{ \phi_q^2 }_\mu$ contribution depends on $\CJ_\chi(\mu_H) $ through $\widetilde m^2$, but the last term does not, since it depends on
$\widetilde m^2_-$. To LL order, the $ \VEV{\left[ \phi_q^2 \right]}_{\mu_L}$ term can be dropped, and Eq.~\eqref{9.33b} explicitly gives $\CJ_\chi(\mu_H) $. To LL plus one-loop accuracy, $\CJ_\chi(\mu_H) $ is given by substituting the LL value of $\CJ_\chi(\mu_H) $ into the one-loop term from $ \VEV{\left[ \phi_q^2 \right]}_{\mu_L}$. 
Expanding Eq.~\eqref{9.33b} in $t$, and only keeping the single log in the LL series gives
\begin{align}
\CJ_\chi(\mu_H)  & \approx \lambda_3(\mu_H) \hat \chi\, \widetilde m_-^2 (\mu_H) t\ .
\label{9.33}
\end{align}

\subsection{Comparison with the Two-loop Result}
\label{sec:twoloopunbroken}

We can now compare our RG improved result to the known two-loop calculation by expanding the RG series in a power series in $t = ( \ln {\mu_L}/{\mu_H})/(16 \pi^2)$. The cosmological constant contribution is
\begin{align}
\widetilde \Lambda(\mu_L) &=\widetilde \Lambda(\mu_H) + \frac12 \widetilde m^4(\mu_H) t + \frac12 \widetilde m^2(\mu_H) \left[ \widetilde \lambda(\mu_H) \widetilde m^2(\mu_H) 
+\widetilde \rho^2 (\mu_H) \right] t^2 + \ldots
\end{align}
using Eq.~\eqref{12.4} to evaluate the RG evolution in a series in $t$. The entire LL contribution is from this term.
The logarithmic terms are
\begin{align}
\begin{split}
\widetilde \Lambda(\mu_L)&=\frac12 \widetilde m_-^4(\mu_H) t + \frac12 \widetilde m_-^2(\mu_H) \left[ \widetilde \lambda(\mu_H) \widetilde m_-^2(\mu_H) 
+\widetilde \rho^2 (\mu_H) \right] t^2 \\
&+  \frac{\widetilde m_-^2(\mu_H) \lambda_3(\mu_H) \hat \chi}{m_\chi^2(\mu_H)} \CJ_\chi(\mu_H) t 
- \frac{ \left(\CJ_\chi(\mu_H) \right)^2}{2 m_\chi^2(\mu_H)} + \ldots
\end{split}
\label{9.22}
\end{align}
rewriting $\widetilde m^2$ in terms of $\widetilde m^2_-$ and $\CJ_\chi(\mu_H)$.
The first term is the $\lambda \LL$ term and everything else is the $\left(\lambda \LL\right)^2$ contribution, since $\CJ_\chi(\mu_H)$ contains a log. This expression simplifies using $\widetilde \rho(\mu_H) = \widetilde \lambda(\mu_H) \hat \phi$ and Eq.~\eqref{9.33},
\begin{align}
\widetilde \Lambda(\mu_L)&=\frac12 \widetilde m_-^4(\mu_H) t + \frac12 \left\{ \widetilde \lambda \widetilde m_-^4 + {\widetilde \lambda^2} \hat \phi^2 \widetilde m_-^2
+ \frac{\lambda_3^2 \hat \chi^2}{m_\chi^2} \widetilde m_-^4 \right\} t^2\,.
\label{9.23}
\end{align}
The last term is the tadpole contribution. Using Eq.~\eqref{9.13},
\begin{align}
\widetilde \Lambda(\mu_L)&=\frac12 \widetilde m_-^4(\mu_H) t + \frac12 \left\{ \left[\lambda_\phi - 3 \frac{\lambda_3^2 \hat \chi^2}{m_\chi^2} \right] \widetilde m_-^4 + \left[\lambda_\phi -3 \frac{\lambda_3^2 \hat \chi^2}{m_\chi^2} \right]^2  \hat \phi^2 \widetilde m_-^2
+ \frac{\lambda_3^2 \hat \chi^2}{m_\chi^2} \widetilde m_-^4 \right\} t^2  \nn
&=\frac12 \widetilde m_-^4(\mu_H) t + \frac12 \left\{ \left[\lambda_\phi - 2 \frac{\lambda_3^2 \hat \chi^2}{m_\chi^2} \right] \widetilde m_-^4 + \left[\lambda_\phi - 6 \frac{\lambda_3^2 \hat \chi^2}{m_\chi^2} \right] \lambda_\phi \hat \phi^2 \widetilde m_-^2
 \right\} t^2 \,.
\label{9.34}
\end{align}
The $( \lambda_3^2 \hat \chi^2 \widetilde m_-^4/m_\chi^2) t^2$ term's coefficient is modified from $-3/2$ to $-1$, a factor of $2/3$, by the tadpole for the reasons discussed below Eq.~\eqref{9.14}. In the second line, we have only retained terms to order $z^6$ in the expansion of $\widetilde \lambda^2$.

Equation~\eqref{9.34} agrees with the explicit two-loop result expanded in $z$ to order $z^6$ using the procedure given below Eq.~\eqref{7.44}. 
One difference between the $O(N)$ model and the two-scalar theory of this section is that the scalar mass matrix $W$  is not diagonal. The expression for the two-loop potential in Ref.~\cite{Ford:1991hw} is given in terms of mass eigenstates of the tree-level Lagrangian.
To use the two-loop result in  Ref.~\cite{Ford:1991hw}, we first diagonalize the scalar mass matrix $W$ and rewrite the Lagrangian in terms of the rotated fields. The heavy mass is $m_\chi^2$ to lowest order in $z$, and the light mass is $\widetilde m_-^2$ given in Eq.~\eqref{9.13} to order $z^2$. Using the rotated masses and couplings, and expanding the result to order $z^6$, we find agreement with our result Eq.~\eqref{9.23}.

\section{Two Scalar Fields in the Broken Phase}
\label{sec:broken}

In this section we study the two-scalar field model of the previous section, but now with a power counting appropriate to the broken phase of the theory where $m_\chi^2$ and $m_\phi^2$ are both negative. We will be brief, since most of the discussion carries over from Section~\ref{sec:unbroken}. This example illustrates the procedure for minimizing the potential for a theory such as a unified theory, with two widely separated symmetry breaking scales.

\subsection{Power Counting}
\label{sec:pcbroken}

In this case, we start with the Lagrangian 
\begin{align}
\begin{split}
	\CL  &= \frac{1}{2} (\partial_\mu \chi)^2 + \frac{1}{2} (\partial_\mu \phi)^2 -  \frac{\lambda_\phi}{24} (\phi^2 - v_\phi^2)^2 - \frac{\lambda_\chi}{24} (\chi^2 - v_\chi^2)^2\\
	& - \frac{\lambda_3}{4} (\phi^2 - v_\phi^2)(\chi^2 - v_\chi^2)  - \Lambda\ ,
\label{5.1}
\end{split}
\end{align}
evaluated at some high reference scale $\mu_0$, and assume widely separated scales $v_\chi \gg v_\phi$. The classical VEVs $v$ are related to the mass-squareds by Eq.~\eqref{A5.16}. The classical theory is stable if $\lambda_\phi \lambda_\chi > 9 \lambda_3^2$, and the minimum is at $\phi=v_\phi$, $\chi=v_\chi$.

The shifted Lagrangian $\hat{\CL}$ takes the same form as Eq.~\eqref{9.2}, but we rewrite the $W$-matrix in terms of the $v_\phi,v_\chi$ variables as
	\begin{align}
	\begin{split}
	W_{\chi\chi}&= \frac13 \lambda_\chi  v_\chi^2 + \frac12 \lambda_\chi \left( \hat \chi^2- v_\chi^2 \right) + \frac12  \lambda_3 \left( \hat \phi^2 - v_\phi^2 \right)  \ , \\
	W_{\phi\phi}&= \frac16 \lambda_\phi \left(3 \hat \phi^2- v_\phi^2 \right) + \frac12  \lambda_3 \left( \hat \chi^2 - v_\chi^2 \right) \ , \\
	W_{\phi\chi}&=W_{\chi\phi} = \lambda_3 \hat{\phi} \hat{\chi}\ .
	\end{split}
		\label{5.3}
	\end{align}
	The cosmological constant is
	\ba{
\hat \Lambda&= \frac{\lambda_\phi}{24} (\hat \phi^2 - v_\phi^2)^2 + \frac{\lambda_\chi}{24} (\hat \chi^2 - v_\chi^2)^2 + \frac{\lambda_3}{4} (\hat \phi^2 - v_\phi^2)(\hat \chi^2 - v_\chi^2)  +\Lambda\,.
\label{5.4}
}

Introduce a power counting 
 $v_\chi \sim 1$ and $v_\phi \sim z$, so that the two VEVs are widely separated, and assume $\lambda_\chi$, $\lambda_\phi$, $\lambda_3$ are order 1. The $\chi$ and $\phi$ mass-squareds at the minimum of the potential are $W_{\chi \chi} = \lambda_\chi v_\chi^2/2 \sim 1 $ and 
 $W_{\phi \phi} = \lambda_\phi v_\phi^2/2 \sim z^2$. As $\hat\chi$ and $\hat \phi$ move away from the minimum of the potential, $W_{\chi \chi}$ and $W_{\phi \phi}$ change. To retain the mass hierarchy, we assume that
\begin{align}
\hat \chi^2 - v_\chi^2 & \sim z^2 \,, \qquad
\hat \phi^2 \sim v_\phi^2  \sim \hat \phi^2 - v_\phi^2 \sim z^2
\label{5.5}
\end{align}
so that
\begin{align}
W_{\chi \chi}  \sim 1\,, \qquad
W_{\phi \phi}   \sim z^2\,, \qquad
W_{\phi \chi} = W_{\chi \phi} \sim z\,.
\label{5.6}
\end{align}
The first term for $W_{\chi \chi}$ in Eq.~\eqref{5.3} is order 1, and the remaining terms are order $z^2$.\footnote{If $\hat \chi^2 - v_\chi^2  \sim 1$, $W_{\phi \phi} \sim 1$, both fields have comparable masses, and there is no need for RG improvement.} In this example, we will keep terms to order $z^4$, since non-trivial effects in the unbroken sector now occur by order $z^4$.

\subsection{The Result}
\label{sec:matchbroken}

The one-loop matching contribution to $\widetilde{\Lambda}$ in the EFT from integrating out $\chi_q$ is given by computing the one-loop $\chi$-bubble. The graphs are shown in Table~\ref{tab:1}. In particular, the $z$-scaling of the diagrams is different from the unbroken case, since now $W_{\phi\chi}\sim z$ rather than $z^2$, and there is one extra diagram that contributes. The result is 
\begin{align}
\begin{split}
V_{\text{match}}(\mu_H) &= \frac{1}{64\pi^2} \Bigg\{ 
\left[ 
 \ln \frac{W_{\chi \chi}}{\mu_H^2}     -\frac32   \right]   W_{\chi \chi}^2  +  \left[ 
\ln \frac{W_{\chi \chi}}{\mu_H^2}   - 1  \right]2 W_{\chi \phi}W_{\phi \chi} \\
& \qquad+  \left[ 
 \ln \frac{W_{\chi \chi}}{\mu_H^2}   - 1  \right]  \frac{2 W_{\chi \phi} W_{\phi \chi} W_{\phi \phi} }{W_{\chi \chi}}  
 -   \left[ 
 \ln \frac{W_{\chi \chi}}{\mu_H^2}   -2  \right]  \frac{W_{\chi \phi}^2 W_{\phi \chi}^2  }{W_{\chi \chi}^2}  \Bigg\} \,,
 \end{split}
 \label{5.7}
\end{align}
where the first term is order $1$, the second term is order $z^2$, and the last two terms are order $z^4$.

The tree level matching proceeds as in the unbroken case, and the EFT couplings defined in Eq.~\eqref{9.12a} are given at the matching scale $\mu_H$ by 
\begin{align}
\begin{split}
\widetilde \lambda  &\overset{\mu=\mu_H}{=}  \lambda_\phi  -  \frac{9 \lambda_3^2 \hat \chi ^2 }{ \lambda_\chi  v_\chi^2}  + \CO(z^2)  \, , \\
\widetilde \rho  &\overset{\mu=\mu_H}{=}   \hat \phi \left[ \lambda_\phi - \frac{9 \lambda_3^2 \hat \chi^2}{ \lambda_\chi  v_\chi^2 }\right] + \CO(z^3) \,,\\
\widetilde m^2  &\overset{\mu=\mu_H}{=}   \frac16 \lambda_\phi   \left(3 \hat \phi^2- v_\phi^2 \right) + \frac12  \lambda_3\left( \hat \chi^2 - v_\chi^2 \right)  - \frac{3 \lambda_3^2 \hat \chi^2 \hat \phi^2}{ \lambda_\chi v_\chi^2} + \frac{3 \lambda_3 \hat \chi  \CJ_\chi }{\lambda_\chi v_\chi^2}   + \CO(z^4) \,, \\
&\equiv\widetilde m_{-}^2  +  \frac{3 \lambda_3 \hat \chi  \CJ_\chi }{\lambda_\chi v_\chi^2}   + \CO(z^4) \,, \\
\widetilde \sigma  &\overset{\mu=\mu_H}{=}   -\CJ_\phi + \frac{3 \lambda_3 \hat \chi \hat \phi \CJ_\chi}{\lambda_\chi v_\chi^2} + \CO(z^5) \,,    \\
\widetilde \Lambda & \overset{\mu=\mu_H}{=}     \hat \Lambda  + V_{\text{match}}-  \frac{3 \left(\CJ_\chi  \right)^2}{2 \lambda_\chi  v_\chi^2} + \CO(z^6) \,.
\end{split}
\label{5.13a}
\end{align}
The resulting \cwp\ again takes the form 
\begin{align}
V_{\text{CW}} &= \widetilde \Lambda(\mu_L) + \frac{1}{64 \pi^2} \widetilde m^4(\mu_L) \left[ \ln \frac{ \widetilde m^2(\mu_L) }{\mu_L^2} - \frac32 \right]\ , \label{eq:145}
\end{align}
with the couplings running according to Eq.~\eqref{B.2a}-\eqref{B.3a}.
The RG improvement of the tadpole is identical to that given in the previous section, with the only change that the parameters are those of the broken phase values in Eq.~\eqref{5.13a} for the low energy theory. We will not repeat the analysis, and simply refer to the results of Section~\ref{sec:unbroken}.     

We have again checked that our result for the $\LL^2$ term agrees with the explicit two-loop calculation of Ref.~\cite{Ford:1991hw}, and as in the other examples, the matching contribution in Eq.~\eqref{5.13a} and the tadpole terms are necessary for agreement. Since the $z$ power counting is different than in the unbroken phase, all terms in $\widetilde \lambda^2$ must be retained in Eq.~\eqref{9.34}.

\subsection{Minimizing the RG Improved Potential}

\begin{figure}
\begin{center}
\includegraphics[width=8cm]{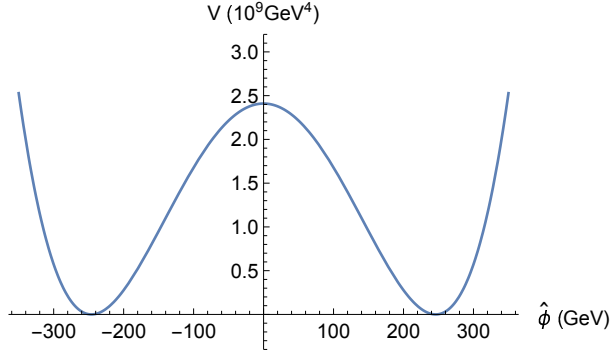}
\end{center}
\caption{The tree level \cwp\ for $\Lambda=0$, plotted for the values of couplings in Eq.~\eqref{5.10}.  \label{fig:treelevel}}
\end{figure}

The minimum of the effective potential is the quantum vacuum of the theory. In this case, it is interesting to plot the RG improved potential and compare it to the fixed one-loop order expression, in order to see the effect of RG improvement on the location of the minimum. We use the couplings
\begin{align}
\frac{\lambda_\phi}{16 \pi^2} &= 0.1\,,  \qquad  \frac{\lambda_\chi}{16 \pi^2} = 0.25 \,, \qquad \frac{\lambda_3}{16 \pi^2} = 0.04 \,,
\label{5.10}
\end{align}
at the scale $\mu_H$ and 
\begin{align}
v_\phi &= 246\,\text{GeV} \,,  \qquad v_\chi = 500\, \text{TeV} \,,
\label{5.11}
\end{align}
which have a large hierarchy, and study the effect on the potential near the classical minimum $\hat \phi = v_\phi$, $\hat \chi=v_\chi$. We will use the variables $\hat \phi$ and $\Delta_\chi \equiv \hat \chi^2-v_\chi^2$, which are order $z$ and $z^2$, respectively. To orient ourselves with these couplings, we have plotted the tree level \cwp \ $V_{\text{tree}}$ as a function of $\hat{\phi}$ at the value $\hat{\chi}=v_\chi$ in Figure~\ref{fig:treelevel}. In later plots we will focus on the region around the right-hand minimum.

Naively plotting the result leads to numerical instabilities from the matching condition Eq.~\eqref{5.7}.  To discuss the instability in more detail, it is convenient to break up $W_{\chi \chi}$ into its order 1 and order $z^2$ pieces,
	\begin{align}
	W_{\chi\chi}&=W_{\chi\chi}^{(0)} + W_{\chi\chi}^{(2)} \,, \nn
	W_{\chi\chi}^{(0)}  &=  \frac13 \lambda_\chi  v_\chi^2 \,, \nn
	W_{\chi\chi}^{(2)}  &= \frac12 \lambda_\chi \left( \hat \chi^2- v_\chi^2 \right) + \frac12  \lambda_3 \left( \hat \phi^2 - v_\phi^2 \right) 
	=  \frac12 \lambda_\chi \Delta_\chi + \frac12  \lambda_3 \left( \hat \phi^2 - v_\phi^2 \right) \,,
		\label{5.12}
	\end{align}
and reexpand Eq.~\eqref{5.7} in $z$,
\begin{align}
\renewcommand{\arraycolsep}{0.25cm}
& V_{\text{match}}(\mu_H) = \frac{1}{64 \pi^2} \bigg\{ \nn
& \begin{array}{ccc}
\left[ \ln \frac{W^{(0)}_{\chi \chi}}{\mu_H^2}     -\frac32 \right] \left[ W_{\chi \chi}^{(0)} \right]^2 &
+\left[ \ln \frac{W^{(0)}_{\chi \chi}}{\mu_H^2}     -1  \right] 2 W_{\chi \chi}^{(0)} W_{\chi \chi}^{(2)}  &
+\left[ \ln \frac{W^{(0)}_{\chi \chi}}{\mu_H^2}   \right] \left[ W_{\chi \chi}^{(2)} \right]^2 \\[10pt]
+ 0 & +\left[ \ln \frac{W^{(0)}_{\chi \chi}}{\mu_H^2}     -1  \right] 2  W_{\phi \chi} W_{\chi \phi} & +\frac{2 W_{\phi \chi} W_{\chi \phi} W_{\chi \chi}^{(2)} }{W_{\chi \chi}^{(0)}  } \\[10pt]
 +0 & +0 & + \left[ \ln \frac{W^{(0)}_{\chi \chi}}{\mu_H^2}     -1  \right]  \frac{2 W_{\phi \chi} W_{\chi \phi} W_{\phi \phi} }{W_{\chi \chi}^{(0)}  } \\[10pt]
 +0 & +0 & -\left[ \ln \frac{W^{(0)}_{\chi \chi}}{\mu_H^2}     -2  \right] \left( \frac{W_{\phi \chi} W_{\chi \phi} }{W_{\chi \chi}^{(0)}  } \right)^2 \biggr\}\,.
\end{array}   
\label{5.13}
\end{align}
The rows of Eq.~\eqref{5.13} are the four terms in Eq.~\eqref{5.7}, and the columns are the contributions at order $1$, $z^2$ and $z^4$ after expanding $W_{\chi \chi}$.  $W_{\chi\chi}^{(0)}$ is independent of $\hat \phi, \, \hat \chi$. We start by choosing $\mu_H^2 = W_{\chi\chi}^{(0)}$.

The first problem is that $V_{\text{match}}$ produces a large shift in the potential. For example, the single term in the first column of Eq.~\eqref{5.13} is an order 1 constant and gives a shift in the cosmological constant of $\simeq -2.6 \times 10^{22}\,\text{GeV}^4$.  The potential for $\hat \phi$ is order $z^4$, and smaller by a factor $\sim  (16 \pi^2/\lambda_\phi)(v_\phi/v_\chi)^4 \sim 10^{-14}$ relative to this contribution. This large constant shift in the potential is the cosmological constant problem. To be able to show the various contributions on the same plot, we shift all terms  in the matching by field-independent constants, by subtracting their values at the classical minimum $\hat \phi=v_\phi$, $\hat \chi=v_\chi$ in plots.

\begin{figure}
\centering
\begin{subfigure}[t]{0.48\textwidth}
\centering
\includegraphics[width=7.8cm]{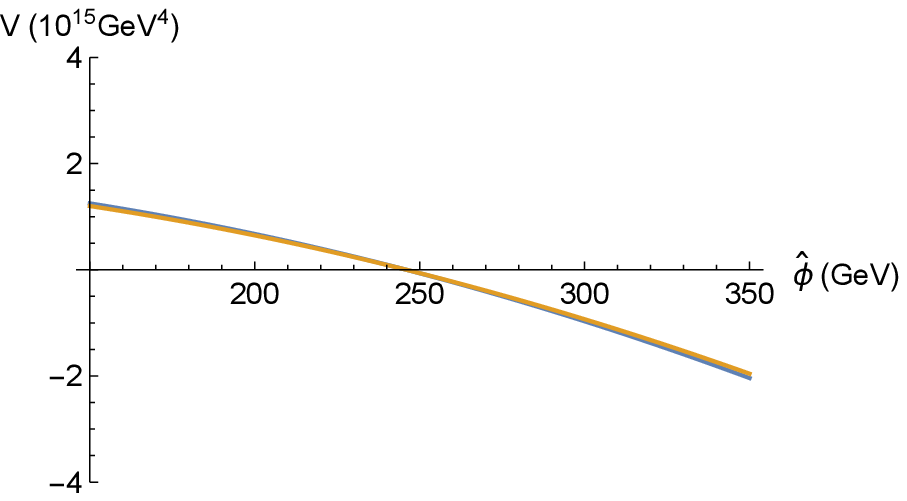}
\caption{\label{fig:plot1} The order $z^2$ contributions. The blue and orange curves are from the first and second rows of Eq.~\eqref{5.13}, respectively. The two values happen to be nearly equal for our particular choice of input parameters.}
\end{subfigure}\hfill
\begin{subfigure}[t]{0.48\textwidth}
\centering
\includegraphics[width=7.8cm]{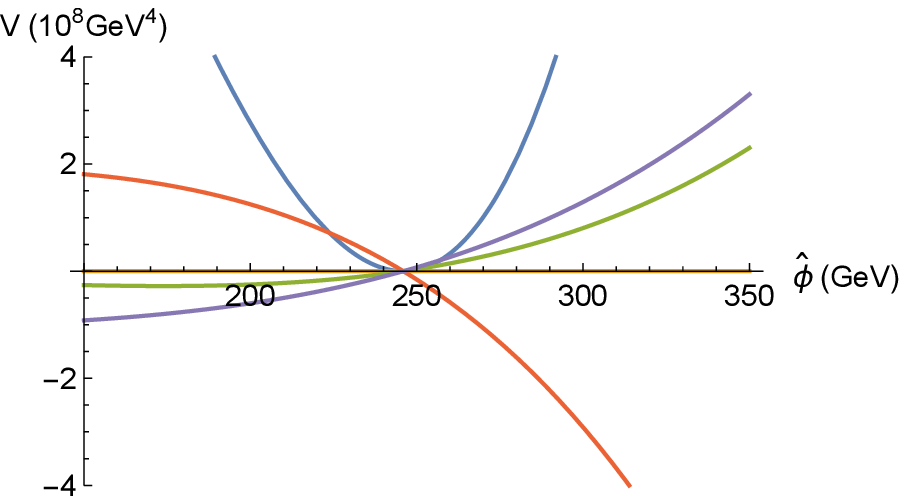}
\caption{\label{fig:plot2} The tree-level potential and the order $z^4$ contributions  The blue curve is the tree-level potential, and the orange, green, red and purple curves are the $z^4$ terms in the four rows of Eq.~\eqref{5.13}.}
\end{subfigure}
\caption{Plots of different contributions  to $V_{\text{match}}$ as a function of $\hat \phi$ for $\hat \chi= v_\chi$ at the matching scale $\mu_H^2 =  W^{(0)}_{\chi \chi}$. Note the $y$ axes' scaling differ by a factor of $10^7$. }
\end{figure}

The two terms in the second column of order $z^2$ are plotted in Fig.~\ref{fig:plot1}.
The two terms are both proportional to $\hat \phi^2$, and pass through $0$ at $\hat \phi=v_\phi$ because of our subtraction for the plots. They are almost equal, which is a numerical coincidence for our choice of input parameters.

The four terms in the third column are order $z^4$ and are plotted in Fig.~\ref{fig:plot2} along with the tree-level potential, which is also of order $z^4$.
Note that the scale in Fig.~\ref{fig:plot1} is $10^{7}$ times the scale in Fig.~\ref{fig:plot2}. As a result, the $z^2$ contribution would overwhelm the $z^4$ terms, including the tree-level potential. The reason for this numerical instability is that $v_\phi$ and $v_\chi$ mix under renormalization, and $v_\phi \ll v_\chi$. Unless the parameters in the high-energy Lagrangian are chosen properly, the physical $\phi$ mass is of order the high scale $v_\chi$, not the low scale $v_\phi$. This phenomenon is  the hierarchy or fine-tuning problem. Our starting assumption is that we have a hierarchy of physical mass scales, so we have to choose input parameters consistent with a large mass ratio. A simple way to do this is to use the matching scale
\begin{align}
\mu_H^2 &= \frac{1}{e} W_{\chi \chi}^{(0)}
\label{5.14}
\end{align}
with Eq.~\eqref{5.10} as the input parameters at this scale. The input parameters at the scale $\mu^2 = W_{\chi \chi}^{(0)}$ or any other high scale $\mu_0$ are given by running the values in Eq.~\eqref{5.10} from $\mu^2=W_{\chi \chi}^{(0)}/e$ to the high scale. The resulting values of $v_\phi$ and $v_\chi$  are almost equal, but careful adjusted so that the $\phi$ mass is light. We instead avoid any fine-tunings by using Eq.~\eqref{5.14} as the matching scale. We are free to choose any matching scale since the physical results are independent of the matching scale. Another way to say the same thing --- the parameters of the high-energy theory live on a renormalization group trajectory. The input parameters are such that when the couplings flow to $\mu_H$ given by Eq.~\eqref{5.14}, we have a hierarchy in the values of $v$. This procedure is a simple solution to the {hierarchy and}  fine-tuning problems.

With the choice Eq.~\eqref{5.14} for $\mu_H$, Eq.~\eqref{5.13} becomes
\begin{align}
 V_{\text{match}}(\mu_H) = \frac{1}{64 \pi^2} \left\{
\begin{array}{ccc}
-\frac12 \left[ W_{\chi \chi}^{(0)} \right]^2 &
+ 0  &
+ {\left[ W_{\chi \chi}^{(2)} \right]^2} \\[10pt]
+ 0 & +0 & +\frac{2 W_{\phi \chi} W_{\chi \phi} W_{\chi \chi}^{(2)} }{W_{\chi \chi}^{(0)}  } \\[10pt]
 +0 & +0 & + 0  \\[10pt]
 +0 & +0 & + \left( \frac{W_{\phi \chi} W_{\chi \phi} }{W_{\chi \chi}^{(0)}  } \right)^2
\end{array}
\right\}\,.
\label{5.15}
\end{align}
Most of the matching terms have disappeared, including the entire problematic $z^2$ contribution. The entire matching contribution other than the order 1 cosmological constant shift is order $z^4$.

\begin{figure}
\begin{center}
\includegraphics[width=8.8cm]{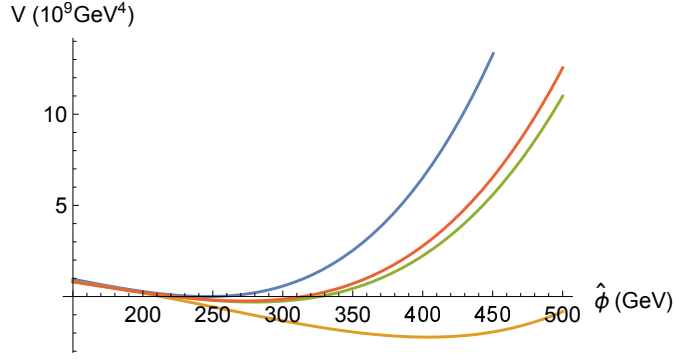}
\end{center}
\caption{\label{fig:plot5} Plot of the \cwp\ as a function of $\hat \phi$ for $\hat \chi= v_\chi$. The curves are the tree-level potential (blue), the fixed order potential at one-loop (orange), the RG improved potential neglecting the tadpole (green), and the RG improved potential (red).}
\end{figure}

Having taken care of the hierarchy and fine-tuning problems, we can now plot the total \cwp, using $\mu_H$ from Eq.~\eqref{5.14}. The value of the tadpole $\CJ_\chi(\mu_H)$ is in the range of approximately $-10^{10}$ to $-5\times 10^{10}$ GeV$^3$ for the values of $\hat{\phi}$ in the plots.
The fixed order and RG improved results for the \cwp\ are plotted in Fig.~\ref{fig:plot5}, and compared with the tree-level potential.  The fixed order potential is evaluated at the scale $\mu_H$. The scalar self-couplings run to zero in the infrared. As a result, RG improvement in this model reduces the size of the quantum corrections, and makes the potential closer to its tree-level value. We have shown the RG improved potential with the tadpole $\CJ_\chi(\mu_H)=0$, and with the value from Eq.~\eqref{9.33b}.
The RG improvement changes the potential substantially.  The minimum of the \cwp\ is shifted from $\hat \phi =246\,\text{GeV},\ \Delta_\chi=0$ at tree-level to 
$\hat \phi =337\,\text{GeV},\ \Delta_\chi=-(135\,\text{GeV})^2$ using the RG improved potential and ignoring the tadpole by setting $\CJ_\chi(\mu_H)=0$.
Including the tadpole to get the correct RG improved potential, the minimum shifts to $\hat \phi =319\,\text{GeV},\ \Delta_\chi=-(122\,\text{GeV})^2$. As we have emphasized, the tadpole terms are important.

\begin{figure}
\centering
\begin{subfigure}[b]{0.48\textwidth}
\centering
\includegraphics[width=7.7cm]{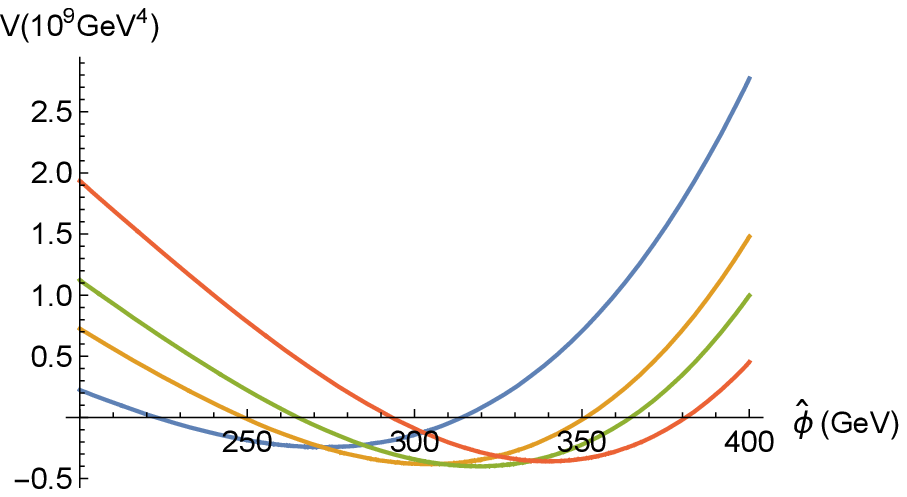}
\caption{ \label{fig:plot6}  $\Delta_\chi=0$ (blue), $\Delta_\chi =-(100\,\text{GeV})^2 $ (orange), $\Delta_\chi$ at the true minimum of the potential (green), and $\Delta_\chi =-(150\,\text{GeV})^2 $ (red). }
\end{subfigure}\hfill
\begin{subfigure}[b]{0.48\textwidth}
\centering
\includegraphics[width=7.7cm]{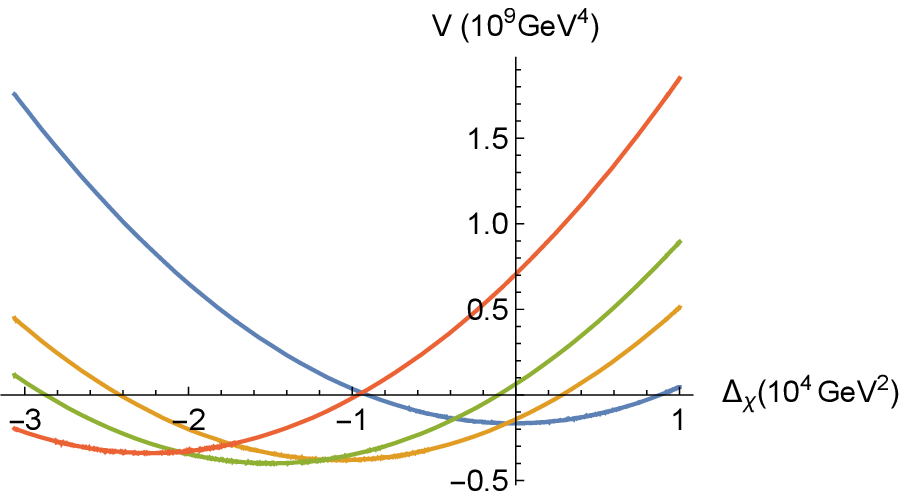}
\caption{\label{fig:plot7}  $\hat \phi = 246\,\text{GeV}$ (blue), $\hat \phi = 300\,\text{GeV}$ (orange), $\hat \phi$ at the true minimum of the potential (green), and $\hat \phi = 350\,\text{GeV}$ (red). }
\end{subfigure}
\caption{Plot of the RG improved \cwp. { The left plot is vs. $\hat \phi$ for various fixed values of $\Delta_\chi$, and the right plot is vs. $\Delta_\chi$ for fixed values of $\hat{\phi}$.}}
\end{figure}

Finally in Fig.~\ref{fig:plot6}, we plot the full RG improved \cwp\ including tadpole contributions vs. $\hat \phi$ for various values of $\Delta_\chi$, and
in Fig.~\ref{fig:plot7}, plot it vs. $\Delta_\chi$ for various values of $\hat \phi$.

As a final comment, note that 
in all of the above plots we have chosen fixed values of $\mu_H$ and $\mu_L$ that do not depend on the background fields. One could instead choose these values to vary with $\hat{\phi}$ and $\hat{\chi}$, and e.g. choose $\mu_L$ to cancel the entire log in \eqref{eq:145}. The difference that this dependence $\mu(\hat{\phi},\hat{\chi})$  makes to the minimum is higher order; this is because $\rd V/\rd \phi$ will now pick up a $\rd V/\rd \mu$ term, but by the RG equations that term is higher order in perturbation theory, {and vanishes if $V$ is computed to all orders.}

\section{Higgs-Yukawa Model}
\label{sec:higgs-yukawa}

Our last test case is the Higgs-Yukawa Model. This model has the new feature of non-zero scalar wavefunction renormalization at one loop. 

\subsection{Power Counting}
\label{sec:pchiggs}

The Higgs-Yukawa model considered in Ref.~\cite{Casas:1998cf} has the Lagrangian
\begin{align}
\CL &= \frac12 (\partial_\mu \phi)^2 + \sum_{k=1}^{N_F} i \, \overline \psi_k\,  \slashed{\partial}\, \psi_k -g \phi \overline \psi_k \psi_k   -  \frac{m_B^2}{2} \phi^2 - \frac{\lambda}{24}  \phi^4 -  \Lambda \,,
\label{14.1}
\end{align}
and is a special case of the more general theory Eq.~\eqref{2.2}, \eqref{2.4}. Eq.~\eqref{14.1} has the discrete chiral symmetry $\psi_L \to -\psi_L$, $\psi_R \to \psi_R$, $\phi \to -\phi$ which forbids odd powers of $\phi$ in the scalar potential.

Following the usual procedure, we shift the scalar field $\phi = \hat \phi + \phi_q$, drop the linear term in $\phi_q$, and introduce a source $\CJ$ for $\phi_q$.
The resulting Lagrangian $\hat{\CL}$ is
\begin{align}
\begin{split}
\hat{\CL} &= \frac12 (\partial_\mu \phi)^2 + i \, \overline \psi_k\,  \slashed{\partial}\, \psi_k  - M_F  \overline \psi_k \psi_k   -   g \phi_q \overline \psi_k \psi_k 
 -  \frac12 M_B^2 \phi_q^2 - \frac{\lambda}{6}  \hat \phi\, \phi_q^3 \\
& - \frac{\lambda}{24}  \phi_q^4 -  \hat \Lambda + \CJ \phi_q \,, \\
\hat \Lambda &= \Lambda +  \frac12 m_B^2 \hat \phi^2 + \frac{\lambda}{24}  \hat\phi^4  \,,
\end{split}
\label{14.2}
\end{align}
and the coefficient $\sigma$ of the linear term  in $\phi_q$  is zero at tree level. 
The boson mass-squared matrix and fermion mass matrix are
\begin{align}
M_B^2(\hat \phi) \equiv W &= m_B^2 + \frac12 \lambda \hat \phi^2\, ,\qquad  M_F(\hat \phi) = g \hat \phi\,.
\label{14.3}
\end{align}
The \cwp\ at one-loop order is
	\begin{align}
	\begin{split}
	V_{\text{tree}} &= \hat{\Lambda} = \Lambda +  \frac12 m_B^2 \hat \phi^2 + \frac{\lambda}{24}  \hat\phi^4  \,,\\
	V_{1{\text{-loop}}} &= \frac{M_B^4(\hat{\phi})}{64\pi^2} \left[  \ln \frac{M_B^2(\hat{\phi})}{\mu^2} - \frac{3}{2}  \right]  - \frac{N_F M_F^4(\hat{\phi})}{16\pi^2}  \left[     \ln \frac{M_F^2(\hat{\phi})}{\mu^2} - \frac{3}{2}  \right] \,.
	\end{split} \label{eq:cwhy}
	\end{align}

We assume the parameters and $\hat \phi$ are chosen so that there is a hierarchy of mass scales, either (a) $M_F \gg M_B$ or (b)   $M_B \gg M_F$ , and our expansion parameter $z \ll 1$ is the ratio of the smaller mass to the larger mass. We consider each of these cases in turn.

\subsection{The Case $M_F \gg M_B$} \label{sec:heavyF}

We use the power counting $m_B \sim z$, $\hat \phi \sim 1$, $\lambda \sim z^2$ and $g \sim 1$, and work to order $z^6$.
Since the fermion is heavier than the boson, the matching at $\mu_H$ involves integrating out the fermion. The one-loop contribution to the cosmological constant
from Fig.~\ref{fig:3} is
\begin{align}
V_{\text{match}}(\mu_H) &=  \frac{1}{64 \pi^2}\left(-4 N_F \right) M_F^4(\hat \phi)  \left[ 
 \ln \frac{M_F^2(\hat \phi) }{\mu_H^2}     -\frac32   \right]  \,,
 \label{14.4}
\end{align}
which, as usual, is precisely the contribution of the heavy particle to the \cwp\ in Eq.~\eqref{eq:cwhy}.
The tree-level matching vanishes in this case, since there is no fermionic analog of the $\chi_q$ exchange graph Fig.~\ref{fig:8}(a).

The EFT Lagrangian is a scalar theory
\begin{align}
\CL_{\text{EFT}} &= \frac12 (\partial_\mu \phi_q)^2  -  \frac12 \widetilde M_B^2 \phi_q^2 - \frac{1}{6} \widetilde \rho \phi_q^3 - \frac{1}{24} \widetilde \lambda \phi_q^4 -  \widetilde \Lambda + \CJ \phi_q \,,
\label{14.5}
\end{align}
with parameters at the scale $\mu_H$ given by
\begin{align}
\begin{split}
\widetilde \lambda  &\overset{\mu=\mu_H}{=} \lambda  + \CO(z^3) \,, \\
\widetilde \rho & \overset{\mu=\mu_H}{=} \lambda \hat \phi+ \CO(z^4) \,, \\
\widetilde m^2 (\hat \phi) &\overset{\mu=\mu_H}{=}  M_B^2 (\hat \phi) + \CO(z^5)\,, \\
\widetilde \sigma &\overset{\mu=\mu_H}{=} -  \CJ+ \CO(z^6) \,,\\
\widetilde \Lambda &\overset{\mu=\mu_H}{=} \hat \Lambda +  V_{\text{match}}(\mu_H)  + \CO(z^7)\,.
\end{split}
\label{14.6}
\end{align}

The theory is sufficiently simple that there are no matching corrections  to the couplings in Eq.~\eqref{14.6}. As a result, one can reabsorb the shift $\hat \phi$ back into $\phi$, evolve down to a low scale $\mu_H$ of order the scalar mass $M_B$, and then shift the scalar field back by $\hat \phi$. The anomalous dimension $\gamma_\phi$ vanishes since the fermion has been integrated out, so we do not need to rescale $\hat \phi$. This observation is useful in comparing with the results of Ref.~\cite{Casas:1998cf}.

The couplings in the low-energy theory are evolved using the RGE to a low scale $\mu_L$ of order the scalar mass $M_B$. The RGE are those of the pure scalar theory given in Eq.~\eqref{B.2}, since the fermion has been integrated out. At the scale $\mu_L$, the quantum field $\phi_q$ is integrated out. This gives the \cwp\
\begin{align}
V_{\text{CW}} &= \widetilde \Lambda(\mu_L) + \frac{1}{64 \pi^2} \widetilde m^4(\mu_L) \left[ \ln \frac{ \widetilde m^2(\mu_L) }{\mu_L^2} - \frac32 \right]\,.
\label{14.7}
\end{align}

Eq.~\eqref{14.7} still contains $\CJ$ which has not been determined. It is  fixed by requiring that  $\VEV{\phi_q}$ vanish. This expectation value is determined by differentiation w.r.t.\ $\CJ$. In this example, the matching Eq.~\eqref{14.6} is particularly simple, and the tadpole condition becomes $\VEV{\phi_q}=0$, i.e.\ the high-energy theory tadpole condition is the same as the EFT tadpole condition. The value of $\widetilde \sigma = -\CJ$ is adjusted so that this condition is satisfied. The terms in Eq.~\eqref{14.6} do not depend on $\widetilde \sigma$, so we can omit this part of the calculation.

We can now compare our RG improved result to the known two-loop calculation by expanding the RG series in a power series in
$t = ( \ln {\mu_L}/{\mu_H})/(16 \pi^2)$.
The leading log contribution is
\begin{align}
\widetilde \Lambda(\mu_L) &=\widetilde \Lambda(\mu_H) + \frac12 \widetilde M_B^4(\mu_H) t + \frac12 \widetilde M_B^2(\mu_H) \left[ \widetilde \lambda(\mu_H) \widetilde M_B^2(\mu_H)
+\widetilde \rho^2 (\mu_H) \right] t^2 + \ldots
\end{align}
using Eq.~\eqref{B.2} to evaluate the RG evolution and expanding in a series in $t$.
The logarithmic terms are, using Eq.~\eqref{14.6} and Eq.~\eqref{14.3}
\begin{align}
\widetilde \Lambda(\mu_L) &\approx\frac12 \widetilde M_B^4(\mu_H) t +  \frac12 \widetilde M_B^2(\mu_H) \left[  \lambda(\mu_H) \widetilde M_B^2(\mu_H) 
+ \lambda^2 (\mu_H) \hat \phi^2 \right] t^2 + \ldots 
\label{14.9}
\end{align}
This result agrees with the explicit two-loop calculation expanded in {$M_B/M_F$.}

\subsection{The Case $M_B \gg M_F$} \label{sec:heavyB}

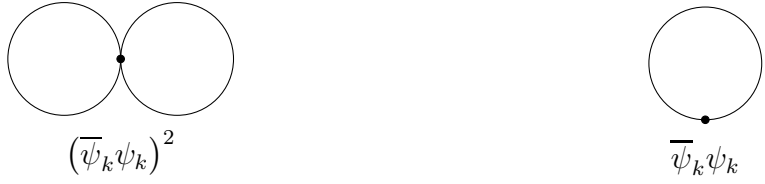
\begin{figure}
\centering
\begin{subfigure}[t]{0.5\textwidth}
\centering
\begin{tikzpicture}
\draw (0,-1.25) node [align=center] {$\left(\overline \psi_k \psi_k\right)^2$};
\draw (0.75,0) circle (0.75);
\draw (-0.75,0) circle (0.75);
\filldraw (0,0) circle (0.05);
\end{tikzpicture}
\caption{\label{fig:hy2} Vacuum graph from the four-fermion operator.}
\end{subfigure}\hfill
\begin{subfigure}[t]{0.5\textwidth}
\centering
\begin{tikzpicture}
\draw (0,-1.25) node [align=center] {$\overline \psi_k \psi_k$};
\draw (0,0) circle (0.75);
\filldraw (0,-0.75) circle (0.05);
\end{tikzpicture}
\caption{\label{fig:Ftadpole} The fermionic tadpole graph.}
\end{subfigure}
\caption{Some relevant fermionic graphs in the Higgs-Yukawa model.}
\end{figure}

We use the power counting $m_B \sim 1$, $\hat \phi \sim z$, and work to order $z^6$.
Since the boson is heavier than the fermion, the matching at $\mu_H$ involves integrating out the boson. The one-loop contribution to the cosmological constant
from Fig.~\ref{fig:3} is
\begin{align}
V_{\text{match}}(\mu_H) &=  \frac{1}{64 \pi^2} M_B^4(\hat \phi)  \left[ 
 \ln \frac{M_B^2(\hat \phi) }{\mu_H^2}     -\frac32   \right]  \,.
 \label{14.10}
\end{align}
The tree-level matching integrating out $\phi_q$ at $\mu_H$ gives
\begin{align}
\CL_{\text{EFT}} &= i \, \overline \psi_k\,  \slashed{\partial}\, \psi_k  - M_F  \overline \psi_k \psi_k  + \frac{1}{2 M_B^2(\mu_H)} \left[ g (\mu_H) \overline \psi_k \psi_k  - \CJ (\mu_H) \right]^2 - \hat \Lambda \,, \nn
&=  i \, \overline \psi_k\,  \slashed{\partial}\, \psi_k  - \widetilde M_F  \overline \psi_k \psi_k  + \frac{g^2(\mu_H)}{2 M_B^2(\mu_H)} \left[\overline \psi_k \psi_k  \right]^2 - \hat \Lambda\,,
\label{14.11}
\end{align}
with a four-fermion interaction, where the EFT couplings at the matching scale are
\begin{align}
\widetilde M_F \overset{\mu=\mu_H}{=}  M_F + \frac{g(\mu_H) \CJ(\mu_H) }{M_B^2(\mu_H)}  + \CO(z^4) \,, \qquad
\widetilde \Lambda \overset{\mu=\mu_H}{=} \hat \Lambda - \frac{ \left( \CJ (\mu_H) \right)^2 }{2 M_B^2(\mu_H)}+ \CO(z^7) \,.
\label{14.12}
\end{align}
The one-loop anomalous dimensions for the Lagrangian Eq.~\eqref{14.11} are,
\begin{align}
\frac{\rd}{\rd t} {\widetilde M_F} =  \frac{2 g^2(\mu_H) \widetilde M_F^3}{M_B^2(\mu_H) } \left(4 N_F - 1 \right)  \,, \qquad \frac{\rd}{\rd t} {\widetilde \Lambda}= - 2 N_F\widetilde M_F^4 \,,
\label{14.13}
\end{align}
using the results of Refs.~\cite{Jenkins:2017dyc,Jenkins:2017jig}. The solution of these equations is
\begin{align}
\widetilde M_F(\mu_L)  &=  \widetilde M_{F}(\mu_H) \left[ 1- \frac{4 (4N_F-1) g^2(\mu_H)  \widetilde M^2_{F}(\mu_H)}{M_B^2(\mu_H) } t \right]^{-1/2} \,, \nn
\widetilde \Lambda (\mu_L) &= \widetilde \Lambda_0 - \frac{2 N_F \widetilde M^4_{F}(\mu_H) t }{ 1- \frac{4 (4N_F-1) g^2(\mu_H)  \widetilde M_{F}^2(\mu_H)}{M_B^2(\mu_H) } t} \,,
\label{14.14}
\end{align}
with the expansion in $t$
\begin{align}
\widetilde M_F(\mu_L)  &=  \widetilde M_{F}(\mu_H) + \frac{2 (4N_F-1) g^2(\mu_H)  \widetilde M_{F}^3(\mu_H) }{M_B^2(\mu_H) } t  + \frac{6 (4N_F-1)^2 g^4(\mu_H)  \widetilde M_{F}^5(\mu_H) }{M_B^4(\mu_H) } t^2 + \ldots \nn
\widetilde \Lambda (\mu_L) &= \widetilde \Lambda(\mu_H)  - 2 N_F \widetilde M_{F}^4(\mu_H) t - \frac{8 N_F(4N_F-1)g^2(\mu_H)  \widetilde M_{F}^6(\mu_H) }{M_B^2(\mu_H) } t^2 + \ldots
\label{14.15}
\end{align}
The fermion $\psi$ is integrated out at a low scale $\mu_L$ of order the fermion mass $M_F$. In addition to the usual contribution from Fig.~\ref{fig:3}, 
\begin{align}
\VEV { \left[ \overline \psi_k \psi_k  \right]}_\mu &= - \frac{4 N_F M_F^3(\mu) }{16 \pi^2} \left[ \ln \frac{M_F^2(\mu)}{\mu^2} - 1 \right]\,,
\label{14.17a}
\end{align}
we also have the two loop contribution from the four-fermion operator from Fig.~\ref{fig:hy2},
\begin{align}
\VEV { \left[ \overline \psi_k \psi_k  \right] ^2 }_\mu &= \frac14 N_F\left(4N_F-1\right) \biggl\{  \frac{4M_F^3(\mu) }{16 \pi^2} \left[ \ln \frac{M_F^2(\mu)}{\mu^2} - 1 \right]
\biggr\}^2 \,.
\label{14.17b}
\end{align}
Combining the pieces gives the \cwp 
\begin{align}
V_{\text{CW}} &= \widetilde \Lambda(\mu_L) + \frac{1}{64 \pi^2}(-4N_F) \widetilde M_F^4(\mu_L) \left[ \ln \frac{ \widetilde M_F^2(\mu_L) }{\mu_L^2} - \frac32 \right]\nn
& -  \frac{g^2(\mu_H)}{8 M_B^2(\mu_H)} N_F \left(4N_F-1\right) \left\{  \frac{4  \widetilde M_F^3(\mu_L) }{16 \pi^2} \left[ \ln \frac{ \widetilde M_F^2(\mu_L) }{\mu_L^2} - 1 \right] \right\}^2 \,.
\label{14.16}
\end{align}
The two-loop contribution is LL order if evaluated at $\mu_H$. By evaluating it at $\mu_L$, we have eliminated the large logarithms, and the contribution can be dropped. The LL piece now arises from the running of the fermion mass proportional to the four-fermion operator, Eq.~\eqref{14.13}.

Eq.~\eqref{14.16} still contains $\CJ$, which is  fixed by requiring that  $\VEV{\phi_q}$ vanish. From Eq.~\eqref{eq:z3}, one sees that this expectation value is determined by differentiation w.r.t.\ $\CJ$ to give the EFT tadpole condition
\begin{align}
-\frac{g}{M_B^2} \VEV { \left[  \overline \psi_k \psi_k\right] }_{\mu_H} + \frac{1}{M_B^2} \CJ  =0 \implies \CJ(\mu_H) = g(\mu_H) \VEV { \left[ \overline \psi_k \psi_k  \right]}_{\mu_H}\ .
\label{14.17}
\end{align}
$ \VEV { \left[ \overline \psi_k \psi_k  \right]}$ has no large logarithms if evaluated at a scale $\mu_L$ of order the fermion mass.
 The RG improved tadpole can be obtained by differentiating the fermion Lagrangian at $\mu=\mu_L$ w.r.t.\ $M_F(\mu_H)$ to bring down a factor of $\left[ \overline \psi_k \psi_k \right]_{\mu_H}$,
\begin{align}
\begin{split}
\left[ \overline \psi_k \psi_k \right]_{\mu_H} &= \left[ \overline \psi_k \psi_k \right]_{\mu_L} \left[ 1- \frac{4 (4N_F-1) g^2(\mu_H) \widetilde M_F^2(\mu_H)}{M_B^2(\mu_H) } t \right]^{-3/2} \\
&
- 8 \widetilde M_F^3 (\mu_H)  N_F t \left[1 - \frac{2 (4N_F-1) g^2(\mu_H) \widetilde M_F^2(\mu_H) }{M_B^2(\mu_H) } t  \right]\,,
\end{split}
\label{14.18}
\end{align}
which also follows from the anomalous dimension of $\left[ \overline \psi_k \psi_k\right]  $  in the EFT.  The RG improved tadpole is given by substituting this result in
Eq.~\eqref{14.17}.
$\VEV{ \left[ \overline \psi_k \psi_k \right] }_{\mu_L} $ has no large logarithm, so the entire LL piece of the tadpole comes from expanding the second term
\begin{align}
\CJ(\mu_H) &=g(\mu_H)  \VEV { \left[ \overline \psi_k \psi_k \right] }_{\mu_H} \approx -8 N_F g(\mu_H)  \widetilde M_F^3(\mu_H) t\,.
\label{14.19}
\end{align}

We can now compare our RG improved result to the known two-loop calculation by expanding the RG series in a power series in $t$.
As usual, the entire leading-log contribution comes from the cosmological constant in the low-energy theory at $\mu_L$, 
\begin{align}
\begin{split}
\widetilde \Lambda(\mu_L) &= \widetilde \Lambda(0)  - 2 N_F \widetilde M_F^4(0) t - \frac{8 N_F(4N_F-1) g^2(\mu_H)  \widetilde M_F^6(0) }{M_B^2} t^2 + \ldots \\
&= \hat \Lambda - \frac{ \left( \CJ (\mu_H) \right)^2 }{2 M_B^2(\mu_H) } 
  - 2 N_F \left[  M_F(\mu_H) +  g(\mu_H)  \frac{ \CJ (\mu_H) }{M_B^2(\mu_H) } \right]^4 t  \\
  &-  \frac{8 N_F(4N_F-1)  g^2(\mu_H) }{M_B^2(\mu_H) } \left[  M_F + g(\mu_H)  \frac{ \CJ(\mu_H) }{M_B^2(\mu_H) } \right]^6 t^2  + \ldots
  \end{split}
\label{14.21}
\end{align} 
The logarithmic terms to order $\LL^2$ are
\begin{align}
\widetilde \Lambda(\mu_L) &\approx
    - 2 N_F M_F^4(0) t   - 8 N_F g(\mu_H)   M_F^3(0)  \frac{ \CJ(\mu_H) }{M_B^2(\mu_H) } t  - \frac{ \left( \CJ(\mu_H)  \right)^2 }{2 M_B^2(\mu_H) } \nn
 & -  \frac{8 N_F(4N_F-1)  g^2(\mu_H)  }{M_B^2(\mu_H) }M_F^6(0)  t^2 
+ \ldots  
\label{14.22}
\end{align}  
Using Eq.~\eqref{14.19} for the tadpole contribution gives
\begin{align}
\widetilde \Lambda(\mu_L) &\approx
  - 2 N _F M_F^4(0) t  + \left[ 64 N_F^2  -  32 N_F^2 -  8 N_F(4N_F-1) \right] \frac{g^2(\mu_H)  M_F^6(0) }{M_B^2}  t^2 + \ldots
\label{14.22b}
\end{align} 
As before, the $\CJ(\mu_H)$ flips the sign of the $\left( \CJ(\mu_H) \right)^2$ term from the cosmological constant. The $N_F^2$ terms cancel, leaving
\begin{align}
\widetilde \Lambda(\mu_L) &\approx
  - 2 N_F  M_F^4(0) t  + 8 N_F g^2(\mu_H)  \frac{ M_F^6(0) }{M_B^2}  t^2 + \ldots
\label{14.23}
\end{align} 
which agrees with the two-loop result {obtained from} Ref.~\cite{Martin:2001vx}, and provides a non-trivial check of our method. The $N_F^2$ term arises in the high-energy theory from graph Fig.~\ref{fig:10}(b) with the dotted scalar replaced by a fermion. This is not 1PI, and is cancelled in the EFT by the tadpole. The $N_F$ term arises in the high-energy theory from Fig.~\ref{fig:10}(a), which is 1PI, and survives in the EFT.

\section{Comparison with Previous Work} 
\label{sec:gb}

In this section we compare our results for various examples with previous results  in the literature.

\subsection{$O(N)$ Model}

The \cwp\  of the $O(N)$ model in the broken phase was considered previously in Ref.~\cite{Ford:1994dt}. This paper assumes a matching condition for the Lagrangian at $\mu_L$, and uses running couplings computed in the full theory to evolve the parameters from $\mu_H$ to $\mu_L$. However, the method is not a systematic procedure (as was noted in Ref.~\cite{Ford:1994dt}), and cannot be extended to higher orders. It also does not  include  power corrections in the LL summation, and misses the $1/m_\chi^2$ term in Eq.~\eqref{7.44}.

\subsection{Two Scalar fields in the Broken Phase}

The \cwp\ of the two-scalar theory with $m_\phi^2=m_\chi^2=0$ was considered previously in Ref.~\cite{Iso:2018aoa} using a different method.  This paper computes the exact eigenvalues of the scalar mass matrix, $w_\pm$, the expressions for which contain square-roots (see Eq.~\eqref{eq:eigs}). The running of the scalar potential, given in Eq.~\eqref{2.6}, implies for the high-energy theory that $2 \dot V = w_+^2 + w_-^2$. The square-roots cancel in this combination, leading to the RGE Eq.~\eqref{A5.6} with $N_\phi=N_\chi=1$. Ref.~\cite{Iso:2018aoa} then defines $\beta$-functions in the theory between $\mu_H$ and $\mu_L$ by treating it as a single scalar theory with a mass given by $w_-$ so that $2 \dot V =  w_-^2$, a non-analytic expression containing square-roots. They then describe a method
Ref.~\cite[above (3.16)]{Iso:2018aoa} to convert this to a polynomial, obtaining their $\beta$-functions Ref.~\cite[(3.17)--(3.19)]{Iso:2018aoa}.  Their $\beta$-functions are then integrated to sum the logarithms.

If we take the special case of their results with $\lambda_h=\lambda_\phi$ and $\lambda_\kappa = \lambda \phi - \lambda_1$, with $\lambda_1 \ll 1$ and $\VEV{\phi}=r \sqrt{\cos \theta}$, $\VEV{\chi}=r \sqrt{\sin \theta}$, in their notation  Ref.~\cite[(3.14)]{Iso:2018aoa}
the mass matrix has eigenvalues
\begin{align}
M_H^2 &= \sqrt{2} \lambda_\phi r^2 - \frac{3}{2 \sqrt {2}} \lambda_1  r^2 \,, & 
M_L^2 &= \frac{1}{2 \sqrt 2} \lambda_1 r^2 \,,
\label{9.35}
\end{align}
with $M_L^2 \ll M_H^2$, a large mass hierarchy. The $\LL$ and $\LL^2$ terms in the \cwp\ are then
\begin{align}
V \approx  \frac{1}{16} \lambda_1^2  r^4 t + \frac{1}{8} \lambda_1\left(2  \lambda_1  \lambda_\phi-\lambda_\phi^2  - \frac{5}{8} \lambda_1^2 + \ldots \right)   r^4    t^2\,.
\label{9.36}
\end{align}
Our method, treating $\lambda_1/\lambda_\phi$ as the expansion parameter $z$, gives instead
\begin{align}
V \approx  \frac{1}{16} \lambda_1^2  r^4 t + \frac18 \lambda_1 \left( \lambda_1\lambda_\phi - \frac{3}{4} \lambda_1^2  - \frac{\lambda_1^3}{4 \lambda_\phi} + \ldots \right)r^4  t^2\,,
\label{9.36a}
\end{align}
which agrees with the explicit two-loop result expanded in $\lambda_1/\lambda_\phi$.

\subsection{Higgs-Yukawa Model}
\label{subsec:higgs-yukawa}

The Higgs-Yukawa theory was studied in detail in Ref.~\cite{Casas:1998cf}, based on the method proposed in Ref.~\cite{Bando:1992wy}. The one-loop matching contributions at the high and low scales are the usual values given in Ref.~\cite{Coleman:1973jx}, and we obtain the same result. The difference between the two approaches is in the evolution between $\mu_H$ and $\mu_L$.
Ref.~\cite{Casas:1998cf} does not construct an EFT, but instead introduces $\theta$-functions into the RGE. Explicitly, they use Ref.~\cite[(3.4)]{Casas:1998cf} 
\begin{align}
\begin{split}
\dot \phi &= -  2g^2 N \theta_F \phi = -\gamma_\phi \theta_F \phi \,, \\
\dot g &=\left(3  \theta_B \theta_F +2 N  \theta_F\right) g^3  = 3  \theta_B \theta_F g^3 + \gamma_\phi \theta_F g  \,, \\
\dot m_B^2 &=  \lambda m_B^2  \theta_B  + 4g^2 N m_B^2  \theta_F    = \lambda m_B^2  \theta_B  + 2 \gamma_\phi \theta_F  m_B^2 \,, \\
\dot  \lambda &= 3 \lambda^2  \theta_B  -48 g^4 N  \theta_F + 8 g^2 N \lambda  \theta_F   = 3 \lambda^2  \theta_B  -48 g^4 N  \theta_F +4  \gamma_\phi \theta_F   \lambda \,, \\
\dot \Lambda &=  \frac12 m_B^4  \theta_B \,.
\end{split}
\label{14.24}
\end{align}
In the theory with $M_B \gg M_F$, they set $\theta_B=0$ and $\theta_F=1$ in the RGE between $M_B$ and $M_F$, whereas for $M_F \gg M_B$, 
$\theta_B=1$ and $\theta_F=0$ in the RGE between $M_F$ and $M_B$. The calculation follows that in Ref.~\cite{Coleman:1973jx} of treating the \cwp\ as graphs with external scalar lines, rather than the method in Ref.~\cite{Jackiw:1974cv} of shifting the scalar field $\phi \to \hat \phi + \phi_q$ and performing the functional integral over $\phi_q$.

For the case $M_F \gg M_B$, our result agrees to order $\LL^2$ with that of Ref.~\cite{Casas:1998cf}. The matching condition Eq.~\eqref{14.9} is trivial, and the tadpole plays no role in the final result. The $\beta$-functions Eq.~\eqref{14.24} with $\theta_B=1$, $\theta_F=0$ agree with those in our low-energy theory after integrating out the fermion. Our RG evolution uses the Lagrangian after the shift $\phi \to \hat \phi + \phi_q$, whereas Ref.~\cite{Casas:1998cf} uses the original $\phi$ field, and treats $\phi$ as the external field which is equivalent to running and then performing the shift $\phi \to \hat \phi + \phi_q$. As commented earlier, in this example running commutes with shifting, so the two answers are the same.

\begin{figure}
\begin{center}
\begin{tikzpicture}[x=1cm,y=1cm]
\draw (0,0) circle (1);
\draw[dashed]  (-2,0) -- (-1,0);
\draw[dashed]  (1,0) -- (2,0);
\draw[dashed]  (0,1) -- (0,2);
\draw[dashed]  (0,-1) -- (0,-2);
\filldraw (-1,0) circle (0.05);
\filldraw (1,0) circle (0.05);
\filldraw (0,1) circle (0.05);
\filldraw (0,-1) circle (0.05);
\end{tikzpicture}
\end{center}
\caption{\label{fig:20} One-loop contribution to the $\phi^4$ interaction from a fermion loop. The same graph can be interpreted as a contribution to the cosmological constant if the external lines are interpreted as background fields $\hat \phi$. }
\end{figure}
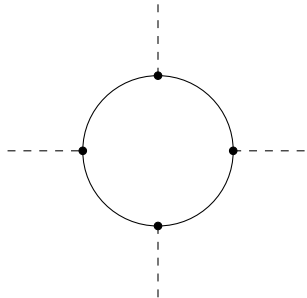

For $M_B \gg M_F$, the $\gamma_\phi$ parts of the $\beta$-function evolution in Eq.~\eqref{14.24} reduce to wavefunction evolution of the background field $\phi$. 
In our method, $\hat \phi$ is the field at $\mu_H$, whereas in the method of Ref.~\cite{Casas:1998cf}, $\phi$ renormalized at $\mu_L$ is considered the external field. Thus we evolve $\hat \phi$ from $ \mu_H$ to $\mu_L$ with its anomalous dimension to compare with the result of Ref.~\cite{Casas:1998cf}.  {A} difference in the two methods is the $\dot \lambda =  -48 g^4 N  \theta_F$ term in the $\beta$-function for $\lambda$. In our method, the EFT has no scalar fields, and no $\lambda \phi^4$ coupling. In Ref.~\cite{Casas:1998cf}, the $\lambda \phi^4$ coupling runs in the theory below $M_B$ by the diagram in Fig.~\ref{fig:20}. 
If one treats the external lines as $\hat \phi$, then this diagram gives the same contribution as the running of the cosmological constant in Eq.~\eqref{14.13} in our method, with a fermion mass $M_F = g \hat \phi$. Thus this contribution agrees, though its origin is different in the two approaches.
The final result for the \cwp\ in Ref.~\cite{Casas:1998cf} differs from our LL result in Eq.~\eqref{14.23} in that it does not contain
 the $t^2$ term that arises from the tadpole.

\subsection{Taming Goldstone Boson Divergences} 
\label{sec:comparison}

\begin{figure}
\begin{center}
\begin{tikzpicture}[x=1cm,y=1cm]
\draw (0:2) circle (0.5);
\draw (60:2) circle (0.5);
\draw (120:2) circle (0.5);
\draw (180:2) circle (0.5);
\draw (240:2) circle (0.5);
\draw (300:2) circle (0.5);
\filldraw (15:2) circle (0.05);
\filldraw (45:2) circle (0.05);
\filldraw (75:2) circle (0.05);
\filldraw (105:2) circle (0.05);
\filldraw (135:2) circle (0.05);
\filldraw (165:2) circle (0.05);
\filldraw (195:2) circle (0.05);
\filldraw (225:2) circle (0.05);
\filldraw (255:2) circle (0.05);
\filldraw (285:2) circle (0.05);
\filldraw (315:2) circle (0.05);
\filldraw (345:2) circle (0.05);
\draw[dashed] (15:2) arc (15:45:2);
\draw[dashed] (75:2) arc (75:105:2);
\draw[dashed] (135:2) arc (135:165:2);
\draw[dashed] (195:2) arc (195:225:2);
\draw[dashed] (255:2) arc (255:285:2);
\draw[dashed] (315:2) arc (315:345:2);
\end{tikzpicture}
\end{center}
\caption{\label{fig:gb} Class of diagrams leading to an infrared divergent \cwp\ in the Standard Model. The fermion loops are top quark loops, and the scalar fields connecting the fermion loops are  Goldstone bosons $\varphi^+$ and $\varphi_Z$, which are propagating degrees of freedom in $R_\xi$ gauge.}
\end{figure}
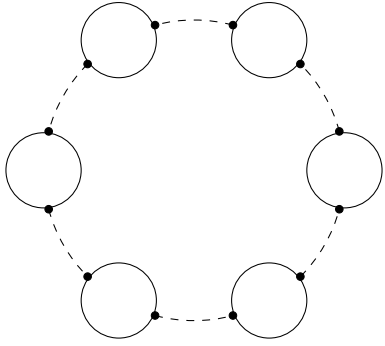

An explicit computation of higher loop corrections to the \cwp\ in the Standard Model shows that the class of graphs in Fig.~\ref{fig:gb} is infrared divergent at three-loop order and beyond~\cite{Martin:2013gka}. This divergence is known as the Goldstone boson catastrophe, whose resolution was given in Refs.~\cite{Martin:2014bca,Elias-Miro:2014pca}.\footnote{We thank the referee for bringing this problem to our attention.} Our method for computing the \cwp\ does not have this infrared divergence problem, and in this subsection we explain the relation of our method to the discussion in  Refs.~\cite{Martin:2014bca,Elias-Miro:2014pca}.

The infrared divergence found in Ref.~\cite{Martin:2013gka} depends on the coupling of Goldstone bosons to a fermion (the top-quark) which gets a mass from spontaneous symmetry breaking. The gauge fields are not important to the discussion, and Ref.~\cite{Martin:2013gka} sets the gauge couplings to zero. We will therefore study the problem in a simplified model given by combining the $O(N)$ theory of Sec.~\ref{sec:on} with the Higgs-Yukawa theory of Sec.~\ref{sec:higgs-yukawa}, keeping the essential features of the problem without irrelevant complications. The Lagrangian we study is
\begin{align}
\mathcal{L} &= \overline \psi i \slashed{\partial} \psi + \partial_\mu \phi^* \partial^\mu \phi - \lambda \left( \phi^* \phi - \frac{v^2}{2} \right)^2 - y \overline \psi_L \psi_R \phi - y \overline \psi_R \psi_L \phi^* - \Lambda\ ,
\label{16.1}
\end{align}
which consists of a complex scalar field (the $O(2)$ model) coupled to a fermion $\psi$. $\phi$ and $\psi_L$ have charge 1 under the $U(1)$ symmetry, and $\psi_R$ has charge zero.
Spontaneous symmetry breaking leads to a massive scalar (the Higgs boson), a massless Goldstone boson, and a massive fermion (the top).

The \cwp\ $V(\hat h)$ for the scalar field is computed by writing
\begin{align}
\phi &= \frac{1}{\sqrt{2}} \left[ \hat h + h_q + i \varphi_q \right].
\label{16.2}
\end{align}
In contrast to previous sections, here the radial mode has been called $h_q$ (for Higgs) instead of $\chi_q$. Including the source term, the shifted Lagrangian $\hat{\CL}$ from \eqref{eq:cl} is
\begin{align}
\hat{\mathcal{L}} &= \overline \psi i \slashed{\partial} \psi +   \frac12 (\partial_\mu h_q)^2 + \frac12 (\partial_\mu \varphi_q)^2 - \frac{y }{\sqrt 2} \left(\hat h + h_q \right) \overline \psi \psi - \frac{y }{\sqrt 2} \varphi_q \overline \psi i\gamma_5 \psi \nn
& - \frac14 \lambda \left( h_q^4 + 2 h_q^2 \varphi_q^2 + \varphi_q^4 \right) - \lambda \hat h \left( h_q^3 + h_q \varphi_q^2 \right)  -\frac12 \lambda \left(3 \hat h^2-v^2\right) h_q^2 \nn
&  -\frac12 \lambda \left( \hat h^2-v^2\right) \varphi_q^2  - \hat \Lambda + \mathcal{J}_h h_q \,,
\label{16.3}
\end{align}
where
\begin{align}
\hat \Lambda &= \frac14 \lambda \left(\hat h^2-v^2\right)^2 \,,
\label{16.4}
\end{align}
is the cosmological constant, and the full source term is split as
\begin{align}
J &= \hat \sigma +  \mathcal{J}_h  \,, & \hat \sigma &= \lambda \hat h \left( \hat h^2-v^2\right) \,,
\label{16.5}
\end{align}
as in Eq.~\eqref{eq:j}. The scalar masses \eqref{eq:w} are
\begin{align}
W_{hh} &= 2  \lambda v^2 +  3 \lambda \left( \hat h^2-v^2\right)\,, & W_{\varphi \varphi} &= \lambda \left(\hat h^2-v^2\right)\,,
\label{16.6}
\end{align}
and the fermion mass is
\begin{align}
m_F &= \frac{y \hat h}{\sqrt 2}\,.
\label{16.7}
\end{align}
At tree-level, with $\hat h^2 - v^2 \sim \mathcal{O}(z^2)$, there is a mass hierarchy with $W_{\varphi\varphi} \sim z^2 \ll W_{hh}, \ m_F^2 \sim 1$. $\varphi$ is a Goldstone boson of  the spontaneously broken $U(1)$ symmetry, and is exactly massless at the minimum of the potential $\hat h = v$. Including quantum corrections, the Goldstone boson remains exactly massless at the minimum of the quantum corrected potential, which is now shifted from $\hat h = v$ to $\hat h = v + \Delta v$. We will treat $\Delta v$ as formally of one-loop order.
We can therefore consider the \cwp\ in a regime with a large mass hierarchy by working near $\hat h = v + \Delta v$. The source $J$ breaks the $U(1)$ symmetry, so $\varphi$ is massless only at the minimum of the potential where the source vanishes, by Eq.~\eqref{eq:variation}.

The Goldstone boson problem studied in Ref.~\cite{Martin:2013gka} arises from one-loop corrections to the $\varphi$ mass, and so is one higher order than the other calculations in this paper. Nevertheless, our method works to all orders, so we can apply it here. In the regime $W_{\varphi\varphi}  \ll W_{hh}, m_F^2$, one can construct a low-energy EFT by integrating out $h_q$ and $\psi$, leaving only $\varphi_q$. The EFT Lagrangian is
\begin{align}
\mathcal{L}_{\text{EFT}} &=   \frac12 (\partial_\mu \varphi_q)^2     -\widetilde{\sigma} \varphi_q -\frac{\widetilde{m}^2}{2} \varphi^2_q  -  \frac{\widetilde{\rho}}{6} \varphi_q^3 - \frac{ \widetilde{\lambda}}{24} \varphi_q^4     
    - \widetilde{\Lambda} + \ldots
\label{16.8}
\end{align}
where the EFT coefficients are determined by one-loop matching from Eq.~\eqref{16.3}. The Goldstone boson problem is related to the $\varphi$ mass, so we focus only on the one-loop corrections we need: the one-loop correction to the $\varphi$ two-point function. 

\begin{figure}
\begin{center}
\begin{tikzpicture}
\begin{scope}[shift={(0,0)}]
\draw[densely dotted] (0,0.5) circle (0.5);
\draw[densely dotted] (-0.75,0) -- (0.75,0);
\filldraw (0,0) circle (0.05);
\draw (0,-1) node [align=center] {(a)};
\end{scope}
\begin{scope}[shift={(2.5,0)}]
\draw[dashed] (0,0.5) circle (0.5);
\draw[densely dotted] (-0.75,0) -- (0.75,0);
\filldraw (0,0) circle (0.05);
\draw (0,-1) node [align=center] {(b)};
\end{scope}
\begin{scope}[shift={(5.0,0)}]
\draw[densely dotted] (-1.0,0) -- (1.0,0);
\draw[dashed] (0.5,0) arc (0:180:0.5);
\filldraw (-0.5,0) circle (0.05);
\filldraw (0.5,0) circle (0.05);
\draw (0,-1) node [align=center] {(c)};
\end{scope}
\begin{scope}[shift={(8.0,0)}]
\draw (0,0) circle (0.5);
\draw[densely dotted] (-1.25,0) -- (-0.5,0);
\draw[densely dotted] (0.5,0) -- (1.25,0);
\filldraw (-0.5,0) circle (0.05);
\filldraw (0.5,0) circle (0.05);
\draw (0,-1) node [align=center] {(d)};
\end{scope}
\end{tikzpicture}

\vspace{1.5cm}

\begin{tikzpicture}
\begin{scope}[shift={(0,0)}]
\draw[densely dotted] (-0.75,0) -- (0.75,0);
\filldraw (0,0) circle (0.05);
\draw[dashed] (0,0) -- (0,1);
\draw[densely dotted] (0,1.5) circle (0.5);
\draw (0,-1) node [align=center] {(e)};
\end{scope}
\begin{scope}[shift={(2.5,0)}]
\draw[densely dotted] (-0.75,0) -- (0.75,0);
\filldraw (0,0) circle (0.05);
\draw[dashed] (0,0) -- (0,1);
\draw[dashed] (0,1.5) circle (0.5);
\draw (0,-1) node [align=center] {(f)};
\end{scope}
\begin{scope}[shift={(5,0)}]
\draw[densely dotted] (-0.75,0) -- (0.75,0);
\filldraw (0,0) circle (0.05);
\draw[dashed] (0,0) -- (0,1);
\draw (0,1.5) circle (0.5);
\draw (0,-1) node [align=center] {(g)};
\end{scope}
\begin{scope}[shift={(7.5,0)}]
\draw[densely dotted] (-0.75,0) -- (0.75,0);
\filldraw (0,0) circle (0.05);
\draw[dashed] (0,0) -- (0,1);
\filldraw[] (-0.1,0.9) rectangle (0.1,1.1);
\draw (0,-1) node [align=center] {(h)};
\end{scope}
\end{tikzpicture}
\end{center}
\caption{\label{fig:twopoint} One-loop contributions to the $\varphi$ two-point function. The dotted lines are $\varphi$, the dashed lines are $h_q$ and the solid line is the fermion $\psi$.}
\end{figure}
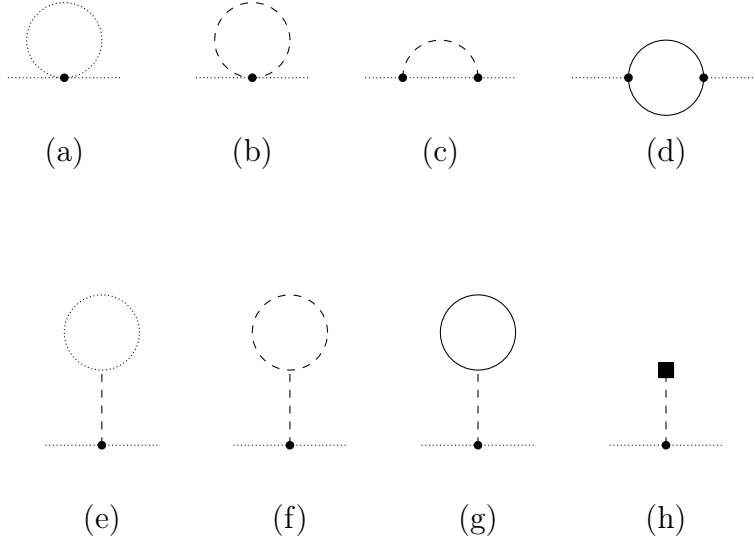

The one-loop correction to the $\varphi$ two-point function is given by the graphs in Fig.~\ref{fig:twopoint}.
The last three graphs in Fig.~\ref{fig:twopoint} are one-particle reducible, but are still included in $\Gamma_{\varphi \varphi}(p)$. It is often stated that $\Gamma$ is given by the sum of one-particle irreducible graphs, but this is not true if there is a non-zero one point function, i.e.\ a non-zero tadpole. Equations~\eqref{eq:background},\ \eqref{eq:variation} give the relation
\begin{align}
\frac{\delta^2 \Gamma}{\delta \varphi(x) \varphi(y)} = -\left[  \frac{\delta^2 W}{\delta J(x) \delta J(y)} \right]^{-1}
\label{16.13}
\end{align}
so that $\Gamma_{\varphi \varphi}(p)$ is the negative of the inverse propagator. It is given by the proper self-energy contribution, i.e.\ graphs which cannot be disconnected by cutting a line \emph{with momentum $p$}.

We can decompose
\begin{align}
-i \Gamma_{\varphi \varphi}(p) &= p^2 - W_{\varphi \varphi} - \Sigma_{\varphi \varphi}(p^2)\,,
\label{16.9}
\end{align}
where $\Sigma_{\varphi \varphi}(p^2)$ is the one-loop correction. In the EFT method, we need the one-loop matching contribution to $\Sigma_{\varphi \varphi}(p^2)$, i.e.\ the difference between the graphs in Fig.~\ref{fig:twopoint} in the high-energy theory with $h_q,\psi,\varphi_q$, and the low-energy theory with only $\varphi_q$. Graphs Fig.~\ref{fig:twopoint}(b,d,f,g,h) contribute to the matching condition. Graph Fig.~\ref{fig:twopoint}(a) is present in the low-energy theory, and does not enter the matching.
The matching contribution from graphs Fig.~\ref{fig:twopoint}(c,e) is given by evaluating the graphs in the high-energy theory, and subtracting the contribution from corresponding graphs in the low-energy theory given by shrinking the $h_q$ line to a point. The latter graphs look like Fig.~\ref{fig:twopoint}(a) in the low-energy theory, but  the $\varphi_q^4$ coupling includes the tree-level correction from integrating out $h_q$ from Fig.~\ref{fig:8}. The matching from Fig.~\ref{fig:twopoint}(c,e) can equivalently be computed by expanding the integrand in low-energy scales (i.e.\ in the power counting parameter $z$) before doing the integral. Note that $p^2 \sim z^2$ since the $\varphi$ momentum is of order the $\varphi$ mass in the low-energy theory.

The one-loop matching is (in terms of Passarino-Veltman functions in the conventions given in Appendix~\ref{app:pv})
\begin{align}
\Sigma_{\varphi \varphi}(p^2) &= \frac{1}{16 \pi^2} \biggl\{- \lambda \overline A_0(W_{hh}) - 4 \lambda^2 \hat h^2  \overline B_0(0,W_{hh}, 0)  + 2  y^2   \overline A_0(m_F^2)  \biggr\} \nn
&+ \frac{1}{16 \pi^2} \left[    \frac{6 \lambda^2 \hat h^2}{W_{h h}}   \overline A_0(W_{hh}) -4 \sqrt 2 \frac{\lambda \hat h}{W_{hh}} y  m_F \overline A_0(m_F^2) \right] + \frac{ 2 \lambda \hat h \mathcal{J}_h}{W_{h h} }  + \mathcal{O}\left(z^2\right)\,,
\label{16.21}
\end{align}
where the second line is from the tadpole graphs in Fig.~\ref{fig:twopoint}. 

At tree-level, the minimum of the potential is at $\hat h = v$. At one-loop, the minimum is at $\hat h = v + \Delta v$, where $\Delta v$ is formally of one-loop order, so that $\hat h^2-v^2 \approx 2 v \Delta v$.  To one-loop order, we can use the tree-level values $\hat h = v$, $W_{hh}=2\lambda v^2$, $W_{\varphi\varphi}=0$, and $m_F=y v/\sqrt{2}$ in the one-loop term $\Sigma_{\varphi \varphi}(p^2)$,
\begin{align}
\Sigma_{\varphi \varphi}(p^2) &= \frac{1}{16 \pi^2} \left\{2  \lambda \overline A_0(2\lambda v^2) - 4 \lambda^2 v^2  \overline B_0(0,2 \lambda v^2, 0)
  \right\} + \frac{\mathcal{J}_h}{v}\,.
\label{16.16}
\end{align}
Using $A_0(m^2)=m^2 B_0(0,m^2,0)$, to one-loop order
\begin{align}
\Sigma_{\varphi \varphi}(p^2=0) &= \frac{\mathcal{J}_h}{v}\,.
\label{16.18}
\end{align}
The tadpole graphs  in the second line exactly cancel the terms in the first line. Including the tree-level term, Eq.~\eqref{16.9} reduces to
\begin{align}
i \Gamma_{\varphi \varphi} (0) = W_{\varphi \varphi} + \Sigma_{\varphi \varphi}(p^2=0) &= \frac{\hat \sigma}{v} + \frac{\mathcal{J}_h}{v} = \frac{J}{v}.
\label{16.18}
\end{align}
Consequently, at the minimum of the potential where $J=0$, at one-loop order
\begin{align}
\Gamma_{\varphi \varphi}(p=0) &=0\,,
\label{16.15}
\end{align}
and the Goldstone boson is exactly massless, as guaranteed by the Ward identity. Near the minimum of the potential with $\hat h^2-v^2 \sim \mathcal{O}(z^2)$,
$ \Gamma_{\varphi \varphi} \sim \mathcal{O}(z^2)$, and $\varphi$ is light.
When the heavy $h_q$ and $\psi$ fields are integrated out, the EFT contains a massless Goldstone boson  which is derivatively coupled at the minimum of the potential. Near the minimum, its mass-squared and couplings are $\mathcal{O}(z^2)$. As a result loop diagrams in the low-energy theory are of order $W_{\varphi\varphi}  \ln W_{\varphi\varphi}  \sim z^2 \ln z^2$, rather than $\ln W_{\varphi\varphi}  \sim \ln z^2$, and so do not have infrared divergences.

We can now compare with the earlier work in Refs.~\cite{Martin:2013gka,Martin:2014bca,Elias-Miro:2014pca}.
The fermion loop diagram Fig.~\ref{fig:twopoint}(c) gives a contribution
\begin{align}
\Sigma_{\varphi \varphi}(p^2) &= - \frac{y^2}{8 \pi^2} m_F^2  \ln \frac{m_F^2}{\mu^2}
\label{16.11}
\end{align}
to the $\varphi$ mass. Fig.~\ref{fig:gb} at $\ell$ loop order --- i.e.\ with $\ell-1$ fermion loops ---  is of the form (see Ref.~\cite[Eq.~(5.1)]{Martin:2013gka})
\begin{align}
V \sim (y^2)^{\ell -1} \left(m_F^2\right)^{\ell-1} W_{\varphi \varphi}^{3-\ell} \left[ \ln \frac{W_{\varphi\varphi}}{m_F^2} + \ldots \right]\,,
\label{16.20}
\end{align}
and leads to an infrared divergence as $W_{\varphi \varphi} \to 0$ for $l \ge 3$ loops.  This is the Goldstone boson infrared divergence found in Ref.~\cite{Martin:2013gka}.

The fermion bubble Eq.~\eqref{16.11}  is of order the heavy mass scale, and order $1$ in the $z$ power counting, whereas the Goldstone boson mass $W_{\varphi \varphi}  \sim z^2$. One cannot expand Fig.~\ref{fig:gb} in a power series in the fermion bubbles, i.e.\ in the $m_F^2/W_{\varphi \varphi}$ expansion of Eq.~\eqref{16.20}, as this is an expansion in a parameter $1/z^2$ much larger than unity.
Refs.~\cite{Martin:2014bca,Elias-Miro:2014pca} included Fig.~\ref{fig:twopoint}(c) into the $\varphi$ propagator in order to avoid this expansion. The EFT method does this automatically, since the one-loop matching condition to the low-energy theory includes these diagrams in the $\varphi$ mass in the low-energy theory. There are also tadpole graphs Fig.~\ref{fig:twopoint}(e--h) in the two-point function, which have to be included. These cancel the zero momentum part of the Goldstone boson propagator, so that the one-loop correction is order $p^2$. This cancellation must happen, because Goldstone bosons are derivatively coupled. We already saw this cancellation at tree level for the $\widetilde \lambda \varphi_q^4$ coupling of the Goldstone bosons in the $O(N)$ model in Sec.~\ref{sec:on}; there, we found that $\widetilde \lambda$  was order $z^2$ and vanished at the tree-level minimum of the potential. Here $\Gamma_{\varphi \varphi}$ is order $p^2$, and so the Goldstone boson loops give $W_{\varphi\varphi}  \ln W_{\varphi\varphi} $ terms, which are not infrared divergent. Thus, in addition to including $\Sigma_{\varphi \varphi}$ into the $\varphi$ propagator as in Refs.~\cite{Martin:2014bca,Elias-Miro:2014pca}, we find that there is a tadpole contribution to $\Sigma_{\varphi \varphi}$ which exactly cancels the contribution Eq.~\eqref{16.11}, so that there is no order $y^2 m_F^2$ shift in the $\varphi$ mass.

\subsection{Comments on the Standard Model}

The Standard Model is an $O(4)$ scalar theory coupled to fermions and gauge fields. All particle masses are proportional to the VEV of the Higgs field $\VEV{H}$. As a result, the theory is a single-scale theory in that $\VEV{H}$ cancels out in mass ratios. There are still, of course, large mass ratios due to a large hierarchy in the couplings, e.g. the ratio of the electron and top-quark masses which is equal to the ratio of their Yukawa couplings $m_e/m_t = y_e/y_t$. Summing these logs is usually neglected, since the light fermion contribution to the potential is negligible.

The stability analysis of the EW vacuum compares the potential computed at $\VEV{H}_1 \sim v = 246\,\text{GeV}$ with  $\VEV{H}_2 \gg v$. In our method, $V(H)$ is computed in the two cases separately, and each calculation is the same as the computation using the fixed order formula at $\VEV{H}_1$ and $\VEV{H}_2$, with couplings renormalized at $\mu=\VEV{H}_1$ and $\mu=\VEV{H}_2$, respectively. The two sets of couplings are related by RG evolution in the high-energy theory, i.e.\ the full SM. This procedure is exactly what is followed in the literature (for a review, see Ref.~\cite{Sher:1988mj}).

\section{Conclusion}
\label{sec:conclusion}

We have shown how to systematically compute the RG improved \cwp\ using EFT methods, and given several examples. We enforced the constraint of only summing 1PI diagrams using a tadpole condition, which can be matched to the EFT, resulting in tadpole contributions to the EFT Lagrangian and to the \cwp. Our results for the $\LL^2$ terms obtained by integrating the {one-loop} RGE agree with explicit two-loop calculations, and provide a highly non-trivial check of our method. An interesting feature of the method is the source $\CJ$, which is introduced in the high-energy theory, but whose value is determined in the low-energy theory. We have also shown how to efficiently compute the RGE in certain cases, {including in the presence of higher-dimension operators,} and shown that shift invariance provides strong constraints on the RGE. Furthermore, we have demonstrated that the EFT formalism in theories with Goldstone bosons automatically produces a low-energy theory free of infrared divergences, and so does not suffer from the Goldstone boson divergence problem of Ref.~\cite{Martin:2013gka}.

The method used in this paper is not restricted to the \cwp, but also applies to the effective action $\Gamma[\hat \varphi]$, and can be used to compute the RG improved
$\Gamma[\hat \varphi]$ in a derivative expansion.

The examples in this paper do not have gauge fields to avoid more complicated $\beta$-function calculations, but this method applies equally well to the gauge case. There are subtleties in interpreting the \cwp\ in the gauge case, which are explained in detail in Refs.~\cite{Andreassen:2014eha,Andreassen:2014gha}.

\subsection*{Acknowledgements}

The work of AM is supported in part by  DOE grant DE-SC0009919, and the work of EN is supported by DOE grant DE-SC0020421. We would like to thank Thomas Dumitrescu for useful discussions.

\begin{appendix}

\section{Examples of $\beta$ functions}
\label{sec:beta}

In this appendix, we
 give useful examples of $\beta$-function evolution for scalar theories using Eq.~\eqref{2.6}. These theories have no scalar field anomalous dimensions at one loop.
 
  Consider a potential $V(\phi^2,\chi^2)$ where
$\chi^2= \chi_i \chi_i$ is a sum over $N_\chi$ components of a real scalar field $\chi$, and $\phi^2=\phi_a \phi_a$ is a sum over $N_\phi$ components of a real scalar field $\phi$. The $W$ matrix of second derivatives (as defined in Eq.~\eqref{eq:w}) is
\begin{align}
\begin{split}
W &= \begin{bmatrix} \frac{\partial^2 V}{\partial \chi_i \chi_j} &  \frac{\partial^2 V}{\partial \chi_i \phi_b} \\
 \frac{\partial^2 V}{\partial \phi_a \chi_j} &  \frac{\partial^2 V}{\partial \phi_a \phi_b} \end{bmatrix} \\
 &= \begin{bmatrix} 2 \delta_{ij} \left( \partial_{\chi^2} V \right) + 4 \chi_i \chi_j \big( \partial^2_{\chi^2} V \big)  & 4 \chi_i  \phi_b \big(\partial_{\chi^2} \partial_{\phi^2} V \big) \\ 4 \phi_a \chi_i \left(\partial_{\chi^2} \partial_{\phi^2} V \right) & 2 \delta_{ab} \left( \partial_{\phi^2} V \right) + 4 \phi_a \phi_b \big( \partial^2_{\phi^2} V \big)   \end{bmatrix}  \,.
 \end{split}
\label{A5.1}
\end{align}
Breaking the indices into radial directions $\propto \chi_i, \phi_a$, and tangential directions, we see that the two directions do not mix. The radial direction matrix is
\begin{align}
W_{\text{rad}} & = \begin{bmatrix} 2 \left( \partial_{\chi^2} V \right) + 4 \chi^2 \big( \partial^2_{\chi^2} V \big)  & 4 \chi \phi \left(\partial_{\chi^2} \partial_{\phi^2} V \right) \\ 4 \chi \phi \left(\partial_{\chi^2} \partial_{\phi^2} V \right) & 2 \left( \partial_{\phi^2} V \right) + 4 \phi^2 \big( \partial^2_{\phi^2} V \big)   \end{bmatrix}  \,,
\label{A5.2}
\end{align}
where $\partial_{\phi^2}$ is the derivative w.r.t.\ $\phi^2$, etc. The tangential direction matrix is diagonal in $\chi$-$\phi$ space, and given by
\begin{align}
 \left[ W_{\chi \chi} \right]_{\text{tan}} = 2 \left( \partial_{\chi^2} V \right) \, ,\qquad
\left[ W_{\phi \phi}  \right]_{\text{tan}} = 2 \left( \partial_{\phi^2} V \right)  \,,
\label{A5.3}
\end{align}
with $N_\chi-1$ and $N_\phi-1$ diagonal entries, respectively. Using Eq.~\eqref{2.6} gives the RG equation $\rd V/\rd t = \tr W^2/2$ (where recall we have defined $t\equiv ({1}/{16 \pi^2}) \ln {\mu}/{\mu_0}$ to absorb factors of $16\pi^2$), which evaluates to
\begin{align} 
\begin{split}
\frac{\rd V}{\rd t} 
&= \frac12 \left[ W_{\phi \phi}^2 + 2 W_{\phi \chi}^2 + W_{\chi \chi}^2  \right]_{\text{rad}}+ \frac12 (N_\chi-1) \left[  W_{\chi \chi}^2  \right]_{\text{tan}} + \frac12 (N_\phi-1) \left[ W_{\phi \phi}^2   \right]_{\text{tan}} \\
&= 2 N_\chi \left( \partial_{\chi^2} V \right)^2+2 N_\phi \left( \partial_{\phi^2} V \right)^2+8  \chi^2 \left( \partial_{\chi^2} V \right) \left( \partial^2_{\chi^2} V \right)  +8  \phi^2 \left( \partial_{\phi^2} V \right) \left( \partial^2_{\phi^2} V \right) \\
& \quad + 8\left(\chi^2\right)^2 \left( \partial^2_{\chi^2} V \right)^2 + 8\left(\phi^2\right)^2 \left( \partial^2_{\phi^2} V \right)^2 +16 \chi^2 \phi^2 \left(\partial_{\chi^2} \partial_{\phi^2} V\right)^2\,.
\end{split}
\label{A5.4}
\end{align}
We now apply this formula to several examples.

\subsection{$O(N_\chi) \times O(N_\phi)$ Model in the Unbroken and Broken phases}\label{app:onon}

For the $O(N_\chi) \times O(N_\phi)$ model with potential
\begin{align}
\begin{split}
V &=  \frac{\lambda_\chi}{24}  (\bm{\chi \cdot \chi})^2  + \frac{\lambda_\phi}{24}(\bm{\phi \cdot \phi})^2 + \frac{\lambda_3}{4} (\bm{\chi \cdot \chi}) (\bm{\phi \cdot \phi}) \\
&+ \frac{ m_\chi^2}{2} (\bm{\chi \cdot \chi}) + \frac{m_\phi^2}{2}  (\bm{\phi \cdot \phi}) +\Lambda\ ,
\end{split}
\label{A5.5}
\end{align}
Eq.~\eqref{A5.4} gives the RGE
\begin{align}
\begin{split}
\dot \lambda_\phi &= \frac13 (N_\phi+8) \lambda_\phi^2 + 3 N_\chi \lambda_3^2 \,, \\
\dot \lambda_\chi &= \frac13 (N_\chi+8) \lambda_\chi^2 + 3 N_\phi \lambda_3^2 \,, \\
\dot \lambda_3 &= \frac13 (N_\chi+2) \lambda_3 \lambda_\chi + \frac13 (N_\phi+2) \lambda_3 \lambda_\phi + 4\lambda_3^2 \,, \\
\dot m^2_\phi &= \frac13 (N_\phi + 2) \lambda_\phi m_\phi^2 + \lambda_3 N_\chi m_\chi^2 \,, \\
\dot m^2_\chi &= \frac13 (N_\chi + 2) \lambda_\chi m_\chi^2 + \lambda_3 N_\phi m_\phi^2 \,, \\
\dot \Lambda &= \frac12 N_\phi m_\phi^4 + \frac 12 N_\chi m_\chi^4 \,.
\end{split}
\label{A5.6}
\end{align}

In the broken phase we take the potential to be
\begin{align}
\begin{split}
V &=  \frac{\lambda_\chi}{24}  \left(\bm{\chi \cdot \chi} -v_\chi^2 \right)^2 + \frac{\lambda_\phi}{24}  \left( \bm{\phi \cdot \phi}  -v_\phi^2 \right)^2 \\
&+ \frac{\lambda_3}{4}   \left( \bm{\chi \cdot \chi}  -v_\chi^2 \right)  \left( \bm{\phi \cdot \phi} -v_\phi^2 \right) +\Lambda_{\text{broken}}\ ,
\end{split}
\label{A5.15}
\end{align}
and compute the $\beta$-functions from Eq.~\eqref{A5.6} by changing variables,
\begin{align}
\begin{split}
m^2_\phi &=-\frac16 \lambda_\phi v_\phi^2  -\frac12 \lambda_3 v_\chi^2 \,, \\
m^2_\chi &=-\frac16 \lambda_\chi v_\chi^2  -\frac12 \lambda_3 v_\phi^2 \,, \\
\Lambda &= \Lambda_{\text{broken}} + \frac1{24} \lambda_\chi v_\chi^4 + \frac1{24} \lambda_\phi v_\phi^4 + \frac14 \lambda_3 v_\chi^2 v_\phi^2\,,
\end{split}
\label{A5.16}
\end{align}
and thus obtain {Eq.~\eqref{A5.6} for $\dot \lambda_\phi$, $\dot \lambda_\chi$, $\dot \lambda_3$ and}
\begin{align}
\dot  v_\chi^2 &= \frac{1}{\lambda_\chi \lambda_\phi - 9 \lambda_3^2}\biggl\{ \left[36 \lambda_3^3  -2 \lambda_\phi \lambda_\chi^2 +  6 \lambda_3^2 \lambda_\chi \right]v_\chi^2  + \left[  4 \lambda_3 \lambda_\phi^2 -12 \lambda_3^2 \lambda_\phi\right] v_\phi^2 \biggr\} \,, \nn
\dot  v_\phi^2 &= \frac{1}{\lambda_\chi \lambda_\phi - 9 \lambda_3^2}\biggl\{ \left[36 \lambda_3^3  -2 \lambda_\chi \lambda_\phi^2 +  6 \lambda_3^2 \lambda_\phi \right]v_\phi^2  + \left[  4 \lambda_3 \lambda_\chi^2 -12 \lambda_3^2 \lambda_\chi\right] v_\chi^2 \biggr\} \,, \nn
\dot \Lambda_{\text{broken}} &= \frac1{18} \lambda_\chi v_\chi^2 + \frac1{18} \lambda_\phi v_\phi^2 + \lambda_3 v_\chi v_\phi\,.
\end{align}
Note that $v_\chi$ and $v_\phi$ mix under the RGE.

We will need the RGE for the case of the $O(N)$ theory, which are given by dropping the $\chi$ terms (taking $\lambda_\chi = m_\chi = \lambda_3=0$),
\begin{align}
\dot \lambda_\phi &= \frac13 (N_\phi+8) \lambda_\phi^2 \,,\qquad
\dot m^2_\phi = \frac13 (N_\phi + 2) \lambda_\phi m_\phi^2 \,,  \qquad
\dot \Lambda = \frac12 N_\phi m_\phi^4 \,,
\label{A5.19}
\end{align}
or in the broken phase where $m_\phi^2 = - \lambda_\phi v^2/6$, 
\begin{align}
\dot \lambda_\phi &=\frac13 (N_\phi+8) \lambda_\phi^2  \,,  \qquad
\dot v_\phi^2 = -2  \lambda_\phi v_\phi^2 \,, \qquad
\dot \Lambda_{\text{broken}} = \frac1{18}  \lambda_\phi^2 v_\phi^4 \,.
\label{A5.8b}
\end{align}

The solution of the RGE Eq.~\eqref{A5.19} is
\begin{align}
\begin{split}
\lambda_\phi(\mu) &= \lambda_\phi(\mu_0) \eta^{-1} \,, \\
m^2_\phi(\mu) &= m^2_\phi(\mu_0) \eta^{-\frac{N+2}{N+8}} \,, \\
\Lambda(\mu) &= \Lambda(\mu_0) + \frac{m^4_\phi(\mu_0)}{\lambda_\phi(\mu_0)} \frac{3N}{2(4-N)}\left[ 1 -   \eta^{\frac{4-N}{N+8}} \right]  \,,
\end{split}
\label{A5.18}
\end{align}
given in terms of
	\begin{align}
	t &= \frac{1}{16 \pi^2} \ln \frac{\mu}{\mu_0}\,, \qquad
\eta = 1- \frac13 (N+8) \lambda_\phi(\mu_0) \ln \frac{\mu}{\mu_0} \,.
	\end{align}
For $N=4$ (where the denominator in Eq.~\eqref{A5.19} goes to zero), the solution is
\begin{align}
\begin{split}
\lambda_\phi(\mu) &= \lambda_\phi(\mu_0) \eta^{-1} \,, \\
m^2_\phi(\mu) &= m^2_\phi(\mu_0) \eta^{-1/2} \,, \\
\Lambda(\mu) &= \Lambda(\mu_0) - \frac{m^4_\phi(\mu_0)}{2 \lambda_\phi(\mu_0)} \ln \eta  \,.
\end{split}
\label{A5.18a}
\end{align}

\subsection{$O(N)$ Model through Dimension 6}\label{sec:smeft}

For the $O(N)$ model with only the field $\bm{\phi}$, one can use Eq.~\eqref{A5.4} dropping the $\chi$ terms. The solution through dimension 4 was given above. Including a dimension 6 term,\footnote{This formalism works even for terms with dimension greater than four; there was no restriction to renormalizable terms in the previous section.}
\begin{align}
V &=   \frac{m^2}{2} (\bm{\phi \cdot \phi}) + \frac{\lambda}{24}  (\bm{\phi \cdot \phi})^2 - \frac{c_6}{8}  (\bm{\phi \cdot \phi})^3+\Lambda\,,
\label{A5.7}
\end{align}
the RGE are
\begin{align}
\dot c_6 &=  (N+14)  \lambda c_6\,, &
\dot \lambda &=-18 (N+4) m^2 c_6 + \frac13 (N+8) \lambda^2  \,,  \nn
\dot m^2 &=\frac13 (N+2) m^2 \lambda \,,&
\dot \Lambda &= \frac12 N m^4 \,.
\label{A5.8}
\end{align}
For $N=4$, these can be compared with the RGE for the Higgs sector of the SMEFT, where $c_6$ is the coefficient of the $(H^\dagger H)^3$ operator. Using the complex Higgs doublet written in terms of real scalar fields
\begin{align}
H &=\frac{1}{\sqrt 2} \begin{bmatrix} \phi_2 + i \phi_1 \\ \phi_4 - i \phi_3
\end{bmatrix} \,,
\label{A5.10}
\end{align}
Eq.~\eqref{A5.7} becomes the potential
\begin{align}
V &= \Lambda + m^2 H^\dagger H  +  \frac16 \lambda \left( H^\dagger H \right)^2 - c_6 \left(H^\dagger H \right)^3 \,,
\label{A5.18b}
\end{align}
and we can make the identification
\begin{align}
c_6 &= C_H^{\text{SMEFT}}  \,, \qquad
\frac16 \lambda = \lambda^{\text{SMEFT}}  \,, \qquad
m^2 = -\frac12   \left(m_H^2 \right)^{\text{SMEFT}} \,,
\label{A5.11}
\end{align}
with SMEFT parameters using the notation in Ref.~\cite{Jenkins:2013zja}. Setting $N=4$ in Eq.~\eqref{A5.8} and using Eq.~\eqref{A5.11},
\begin{align}
\begin{split}
 \dot C_H^{\text{SMEFT}}  &=  108  \lambda C_H^{\text{SMEFT}} \,,  \\
 \dot\lambda^{\text{SMEFT}}  &=  12 \left(m_H^2 \right)^{\text{SMEFT}}C_H^{\text{SMEFT}}  + 24 (\lambda^{\text{SMEFT}})^2 \,, \\
 \dot { \left(m_H^2 \right)^{\text{SMEFT}} } &=  12  \lambda^{\text{SMEFT}} \left(m_H^2 \right)^{\text{SMEFT}} \,, \\
\dot \Lambda &= \frac12 \left(m_H^4 \right)^{\text{SMEFT}} \,,
\end{split}
\label{A5.9}
\end{align}
which agrees with Refs.~\cite{Jenkins:2013zja,Jenkins:2013wua,Alonso:2013hga}.
Eq.~\eqref{A5.4} provides a quick way of obtaining the RGE for a scalar theory with arbitrary higher dimension operators.

\subsection{$O(N)$ model with a Singlet}

Finally we consider the case of the $O(N_\phi)$ $\lambda \phi^4$ theory with an additional scalar singlet $\chi$, with potential 
\begin{align}
V &= \Lambda  + \sigma \chi   + \frac{m_\chi^2}{2} \chi^2  + \frac{m_\phi^2}{2} \phi^2 + \frac{\rho_\chi}{6} \chi^3 + \frac{\rho_\phi}{2} \chi \phi^2+ \frac{\lambda_\chi}{24} \chi^4 +\frac{\lambda_3}{4} \chi^2 \phi^2 +  \frac{\lambda_\phi }{4}\left(\phi^2\right)^2 
\label{A5.13}
\end{align}
for which Eq.~\eqref{A5.4} gives the RGE
\begin{align}
\dot \lambda_\phi &= \frac13 (N_\phi+8) \lambda_\phi^2 + 3 \lambda_3^2 \,, &
\dot \lambda_\chi &= 3 \lambda_\chi^2 + 3 N_\phi \lambda_3^2 \,, \nn
\dot \lambda_3 &= \lambda_3 \lambda_\chi + \frac13 (N_\phi+2) \lambda_3 \lambda_\phi + 4\lambda_3^2 \,, \nn
\dot \rho_\phi &=    \lambda_3 \rho_\chi + 4 \lambda_3 \rho_\phi +  \frac13 (N_\phi+2) \lambda_\phi  \rho_\phi  \,, &
\dot \rho_\chi &=3 \lambda_\chi \rho_\chi +  3 \lambda_3 N_\phi \rho_\phi  \,, \nn
\dot m^2_\phi &=  2 \rho_\phi^2 + \lambda_3 m_\chi^2 + \frac13 (N_\phi+2) \lambda_\phi m_\phi^2 \,, &
\dot m^2_\chi &= \rho_\chi^2 + \lambda_\chi m_\chi^2 + N_\phi \rho_\phi^2 + N_\phi \lambda_3 m_\phi^2 \,, \nn
\dot \sigma &= \rho_\chi m_\chi^2 + N_\phi \rho_\phi m_\phi^2 ,, &
\dot \Lambda &= \frac12 N_\phi m_\phi^4 + \frac 12  m_\chi^4\,.
\label{A5.14}
\end{align}

\section{RGE Solution for the Scalar Theory}\label{app:soln}

The solution of the RGE for a scalar theory with the potential
\begin{align}
V(\varphi) &= \Lambda + \sigma \varphi + \frac{m^2}{2} \varphi^2 + \frac{ \rho}{6} \varphi^3 + \frac{\lambda}{24}  \varphi^4 
\end{align}
is used repeatedly in this paper.
The RGE equations are a special case of Eq.~\eqref{2.8}, 
\begin{align}
\gamma_\varphi &= 0 \,, &
\beta_\Lambda &= \frac12 m^4\,, &
\beta_\sigma &=  m^2 \rho\, , \nn
\beta_{m^2} & =\lambda   m^2 +  \rho^2\,, &
\beta_{\rho}  &= 3 \lambda   \rho \,, &
\beta_{\lambda} &= 3 \lambda^2 \,,
\label{B.1}
\end{align}
with solution
\begin{align}
\begin{split}
\lambda(\mu) & = \lambda(\mu_0) \ \eta^{-1}  \,, \\
\rho(\mu) &= \rho(\mu_0)\ \eta^{-1} \,,  \\
m^2(\mu)
&=m^2(\mu_0) \left\{ \eta^{-1/3} \left[1- \frac12 \xi \right] +\frac12 \xi  \eta^{-1} \right\} \,,\\
\sigma(\mu) &= \sigma(\mu_0) + \frac{m^2(\mu_0) \rho(\mu_0)}{\lambda(\mu_0)} \left\{ \left(\frac13 \xi - 1 \right) +\frac16 \xi \eta^{-1}  +\left(1-\frac12 \xi \right) \eta^{-1/3} \right\} \,, \\
\Lambda(\mu)   &=  \Lambda(\mu_0)+ \frac{m^4(\mu_0)}{2\lambda(\mu_0)} \biggl\{ \frac13 \left( 3 - 6 \xi +2  \xi^2 \right) -\frac12
   \xi \left(\xi -2 \right) \eta^{-1/3}\\
   &\qquad\qquad\qquad\qquad\qquad\qquad\qquad-\frac{1}{4} \left(\xi -2\right)^2\eta^{1/3} + \frac1{12} \xi^2 \eta^{-1} \biggr\}\,,
   \end{split}
\label{B.2}
\end{align}
where
\begin{align}
t &= \frac{1}{16\pi^2} \ln \frac{\mu}{\mu_0}\,, \qquad
\eta = 1- 3 \lambda(\mu_0) t \,, \qquad
\xi = \frac{ \rho^2(\mu_0) }{ \lambda(\mu_0) m^2(\mu_0)} \,.
\label{B.3}
\end{align}
The expansion of Eq.~\eqref{B.2} in powers of $t$ gives
\begin{align}
\begin{split}
\lambda (t) &= \lambda_0 + 3 \lambda^2_0 t + 9 \lambda^3_0 t^2 + \mathcal{O}(t^3) \,, \\
\rho (t) &= \rho_0 + 3 \lambda_0 \rho_0 t + 9 \lambda^2_0 \rho_0 t^2 + \mathcal{O}(t^3) \,, \\
m^2 (t) &= m_0^2 + \left( \lambda m_0^2 + \rho_0^2 \right)  t + \frac12 \lambda_0 \left(4 \lambda_0 m^2_0 +7 \rho^2_0 \right) t^2 + \mathcal{O}(t^3) \,, \\
\sigma (t) &= \sigma_0 + m^2_0 \rho_0 t + \frac12 \rho_0 \left( 4 \lambda_0 m_0^2 + \rho_0^2 \right) t^2 + \mathcal{O}(t^3) \,, \\
\Lambda (t) &= \Lambda_0 + \frac12 m_0^4  t + \frac12 m_0^2 \left( \lambda_0 m_0^2 + \rho^2_0 \right) t^2 + \mathcal{O}(t^3) \,,
\end{split}
\label{B.4}
\end{align}
where all couplings on the r.h.s.\ are evaluated at $t=0$, i.e.\ $\mu=\mu_0$ (hence the subscripts).

\section{Perturbative Integration of RGE}
\label{app:pert}

To check the leading logarithmic series with the explicit two-loop computation of the Coleman-Weinberg potential, it is useful to have the solution to the RGE as a power series in $t$. Let $\{\lambda(t)\}$ denote the running parameters, with evolution equations
\begin{align}
\frac{\rd}{\rd t} \lambda_i(t) &= \beta_i(\{\lambda(t)\}) \,.
\label{12.1}
\end{align}
The solution to these coupled equations can be written as a series in $t$,
\begin{align}
\lambda_i (t) &= \lambda_i^{(0)} + \lambda_i^{(1)} t + \frac12 \lambda_i^{(2)} t^2 + \frac16 \lambda_i^{(3)} t^3 + \ldots
\label{12.2}
\end{align}
Repeated derivatives of Eq.~\eqref{12.1} give the series coefficients
\begin{align}
\lambda_i^{(1)}  &=  \beta_i  \,, &
\lambda_i^{(2)} &=   \frac{\partial \beta_i}{\partial \lambda_j} \beta_j \,, &
\lambda_i^{(3)} &= 
\frac{\partial \beta_i}{\partial \lambda_j} \frac{\partial \beta_j}{\partial \lambda_k}  \beta_k  +  \frac{\partial^2 \beta_i}{\partial \lambda_j \partial \lambda_k}
\beta_j \beta_k \,,
\label{12.4}
\end{align}
where all quantities are evaluated at the initial point $\{\lambda^{(0)}\}$.

\section{Expanding $\Omega(\Delta)$}\label{app:omega}

The function $\Omega(\Delta)$ that enters the two-loop computation of the Coleman-Weinberg potential is given by \cite{Ford:1991hw}
\begin{align}
\Omega(\Delta) &= \frac{ \sqrt{\Delta(\Delta-4)} }{\Delta + 2 } \int_0^\alpha \ln(2 \cosh x) \ \rd x\ ,\qquad   \cosh \alpha = \frac12 \sqrt{\Delta}\,,
\label{11.1}
\end{align}
for $\Delta \ge 4$, and by
\begin{align}
\Omega(\Delta) &= \frac{ \sqrt{\Delta(4-\Delta)} }{\Delta + 2 } \int_0^\theta \ln(2 \sin x) \ \rd x \ ,\qquad  \sin \theta = \frac12 \sqrt{\Delta} \,,
\label{11.7}
\end{align}
for $\Delta \le 4$, where $\Delta=m_1^2/m_2^2$ is the ratio of two masses.  When this ratio is large, we need the expansion of $\Omega(\Delta)$ in the $\Delta \to \infty$ and $\Delta \to 0$ limits to order $1/\Delta^2$ or $\Delta^2$, respectively, in order to obtain terms of order $m_1^4/m_2^4$.

Let us first expand $\Omega(\Delta)$ in \eqref{11.1} for  $\Delta \to \infty$. The prefactor is expanded
\begin{align}
\frac{ \sqrt{\Delta(\Delta-4)} }{\Delta + 2 } \overset{\Delta\to\infty}{=} 1 - \frac{4}{\Delta} + \frac{6}{\Delta^2} + \ldots\ .
\label{11.2}
\end{align}
Using $2\cosh x = e^x + e^{-x}$, we can rewrite the integral as 
\begin{align}
I &=\int_0^\alpha \ln(2 \cosh x)  \rd x 
= \int_0^\alpha \left[x + \ln \left(1 + e^{-2x} \right)\right]  \rd x 
= \frac{\alpha^2}{2} +  \frac{ \text{Li}_2\left(-e^{-2\alpha}\right)}{2} + \frac{\pi^2}{24}\,.
\label{11.3}
\end{align}
Using $e^{\alpha} + e^{-\alpha} = \sqrt{\Delta}$, we can expand
\begin{align}\bs{
e^{ \alpha } &= \frac12 \sqrt{\Delta }+ \frac12 \sqrt{\Delta-4} = \sqrt{\Delta} \left[1 - \frac{1}{\Delta} - \frac{1}{\Delta^2} + \ldots \right] \,, \\
e^{-\alpha} &= \frac12 \sqrt{\Delta }- \frac12 \sqrt{\Delta-4} = \frac{1}{\sqrt{\Delta}} \left[1 + \frac{1}{\Delta} + \frac{2}{\Delta^2} + \ldots \right] \,, \\
\Rightarrow e^{-2\alpha} &=  \frac{1}{{\Delta}} \left[1 + \frac{2}{\Delta} + \frac{5}{\Delta^2} + \ldots \right]\,. } \label{expanding}
\end{align}
Since $\text{Li}_2(z) = \sum_{k=1}^\infty \frac{z^k}{k^2}$ and the argument $|z|=e^{-2\alpha}\ll 1$,
\begin{align}
 \text{Li}_2(-e^{-2\alpha}) &= -e^{-2\alpha} + \frac{e^{-4\alpha}}{4} - \frac{e^{-6\alpha}}{9} + \ldots = -\frac{1}{\Delta} - \frac{7}{4 \Delta^2} + \dots\ .
 \end{align}
 We also have from \eqref{expanding} that
 \begin{align}
\alpha &= \frac12 \ln \Delta - \frac{1}{\Delta} - \frac{3}{2\Delta^2} + \ldots \ .
\label{11.5}
\end{align}
Using these expressions, the expansion of the integral defined in \eqref{11.3} is 
\begin{align}
I &= \frac18 \ln^2 \Delta + \frac{\pi^2}{24} - \frac{1}{2 \Delta} \left( \ln \Delta + 1\right)-  \frac{3}{8\Delta^2} \left(2 \ln \Delta + 1\right) + \ldots \ .
\end{align}
Together with \eqref{11.2}, we obtain $\Omega(\Delta)$ to the desired order,
\begin{align}\bs{
\Omega(\Delta) \overset{\Delta \to \infty}{=} &\frac18 \ln^2 \Delta + \frac{\pi^2}{24} - \frac{1}{6 \Delta} \left(3 \ln^2 \Delta+ 3   \ln \Delta + \pi^2 + 3 \right) \\
&+
 \frac{1}{8\Delta^2} \left(6 \ln^2 \Delta + 10 \ln \Delta +2 \pi^2 +1 3 \right) + \ldots\ . }
\label{11.6}
\end{align}
{The first two terms agree with Ref.~\cite{Ford:1991hw}. We have extended the expansion to higher order in $1/\Delta$.}

The other limit $\Delta\to0$ proceeds in a similar fashion. The prefactor is expanded 
\begin{align}
\frac{ \sqrt{\Delta(4-\Delta)} }{\Delta + 2 } \to \sqrt{\Delta} \left[ 1 - \frac{5\Delta}{8} + \frac{39\Delta^2}{108}+ \ldots \right]\,,
\label{11.8}
\end{align}
and the integral becomes
\begin{align}
I &=\int_0^\theta \ln(2 \sin x) \ \rd x  
= \theta \ln \theta + \left(\ln 2-1\right) \theta -\frac{\theta^3}{18} + \ldots\ .
\label{11.9}
\end{align}
$\theta$ can be expanded
\begin{align}
\theta &= \arcsin \left( \frac{\sqrt{\Delta}}{2}\right) =  \frac12 \sqrt{\Delta} \left[ 1 + \frac{\Delta}{24}+\frac{3 \Delta^2}{640} + \ldots \right]\ .
\label{11.10}
\end{align}
Then, $\Omega(\Delta\to0)$ is given by
\begin{align}
\Omega(\Delta) &\overset{\Delta\to0}{=}  \frac{\Delta}{4} (\ln \Delta -2 ) +\frac{\Delta^2 }{144} \left(44-21 \ln \Delta \right) + \ldots\ .
\label{11.11}
\end{align}
There are no $\Delta^n \ln^2 \Delta$ terms in this limit.

\section{Passarino-Veltman Integrals}\label{app:pv}

We summarize the Passarino-Veltman integrals~\cite{Passarino:1978jh} needed in Sec.~\ref{sec:gb} in the sign conventions of this paper. We  denote the finite parts of the integrals by $\overline A_0$, $\overline B_0$ and $\overline B_1$.

The integrals are
\begin{align}
\mathcal{S} \int \frac{\rd^d k}{(2\pi)^d}\,  \frac{1}{k^2-m^2 + i 0^+} &= i A_0(m^2)\,,
\label{E.2} \\
\mathcal{S} \int \frac{\rd^d k}{(2\pi)^d}\,  \frac{1}{\left[k^2-m_1^2 + i 0^+\right]\left[(k+p)^2-m_2^2 + i 0^+\right]} &= i B_0(p^2,m_1^2,m_2^2)\,,
\label{E.3} \\
\mathcal{S} \int \frac{\rd^d k}{(2\pi)^d}\,  \frac{k_\mu}{\left[k^2-m_1^2 + i 0^+\right]\left[(k+p)^2-m_2^2 + i 0^+\right]} &= i B_1(p^2,m_1^2,m_2^2) p_\mu\,,
\label{E.4} 
\end{align}
where
$\mathcal{S} =(4\pi)^{d/2} e^{\epsilon \gamma_E} \mu^{2 \epsilon}$.
The explicit values are
\begin{align}
A_0(m^2) &= m^2 \left( \frac{1}{\epsilon}- \ln \frac{m^2}{\mu^2} + 1 \right)\,,
\label{E.5}\\
B_0(p^2,m_1^2,m_2^2) &= 
\frac{1}{\epsilon}-\int_0^1{\rm d}x\ \ln \frac{m_1^2x+m_2^2(1-x)-p^2 x (1-x)}{\mu^2}\,,
\label{E.6} \\
2 B_1(p^2,m^2,m^2) &= -B_0(p^2,m^2,m^2) \,.
\label{E.7}
\end{align}

\end{appendix}

\bibliographystyle{JHEP}
\bibliography{refs}

\providecommand{\href}[2]{#2}\begingroup\raggedright\begin{thebibliography}{10}

\bibitem{Coleman:1973jx}
S.~R. Coleman and E.~J. Weinberg, {\it {Radiative Corrections as the Origin of
  Spontaneous Symmetry Breaking}},  {\em Phys. Rev. D} {\bf 7} (1973)
  1888--1910.

\bibitem{Ford:1991hw}
C.~Ford and D.~Jones, {\it {The Effective potential and the differential
  equations method for Feynman integrals}},  {\em Phys. Lett. B} {\bf 274}
  (1992) 409--414. [Erratum: Phys.Lett.B 285, 399 (1992)].

\bibitem{Ford:1992pn}
C.~Ford, I.~Jack, and D.~Jones, {\it {The Standard model effective potential at
  two loops}},  {\em Nucl. Phys. B} {\bf 387} (1992) 373--390,
  [\href{http://arxiv.org/abs/hep-ph/0111190}{{\tt hep-ph/0111190}}]. [Erratum:
  Nucl.Phys.B 504, 551--552 (1997)].

\bibitem{Martin:2001vx}
S.~P. Martin, {\it {Two Loop Effective Potential for a General Renormalizable
  Theory and Softly Broken Supersymmetry}},  {\em Phys. Rev. D} {\bf 65} (2002)
  116003, [\href{http://arxiv.org/abs/hep-ph/0111209}{{\tt hep-ph/0111209}}].

\bibitem{Einhorn:1983fc}
M.~B. Einhorn and D.~Jones, {\it {A new renormalization group approach to
  multiscale problems}},  {\em Nucl. Phys. B} {\bf 230} (1984) 261--272.

\bibitem{Ford:1996hd}
C.~Ford and C.~Wiesendanger, {\it {A Multiscale subtraction scheme and partial
  renormalization group equations in the $O(N)$ symmetric $\phi^4$ theory}},
  {\em Phys. Rev. D} {\bf 55} (1997) 2202--2217,
  [\href{http://arxiv.org/abs/hep-ph/9604392}{{\tt hep-ph/9604392}}].

\bibitem{Ford:1996yc}
C.~Ford and C.~Wiesendanger, {\it {Multiscale renormalization}},  {\em Phys.
  Lett. B} {\bf 398} (1997) 342--346,
  [\href{http://arxiv.org/abs/hep-th/9612193}{{\tt hep-th/9612193}}].

\bibitem{Steele:2014dsa}
T.~Steele, Z.-W. Wang, and D.~McKeon, {\it {Multiscale renormalization group
  methods for effective potentials with multiple scalar fields}},  {\em Phys.
  Rev. D} {\bf 90} (2014), no.~10 105012,
  [\href{http://arxiv.org/abs/1409.3489}{{\tt arXiv:1409.3489}}].

\bibitem{Bando:1992wy}
M.~Bando, T.~Kugo, N.~Maekawa, and H.~Nakano, {\it {Improving the effective
  potential: Multimass scale case}},  {\em Prog. Theor. Phys.} {\bf 90} (1993)
  405--418, [\href{http://arxiv.org/abs/hep-ph/9210229}{{\tt hep-ph/9210229}}].

\bibitem{Casas:1998cf}
J.~Casas, V.~Di~Clemente, and M.~Quiros, {\it {The Effective potential in the
  presence of several mass scales}},  {\em Nucl. Phys. B} {\bf 553} (1999)
  511--530, [\href{http://arxiv.org/abs/hep-ph/9809275}{{\tt hep-ph/9809275}}].

\bibitem{Iso:2018aoa}
S.~Iso and K.~Kawana, {\it {RG-improvement of the effective action with
  multiple mass scales}},  {\em JHEP} {\bf 03} (2018) 165,
  [\href{http://arxiv.org/abs/1801.01731}{{\tt arXiv:1801.01731}}].

\bibitem{Ookane:2019iwq}
H.~Okane, {\it {Construction of a renormalization group improved effective
  potential in a two real scalar system}},  {\em PTEP} {\bf 2019} (2019), no.~4
  043B03, [\href{http://arxiv.org/abs/1901.05200}{{\tt arXiv:1901.05200}}].

\bibitem{Buchbinder:2019bcc}
I.~L. Buchbinder, A.~R. Rodrigues, E.~A. dos Reis, and I.~L. Shapiro, {\it
  {Quantum aspects of Yukawa model with scalar and axial scalar fields in
  curved spacetime}},  {\em Eur. Phys. J. C} {\bf 79} (2019), no.~12 1002,
  [\href{http://arxiv.org/abs/1910.01731}{{\tt arXiv:1910.01731}}].

\bibitem{Ribeiro:2019xgu}
T.~G. Ribeiro and I.~L. Shapiro, {\it {Scalar model of effective field theory
  in curved space}},  {\em JHEP} {\bf 10} (2019) 163,
  [\href{http://arxiv.org/abs/1908.01937}{{\tt arXiv:1908.01937}}].

\bibitem{Ford:1994dt}
C.~Ford, {\it {Multiscale renormalization group improvement of the effective
  potential}},  {\em Phys. Rev. D} {\bf 50} (1994) 7531--7537,
  [\href{http://arxiv.org/abs/hep-th/9404085}{{\tt hep-th/9404085}}].

\bibitem{Nakano:1993jq}
H.~Nakano and Y.~Yoshida, {\it {Improving the effective potential, multimass
  problem and modified mass dependent scheme}},  {\em Phys. Rev. D} {\bf 49}
  (1994) 5393--5407, [\href{http://arxiv.org/abs/hep-ph/9309215}{{\tt
  hep-ph/9309215}}].

\bibitem{Chataignier:2018aud}
L.~Chataignier, T.~Prokopec, M.~G. Schmidt, and B.~Swiezewska, {\it
  {Single-scale Renormalisation Group Improvement of Multi-scale Effective
  Potentials}},  {\em JHEP} {\bf 03} (2018) 014,
  [\href{http://arxiv.org/abs/1801.05258}{{\tt arXiv:1801.05258}}].

\bibitem{Jackiw:1974cv}
R.~Jackiw, {\it {Functional evaluation of the effective potential}},  {\em
  Phys. Rev.} {\bf D9} (1974) 1686.

\bibitem{Kastening:1991gv}
B.~M. Kastening, {\it {Renormalization group improvement of the effective
  potential in massive $\phi^4$ theory}},  {\em Phys. Lett. B} {\bf 283} (1992)
  287--292.

\bibitem{Ford:1992mv}
C.~Ford, D.~R.~T. Jones, P.~W. Stephenson, and M.~B. Einhorn, {\it {The
  Effective potential and the renormalization group}},  {\em Nucl. Phys.} {\bf
  B395} (1993) 17--34, [\href{http://arxiv.org/abs/hep-lat/9210033}{{\tt
  hep-lat/9210033}}].

\bibitem{Manohar:2018aog}
A.~V. Manohar, {\it {Introduction to Effective Field Theories}},  {\em Les
  Houches Lect. Notes} {\bf 108} (2020)
  [\href{http://arxiv.org/abs/1804.05863}{{\tt arXiv:1804.05863}}].

\bibitem{Jenkins:2017dyc}
E.~E. Jenkins, A.~V. Manohar, and P.~Stoffer, {\it {Low-Energy Effective Field
  Theory below the Electroweak Scale: Anomalous Dimensions}},  {\em JHEP} {\bf
  01} (2018) 084, [\href{http://arxiv.org/abs/1711.05270}{{\tt
  arXiv:1711.05270}}].

\bibitem{Jenkins:2017jig}
E.~E. Jenkins, A.~V. Manohar, and P.~Stoffer, {\it {Low-Energy Effective Field
  Theory below the Electroweak Scale: Operators and Matching}},  {\em JHEP}
  {\bf 03} (2018) 016, [\href{http://arxiv.org/abs/1709.04486}{{\tt
  arXiv:1709.04486}}].

\bibitem{Martin:2013gka}
S.~P. Martin, {\it {Three-Loop Standard Model Effective Potential at Leading
  Order in Strong and Top Yukawa Couplings}},  {\em Phys. Rev. D} {\bf 89}
  (2014), no.~1 013003, [\href{http://arxiv.org/abs/1310.7553}{{\tt
  arXiv:1310.7553}}].

\bibitem{Martin:2014bca}
S.~P. Martin, {\it {Taming the Goldstone contributions to the effective
  potential}},  {\em Phys. Rev. D} {\bf 90} (2014), no.~1 016013,
  [\href{http://arxiv.org/abs/1406.2355}{{\tt arXiv:1406.2355}}].

\bibitem{Elias-Miro:2014pca}
J.~Elias-Miro, J.~Espinosa, and T.~Konstandin, {\it {Taming Infrared
  Divergences in the Effective Potential}},  {\em JHEP} {\bf 08} (2014) 034,
  [\href{http://arxiv.org/abs/1406.2652}{{\tt arXiv:1406.2652}}].

\bibitem{Sher:1988mj}
M.~Sher, {\it {Electroweak Higgs Potentials and Vacuum Stability}},  {\em Phys.
  Rept.} {\bf 179} (1989) 273--418.

\bibitem{Andreassen:2014eha}
A.~Andreassen, W.~Frost, and M.~D. Schwartz, {\it {Consistent Use of Effective
  Potentials}},  {\em Phys. Rev. D} {\bf 91} (2015), no.~1 016009,
  [\href{http://arxiv.org/abs/1408.0287}{{\tt arXiv:1408.0287}}].

\bibitem{Andreassen:2014gha}
A.~Andreassen, W.~Frost, and M.~D. Schwartz, {\it {Consistent Use of the
  Standard Model Effective Potential}},  {\em Phys. Rev. Lett.} {\bf 113}
  (2014), no.~24 241801, [\href{http://arxiv.org/abs/1408.0292}{{\tt
  arXiv:1408.0292}}].

\bibitem{Jenkins:2013zja}
E.~E. Jenkins, A.~V. Manohar, and M.~Trott, {\it {Renormalization Group
  Evolution of the Standard Model Dimension Six Operators I: Formalism and
  {$\lambda$} Dependence}},  {\em JHEP} {\bf 10} (2013) 087,
  [\href{http://arxiv.org/abs/1308.2627}{{\tt arXiv:1308.2627}}].

\bibitem{Jenkins:2013wua}
E.~E. Jenkins, A.~V. Manohar, and M.~Trott, {\it {Renormalization Group
  Evolution of the Standard Model Dimension Six Operators II: Yukawa
  Dependence}},  {\em JHEP} {\bf 01} (2014) 035,
  [\href{http://arxiv.org/abs/1310.4838}{{\tt arXiv:1310.4838}}].

\bibitem{Alonso:2013hga}
R.~Alonso, E.~E. Jenkins, A.~V. Manohar, and M.~Trott, {\it {Renormalization
  Group Evolution of the Standard Model Dimension Six Operators III: Gauge
  Coupling Dependence and Phenomenology}},  {\em JHEP} {\bf 04} (2014) 159,
  [\href{http://arxiv.org/abs/1312.2014}{{\tt arXiv:1312.2014}}].

\bibitem{Passarino:1978jh}
G.~Passarino and M.~J.~G. Veltman, {\it {One Loop Corrections for $e^+ e^-$
  Annihilation Into $\mu^+ \mu^-$ in the Weinberg Model}},  {\em Nucl. Phys.}
  {\bf B160} (1979) 151.

\end{thebibliography}\endgroup

\end{document}